\documentclass{aa} 

\usepackage[varg]{txfonts}
\usepackage{epsfig}
\usepackage{natbib}
\usepackage{amsmath}
\usepackage{amssymb}
\bibpunct{(}{)}{;}{a}{}{,} 

\usepackage{graphicx}
\usepackage[citecolor=blue,colorlinks=true,linkcolor=blue,urlcolor=blue]{hyperref}

\usepackage{siunitx}
\usepackage{bm}

\makeatletter
\renewcommand*\aa@pageof{, page \thepage{} of \pageref*{LastPage}}
\makeatother

\begin{document}

\title{A detailed look at the thermal and non-thermal X-ray emission from the Vela supernova remnant with SRG/eROSITA}

\author{Martin G.~F.~Mayer\thanks{\email{mmayer@mpe.mpg.de}}\inst{1} \and Werner Becker\inst{1,2} \and Peter Predehl\inst{1} \and Manami Sasaki\inst{3}}
\institute{Max-Planck Institut f\"ur extraterrestrische Physik, Giessenbachstrasse, 85748 Garching, Germany \and 
Max-Planck Institut f\"ur Radioastronomie, Auf dem H\"ugel 69, 53121 Bonn, Germany \and Dr. Karl Remeis Observatory, Erlangen Centre for Astroparticle Physics, Friedrich-Alexander-Universit\"at Erlangen-N\"urnberg,
Sternwartstrasse 7, 96049 Bamberg, Germany} 

\date{Received XXX /
Accepted YYY }

\abstract
{The Vela supernova remnant (SNR) is one of the most nearby and extended objects in the X-ray sky. It constitutes a unique laboratory for studying the thermal and non-thermal X-ray emission from an evolved SNR and its central plerion at an unprecedented level of detail.}  
{Our goal is the characterization of the hot ejecta and shocked interstellar medium (ISM) associated to the Vela SNR, as well as the synchrotron-emitting relativistic electrons injected into the ambient medium by the central pulsar. To achieve this, we analyze the data set of Vela acquired by {\it SRG}/eROSITA during its first four all-sky surveys.}
{We present and analyze the energy-dependent morphology of Vela using X-ray images extracted in multiple energy bands. A quantitative view of the physical parameters affecting the observed thermal and non-thermal emission is obtained by performing spatially resolved X-ray spectroscopy of over 500 independent regions using multi-component spectral models.}
{Imaging demonstrates that the X-ray emission of the Vela SNR consists of at least three morphologically and energetically distinct components, with shell-like structures dominating below $0.6\,\si{keV}$, radial outward-directed features becoming apparent at medium energies, and the pulsar wind nebula (PWN) dominating the hard emission above $1.4\,\si{keV}$. 
Our spectroscopy reveals a highly structured distribution of X-ray absorption column densities, which intriguingly appears to lack any correlation with optical extinction measurements, possibly due to dust destruction or a clumpy ISM. 
The shock-heated plasma in Vela is found to be comparatively cool, with a median temperature of $0.19\,\si{keV}$, but exhibits several, often ejecta-rich, warmer regions. 
Within the observed ejecta clumps, we find an unexpectedly high concentration of neon and magnesium relative to oxygen, when compared to nucleosynthetic predictions. This includes the bright ``shrapnel D'', in which we can separate shocked ISM in the soft bow-shock from a hot, ejecta-rich clump at its apex, based on the new data.
Finally, we find an extremely extended, smoothly decreasing distribution of synchrotron emission from the PWN, which extends up to three degrees ($14\,\si{pc}$) from the pulsar. The integrated X-ray luminosity of the PWN in the $0.5-8.0\,\si{keV}$ energy band corresponds to $1.5\times10^{-3}$ of the pulsar's present-day spin-down power. The extended PWN emission likely traces the diffusion of a high-energy electron population in an ISM-level magnetic field, which requires the existence of a TeV counterpart powered by inverse Compton radiation. 
}
{}


\titlerunning{{\it SRG}/eROSITA spectro-imaging analysis of Vela}
\maketitle

\section{Introduction}

There are few objects in the X-ray sky as bright and extended as G263.9$-$3.3, the Vela supernova remnant (SNR).
This composite core-collapse SNR was originally discovered as three separate radio structures named Vela X, Y, and Z \citep{Rishbeth58}. In X-rays, it appears as a large bright shell with a diameter of around $8^{\circ}$ with a central pulsar (PSR B0833$-$45, the ``Vela pulsar'') and its associated pulsar wind nebula (PWN), commonly referred to as Vela X.
The distance to the system is quite precisely known to $287^{+19}_{-17}\,\si{pc}$, owing to measurements of optical absorption lines toward Vela and of the pulsar parallax in the radio and the optical bands \citep{Dodson03,Cha99,Caraveo01}.
In contrast, the age of the SNR is not known precisely, and is commonly assumed to be equal to the characteristic spin-down age of the pulsar, around $11\,\si{kyr}$ \citep{Manchester05}. The true age may, however, deviate significantly from this value, and might even be as large as $\sim 30 \,\si{kyr}$, as indicated by the pulsar's distance from the apparent explosion site \citep{Aschenbach95}, and by its very low braking index \citep{Lyne96,Espinoza17}.  
In any case, Vela can likely be considered an evolved SNR, significantly older than, for instance, the overlapping X-ray-bright SNR Puppis A \citep{Winkler88,Mayer20}.

Thanks to its proximity and evolved state, Vela allows us to study the X-ray emission of an SNR at a level of depth and detail that is not attainable for most other SNRs at a larger distance. 
In the soft band, Vela appears as a relatively cool thermally emitting shell with significant substructure \citep{LuAschenbach00}. A revolutionary finding was the discovery of the Vela ``shrapnels'' with ROSAT \citep{Aschenbach95}. These are likely to be the signatures of dense ejecta clumps, produced in an inhomogeneous explosion, overtaking the main blast wave and penetrating the unshocked interstellar medium (ISM). This interpretation is supported by their characteristic bow shocks and apparent trajectories consistent with an origin close to the SNR center, as well as by the enhanced abundances of typical ejecta elements contributing to their X-ray emission. In particular, the shrapnels labeled B and D by \citet{Aschenbach95} seem to be strongly enriched with oxygen, neon, and magnesium \citep{Katsuda05,Yamaguchi09}, whereas shrapnel A and the more recently studied feature G appear to form a bilateral jet-like structure rich in silicon \citep{Tsunemi99,Miyata01,Katsuda06,Garcia17}.

At higher energies ($\gtrsim1.3\,\si{keV}$), a completely different morphology emerges: Apart from the overlapping circular shell of the SNR RX J0852.0$-$4622 \citep{Aschenbach98}, which is most likely unrelated to the Vela SNR, strong non-thermal emission from the PWN is detected. 
The emission appears concentrated on a feature dubbed the ``cocoon'' extending about one degree southward from the pulsar, which was originally interpreted as a pulsar jet \citep{Markwardt95}. However, high-resolution X-ray observations disproved this interpretation. Data from the {\it Chandra} X-ray observatory revealed a complex structure in the immediate surroundings of the pulsar, including an equatorial torus and the actual pulsar jet, which emanates from the pulsar toward the northwest, along its proper motion direction  \citep{Helfand01,Pavlov03}. Recently, observations by the {\it IXPE} mission have shown that the X-ray emission in this region is highly polarized, arguing in favor of a highly ordered magnetic field structure in the region of the torus \citep{Xie22}.  
The question regarding the true nature of the cocoon has not been resolved beyond doubt, but several authors \citep[e.g.,][]{Blondin01,Slane18} have proposed an interesting model in which an asymmetric reverse shock has crushed the PWN and shifted its apparent center away from the pulsar toward the south, in agreement with its observed position and morphology in X-rays.
Evidence for the presence of non-thermal X-ray emission in regions beyond the cocoon has been presented in several studies using pointed observations \citep{Katsuda11,Slane18}, and data from coded-mask instruments at higher X-ray energies \citep{Willmore92,Mattana11}. At the present time, it is however unclear how large the true extent of the Vela PWN in the X-ray regime is.

Emission associated to Vela has been identified and studied across the entire electromagnetic spectrum, from the radio domain \citep[e.g.][]{Duncan96,Frail97,Bock98,Alvarez01} up to very high-energy (VHE) $\gamma$-rays 
\citep{HESS06,Fermi10,HESS12,Grondin13,Tibaldo18,HESS19}. 
The radio band clearly reveals the composite nature of the Vela SNR, with prominent non-thermal emission originating from narrow filaments forming the SNR shell and from Vela X. In the latter region, the population of electrons accelerated in the pulsar wind creates a chaotic extended network of filaments with a flat spectrum \citep{Bock98,Alvarez01}, as typical for PWNe. 
Similarly, at $\gamma$-ray energies $<100\,\si{GeV}$, extended diffuse emission is observed, approximately consistent with the extent of the radio halo \citep{Grondin13,Tibaldo18}. 
In contrast, at TeV energies, primarily the cocoon and its immediate surroundings are visible, 
whereas the SNR shell and a large part of the radio-bright Vela X region are not detected \citep{HESS12, Tibaldo18}. 
These findings may indicate the existence of two separate electron populations in Vela X, where the lower-energy population is responsible for the extended halo emission in the GeV and radio bands, via inverse Compton and synchrotron emission, respectively. A ``younger'', more energetic electron population would accordingly produce the observed TeV and X-ray emission in the cocoon \citep{deJager08}. 

In this work, we present and explore the X-ray data set of the Vela region gathered by the eROSITA telescope \citep{Merloni12, Predehl21} on board the {\it SRG} mission \citep{Sunyaev21} during its first four all-sky surveys.
The observations provide a higher sensitivity and much better spectral resolution (e.g., $\Delta E \sim 80\,\si{eV}$ at $1.5\,\si{keV}$) than available in the ROSAT all-sky survey \citep{Aschenbach93}. Furthermore, the effectively infinite field of view contrasts the comparatively small regions within and around Vela covered by pointings or even mosaics with instruments such as {\it Suzaku} or {\it XMM-Newton} \citep[e.g.,][]{Katsuda05,Katsuda06,Miceli08,Yamaguchi09,Garcia17,Slane18}. 
The ability to resolve spectral emission lines across the entirety of Vela is  crucial  for separating and characterizing the contributions of thermal emission from hot plasma and non-thermal synchrotron emission in imaging and spectroscopy throughout the SNR.

Our paper is organized as follows: after a brief description of data assembly and cleaning procedures (Sect.~\ref{Data}), we perform imaging and spectroscopic analyses of the X-ray emission of the Vela SNR shell, the Vela shrapnels, and the central PWN in Sect.~\ref{Analysis}. The implications of our findings on the distribution and composition of intervening material, the presence of ejecta inside and outside the shell, and the size and multiwavelength properties of Vela X are discussed in Sect.~\ref{Discussion}, and our results summarized in Sect.~\ref{Summary}. 

\begin{figure*}
\centering
\vspace{-0.2cm}
\includegraphics[width=18.0cm]{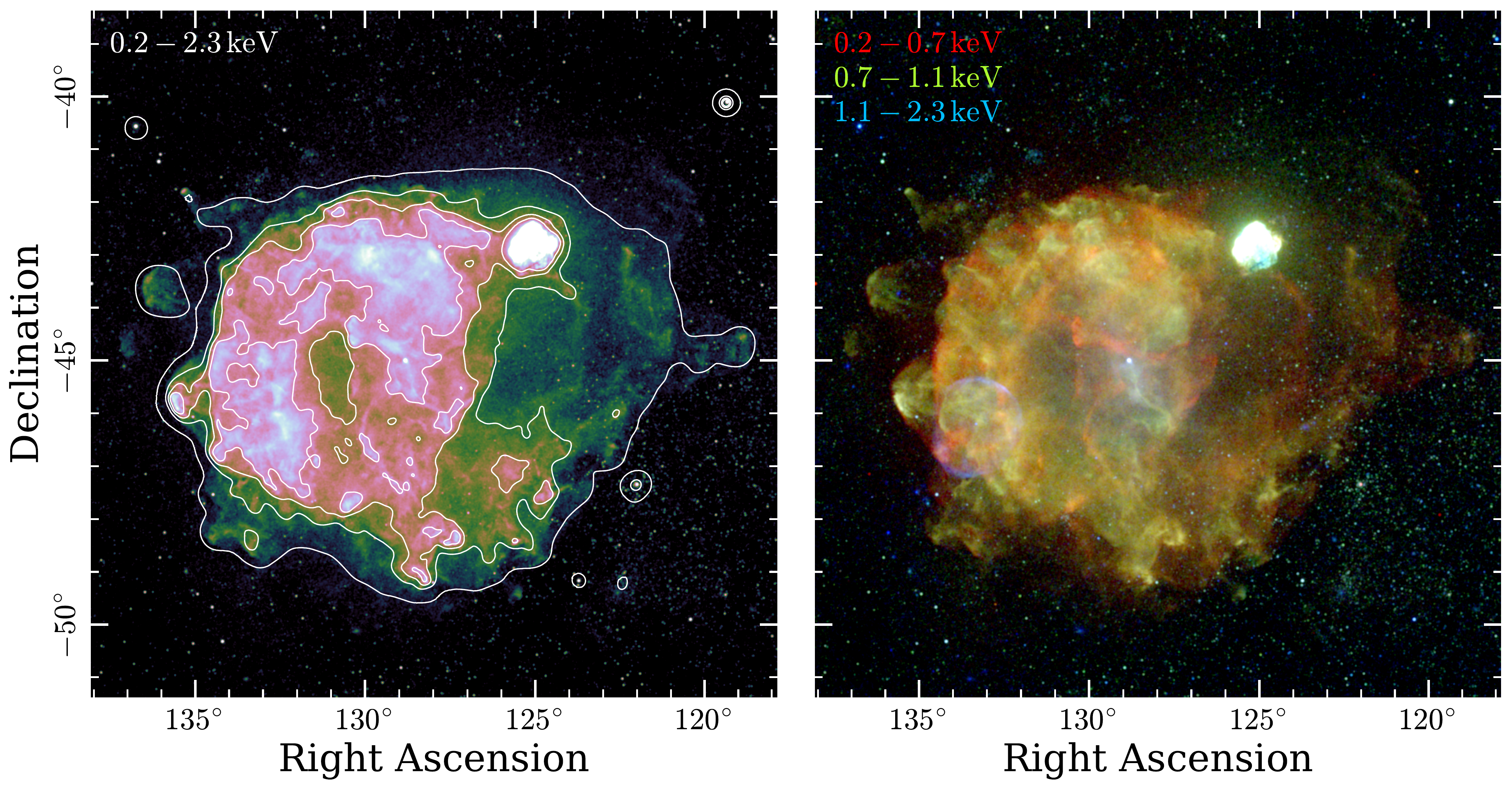} 
\vspace{-0.2cm}
\caption{Exposure-corrected images of the Vela SNR displayed in the broad band (left) and as an RGB false-color image (right) in the energy range $0.2-2.3\,\si{keV}$. Gaussian smoothing with a kernel size of $\sigma=45\arcsec$ was applied to both images, and a logarithmic brightness scale was used. The contours in the left panel trace the heavily smoothed $0.2-2.3\,\si{keV}$ emission at levels of $1.4\times10^{-2}, 4.0\times10^{-2}, 8.0\times10^{-2}, 1.6\times10^{-1}\,\si{ct.s^{-1}.arcmin^{-2}}$. } 
\label{VelaImage}
\end{figure*}

\section{Observations and data preparation\label{Data}}
The region of Vela is observed every six months during the eROSITA all-sky survey (eRASS).\footnote{At the time of writing, science operations of eROSITA are interrupted, in response to the Russian invasion of Ukraine.} eROSITA scans the sky along great circles of constant ecliptic longitude, at a rate of four hours per revolution, accumulating around $40\,\si{s}$ of exposure per scan, while slowly advancing the scanning axis by around one degree per day \citep{Predehl21}. Here, we concentrate on the combined data set of the Vela region taken during the first four surveys (commonly referred to as eRASS:4), between May 2020 and November 2021. The total unvignetted exposure acquired during these four surveys ranges between around $1000$ and $1400\,\si{s}$ over the extent of Vela.

We started our work by merging the data in the {\tt c020} processing version from all eROSITA sky tiles in a $15^{\circ}\times15^{\circ}$ region centered on the Vela pulsar.\footnote{The whole ``German'' eRASS data set will be released incrementally, with the processing {\tt c020} being the most current version of the eRASS:4 data set. Currently, the release of the first all-sky survey is scheduled for September 2023, with the second release, likely including data up to eRASS4, projected for the second quarter of 2024.} 
Compared to the earlier versions {\tt c946}/{\tt c001}, this current eROSITA processing entails strongly suppressed electronic noise at low energies, increased precision of boresight corrections, and an improved handling of different event pattern types (A. Merloni et al., in prep.). 
For the merging of the individual sky tiles, we used the {\tt evtool} task of the latest internal release of the eROSITA science analysis software \citep{Brunner21}, {\tt eSASSusers\_211214}, to combine data from all seven telescope modules (TMs), while using the recommended {\tt flag} and {\tt pattern} filter keywords.\footnote{\url{https://erosita.mpe.mpg.de/edr/DataAnalysis/esasscookbook.html}}

Prior to beginning scientific analysis, we performed a thorough check of the acquired data set for time-variable artifacts visible in imaging, focussing especially on the low- and high-energy ends of the spectral range. While eROSITA is typically not too strongly affected by temporal background variations due to enhanced solar activity \citep{Freyberg20}, we found a few stripes exhibiting an enhanced high-energy count rate oriented along the telescope's scanning direction, typical for such flares. We used the {\tt flaregti} tool, applying a fixed count rate threshold of $1.2\,\si{ct.s^{-1}.deg^{-2}}$ in the $4.0-8.5\,\si{keV}$ range, to filter out those time intervals most strongly affected by enhanced background, while only losing around $1\%$ of all events. 

In the energy range $0.2-0.3\,\si{keV}$, we encountered two further stripe-like artifacts in the south and southeast of Vela, which we found originated in TM4 and during eRASS3 and eRASS4 alone. TM4 was struck by a micrometeorite during the third survey \citep{Freyberg22}, rendering several thousands of pixels bright, hence unusable. Therefore, the likely reason for the occurrence of these low-energy stripes in the data is the variability of a few affected unmasked pixels, which may have temporarily exceeded the on-board threshold, producing spurious low-energy events. We were able to completely filter out the spurious feature in eRASS3 by removing the TM4 good time intervals (GTIs) corresponding to a single scan at around 2021-05-21 18:20:00 (UTC) from the event file. In contrast, the stripe in eRASS4 was found to occur during multiple scans, but to originate only from a single bright row of pixels in TM4 (with the pixel coordinate {\tt RAWX=150}), and was filtered out accordingly. 
The final cleaned event file on which we based our analysis contains around $19$ million entries. 
In the following, all our imaging analysis is based on the combination of events from all seven TMs, while, for spectroscopy, we excluded TMs 5 and 7, leaving a total of around $12$ million events, as their contamination by optical light \citep{Predehl21} makes them less suitable for this purpose.

\begin{figure*}
\centering
\includegraphics[width=0.49\linewidth]{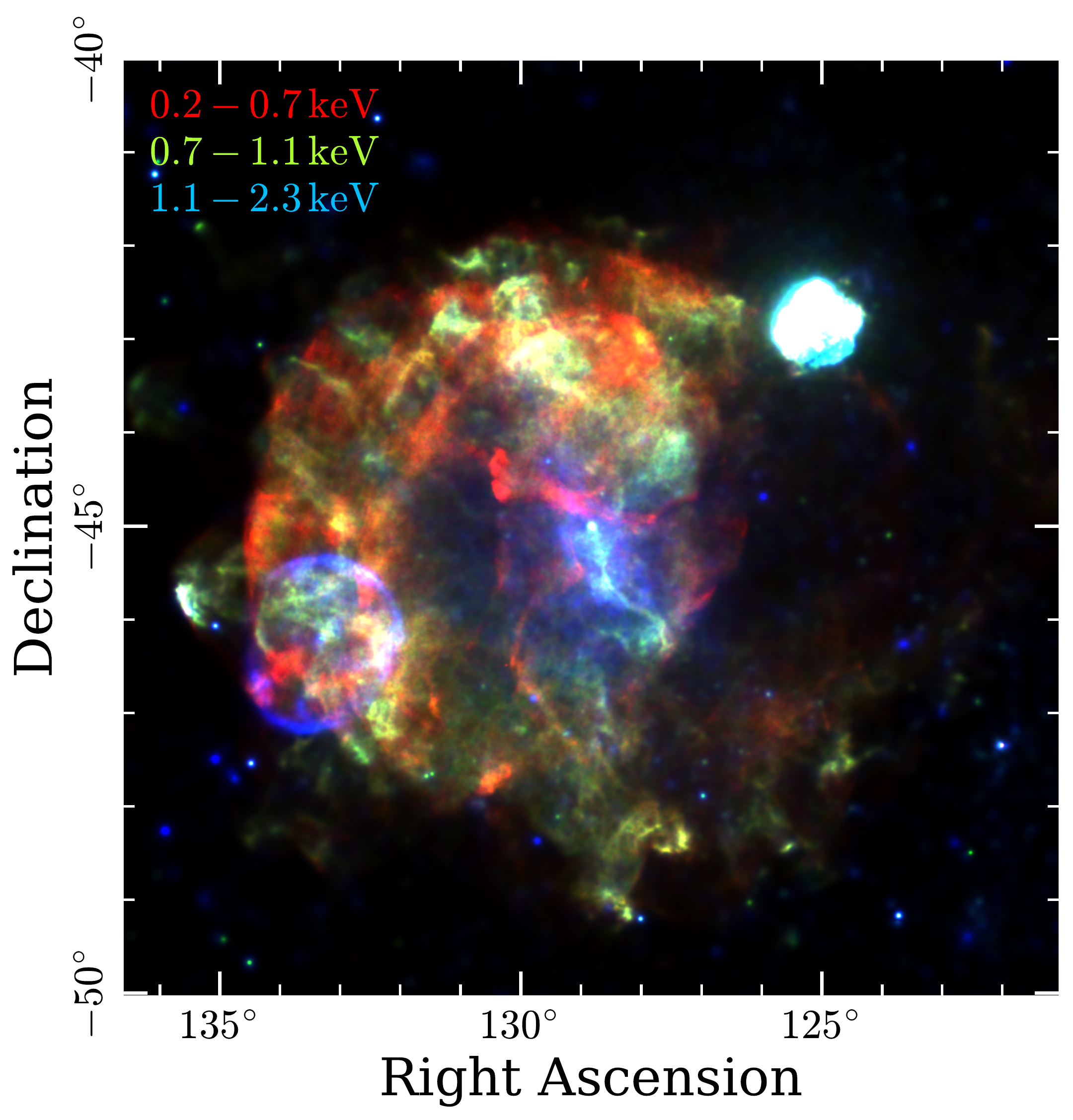}
\includegraphics[width=0.49\linewidth]{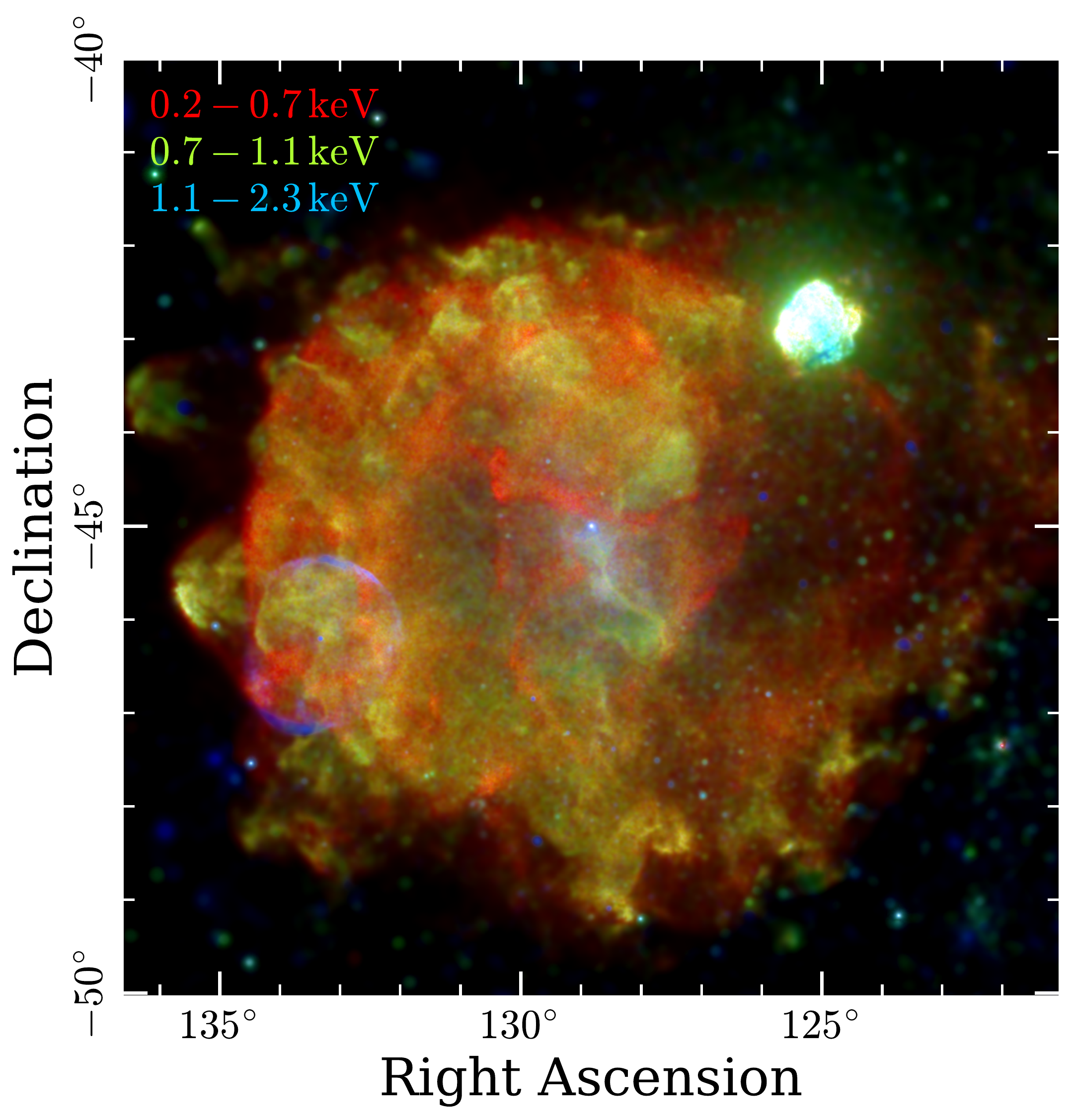} 
\caption{Color-enhanced images of Vela.
The left panel shows the RGB image from Fig.~\ref{VelaImage} in linear brightness scale, with the saturation of each band at the 99.5th percentile of the observed brightness distribution; the right panel keeps the image in a logarithmic brightness scale, but with a quadratic stretch applied to the RGB colors.
}
\label{VelaImage_Linear}
\end{figure*}

\section{Analysis and results \label{Analysis}}
\subsection{Energy-dependent morphology \label{Broadband}}

\begin{figure*}
\centering
\vspace{-0.3cm}
\includegraphics[width=18.0cm]{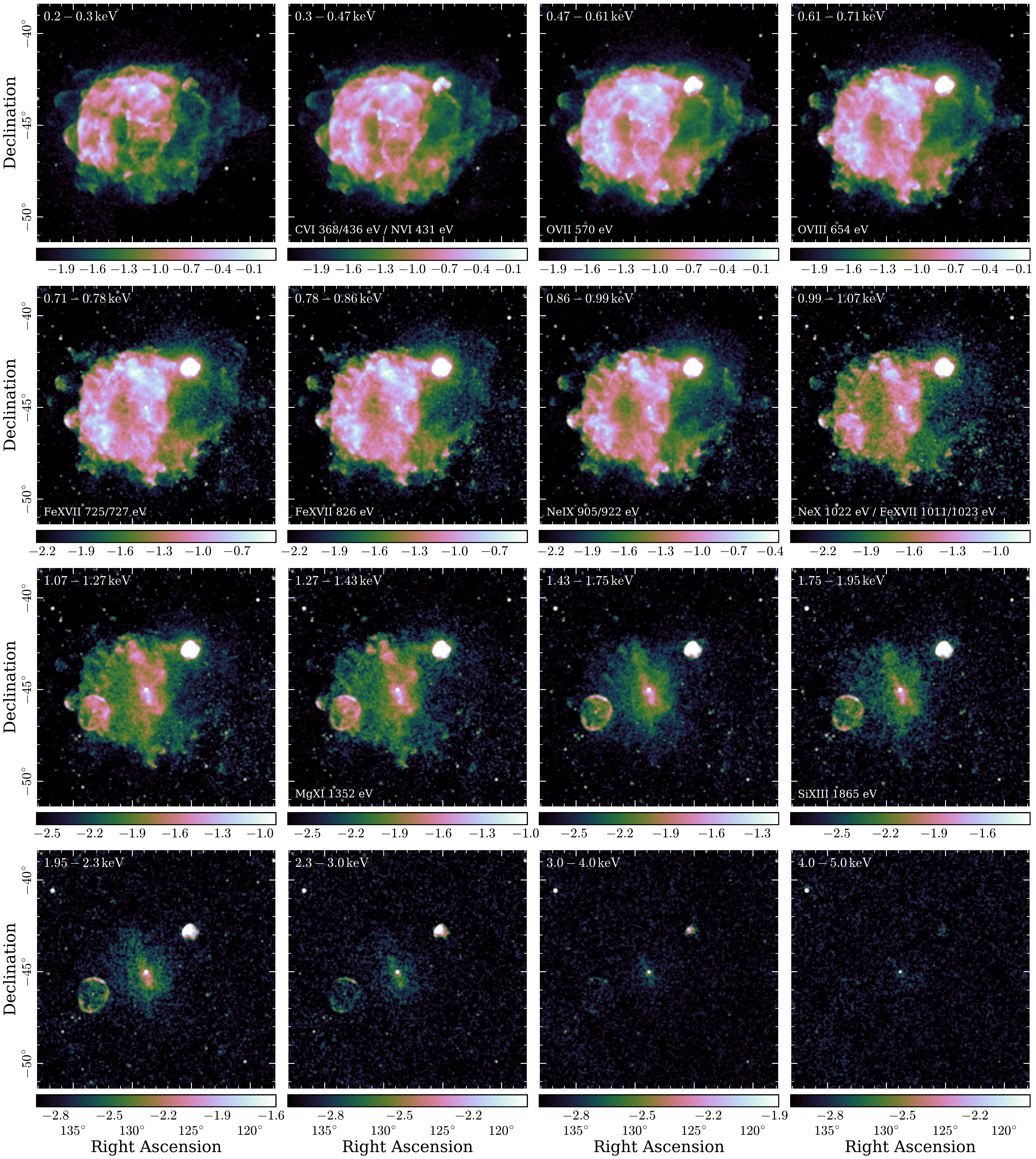} 
\vspace{-0.1cm}
\caption{Exposure-corrected images of Vela in 16 narrow bands of ascending energy. 
The image in each band was smoothed with a $2.5\arcmin$ Gaussian kernel.
The upper right corner of each panel denotes the displayed energy band, whereas, in the lower left corner, we indicate strong spectral lines expected to contribute to the emission in the respective band.
The color bar underneath each panel indicates the logarithmic range of the displayed count rate, specified in units of $\si{ct.s^{-1}.keV^{-1}.arcmin^{-2}}$. 
}
\label{NarrowBand}
\vspace{-0.3cm}
\end{figure*}

While imaging of Vela in the ROSAT era was usually limited to a soft and a hard band, the energy resolution of the eROSITA CCDs \citep{Predehl21, Meidinger21} allows us to study the morphology of the entire SNR across a larger number of independent energy bands for the first time. 
In order to obtain a three-band false-color image of our target, we used the {\tt evtool} and {\tt expmap} tasks to create exposure-corrected images of the Vela region, binned to a pixel size of $30\arcsec$. We used typical energy bands covering the most sensitive range of the eROSITA response, $0.2-0.7$, $0.7-1.1$, and $1.1-2.3\,\si{keV}$, as well as a broad band $0.2-2.3\,\si{keV}$.

Figure \ref{VelaImage} displays the resulting single-band and false-color images of Vela, using a logarithmic brightness scale in order to represent the full dynamic range covered by our data set. In order to emphasize spectral variations in the bright portions of the SNR, this is complemented by Fig.~\ref{VelaImage_Linear}, which displays a false-color image of the same data set after adaptive smoothing of each band to $S/N=30$,\footnote{\url{https://xmm-tools.cosmos.esa.int/external/sas/current/doc/asmooth/index.html}} and using a linear brightness scale.
In addition, we display an image in a logarithmic brightness scale, but with an artificially enhanced color contrast through a quadratic stretch applied to the RGB array, which preserves the visibility of faint features.
The excellent statistics and spectral resolution of our data set permit also the construction of a set of narrow-band images, isolating the contribution of prominent emission lines to the SNR's morphology. In Fig.~\ref{NarrowBand}, we show exposure-corrected images of the Vela region in 16 non-overlapping bands of increasing energy. 

The X-ray emission of Vela exhibits a quite complex morphology, which varies strongly  with energy, with its filled shell being visible across an extent of $10^{\circ} \times 8^{\circ}$. 
A strong horizontal brightness gradient is visible in the emission, with the west generally exhibiting a much lower surface brightness, and only some faint soft filaments tracing what appears to be the western SNR shell. This may be related to a density gradient in the surrounding ISM, which likely exhibits higher densities toward north and east \citep[e.g.][]{Moriguchi01,Sushch11,Slane18}. 
Furthermore, Fig.~\ref{VelaImage_Linear} reveals the presence of multiple morphological components: the soft energy band exhibits a diffuse shell of emission toward north and east, as well as several thick filamentary structures, extending in a tangential direction with respect to the center. In contrast, structures at intermediate energies seem to be preferentially oriented radially, with several ``fingers'' of emission appearing to intersect almost perpendicularly with the soft shell in the northeast.   
As can be seen in Fig.~\ref{NarrowBand}, the transition between the two components seems to occur at around $0.6\,\si{keV}$, between the two bands dominated by line emission from different ionization states of oxygen, \ion{O}{vii} and \ion{O}{viii}, respectively. Therefore, it is very likely that the difference in morphology is at least partly caused by different plasma temperatures in the emitting components.

Apart from thermal emission from the SNR shell, several shrapnels in the east, most prominently those labelled A, B, and D \citep{Aschenbach95} are clearly visible through their characteristic bow shocks, created as the ejecta clumps penetrate into the ISM.   
In addition to the originally established shrapnels, our image reveals several similar structures, preferentially just outside the southern part of the shell, which may also be interpreted as signs of dense clumps of outward-protruding material \citep[see][]{Garcia17}. It is conceivable that these features correspond to ejecta shrapnels at an earlier stage, which do not exhibit an equally visible bow shock.  
A more detailed imaging and spectroscopic study of the known and suspected shrapnels is performed in Sect.~\ref{DetailedFitSection}.

The Vela pulsar is clearly visible as a bright point source across the whole energy range, with the nonthermal emission of its plerion dominating the hard energy band ($> 1.1\,\si{keV}$). The cocoon south of the pulsar constitutes the brightest portion of the extended non-thermal emission, which is spatially coincident with thermal emission seen in the medium energy band \citep{Slane18}. In addition, Fig.~\ref{VelaImage_Linear} clearly reveals the presence of non-thermal emission beyond the cocoon, in particular in the northern direction. While indications for the presence of more extended hard X-ray emission had been found previously in pointed observations \citep[e.g.][]{Katsuda11,Slane18} and weak hints were visible also in the ROSAT hard band \citep[Fig.~1 in][]{Aschenbach98}, our imaging data demonstrates the contiguous nature of the extended hard X-ray emission centered on the pulsar.  
In Fig.~\ref{NarrowBand}, we can see that, above $1.4\,\si{keV}$, the contribution from the presumably thermal morphological components becomes negligible.  
Thus, the emission detected in energy bands without strong expected emission lines, for instance $1.43-1.75\,\si{keV}$ and $1.95-2.30\,\si{keV}$, likely traces the emission from Vela X only. These bands reveal a vast extent of the plerion, with an apparent diameter around five degrees in the north-south direction. 
The fact that the apparent size of Vela X seems to decrease toward higher energies may be interpreted as a sign of synchrotron cooling of the emitting electrons \citep{Tang12}. However, a more quantitative analysis is needed to confirm this hypothesis, in particular since the relative background contribution rises strongly with energy (see Sect.~\ref{DiscVelaX}).   

Multiple physically unrelated objects stand out against the large, soft shell of Vela. Apart from the high-mass X-ray binary Vela X-1 in the northeast, a few star clusters in the southwest, and numerous, mostly stellar, point sources, this includes the extremely bright SNR Puppis A in the northwest, and the hard, almost circular shell of SNR RX J0852.0$-$4622 \citep[``Vela Jr.'',][]{Aschenbach98}, in the southeast. Both Puppis A \citep{Mayer22} and Vela Jr. \citep{Camilloni23} have recently been studied using eROSITA data, and will therefore not be discussed in detail here.
However, the large field of view of our data set allows for two brief observations regarding the periphery of Puppis A: First, there exists a diffuse halo of emission centered on the SNR, visible most clearly in the intermediate energy band ($0.7 - 1.1\,\si{keV}$). Given the considerable amount of intervening material toward Puppis A \citep{Mayer22}, this could be interpreted as a dust-scattering halo. Second, a previously unknown faint arc is visible just outside the northwest rim of Puppis A. At the present time, it is unclear whether this feature is physically associated to Puppis A, to Vela, or to neither of the two.

\begin{figure*}
\centering
\includegraphics[width=0.49\linewidth]{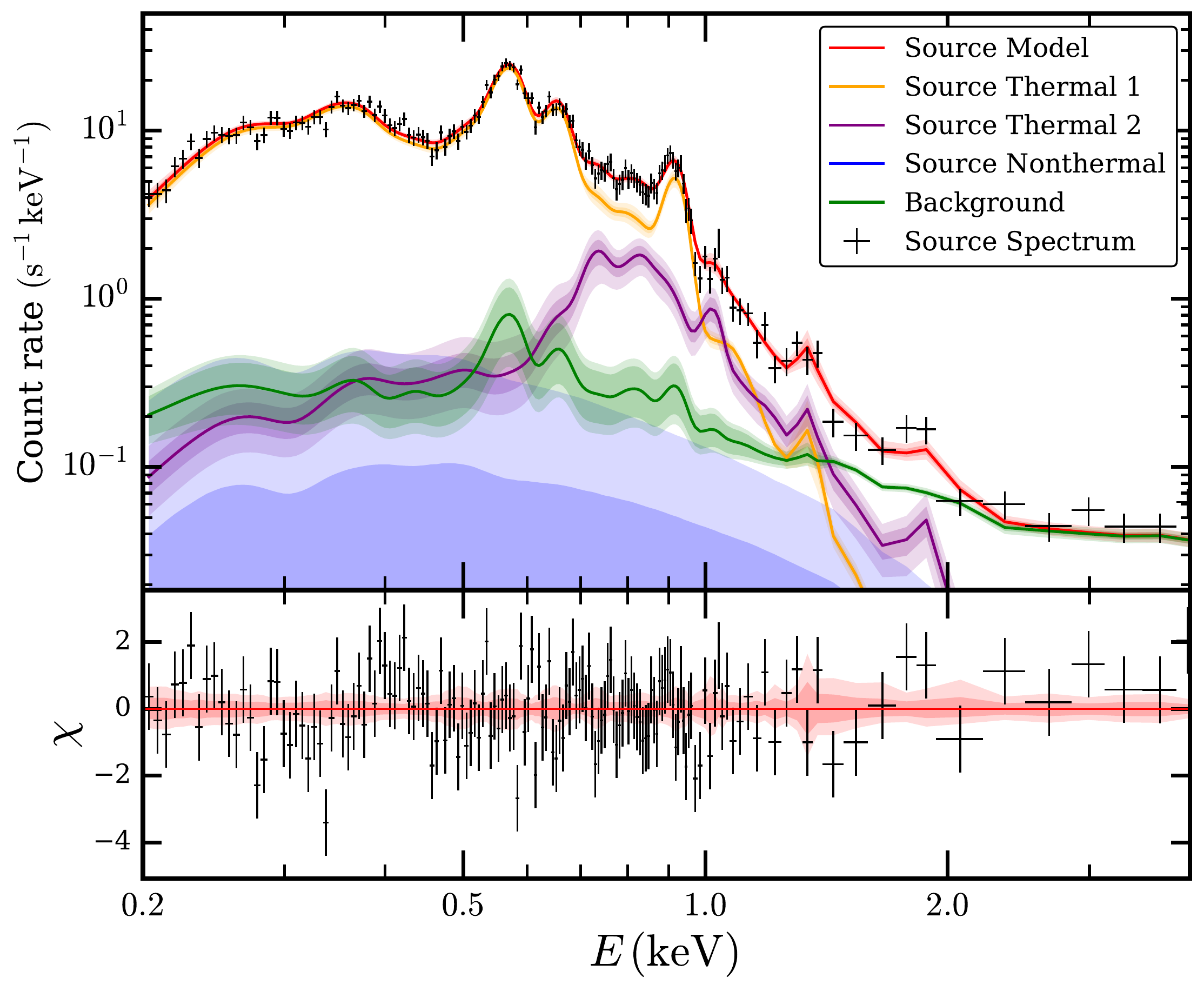} 
\includegraphics[width=0.49\linewidth]{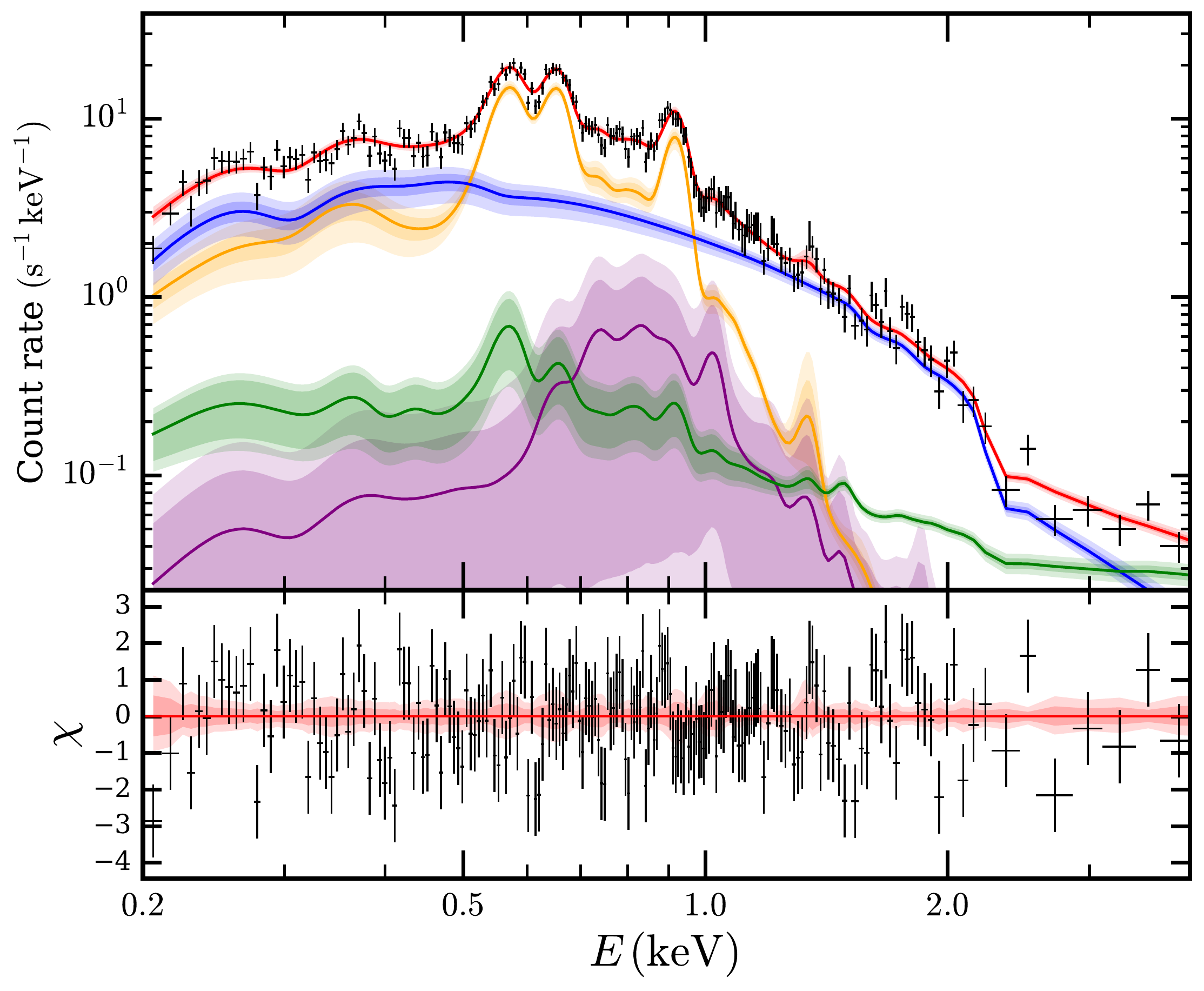} 
\caption{
Fully modelled spectra of two sample regions. The region in the left panel (see Fig.~\ref{Spectroscopy_2TNT}) is dominated only by thermal emission, while the right panel exhibits a strong non-thermal contribution. 
In both panels, the model range predicted by the posterior distribution of the 2TNT spectral model parameters is displayed, with the contributions of background, non-thermal, and the two thermal components indicated in different colors. The background model shown here encompasses both instrumental and astrophysical components, as detailed in the text.  
The dark and light shaded areas correspond to the 68\% and 95\% central intervals of the posterior prediction of the respective model component.
The lower portion of the panels indicate the residuals of the data, normalized by their error bars, with respect to the model range allowed by the posterior. 
}
\label{SampleSpectra}
\end{figure*}

\subsection{Spatially resolved spectroscopy \label{Spectroscopy}}
\begin{figure*}
\centering
\includegraphics[width=18cm]{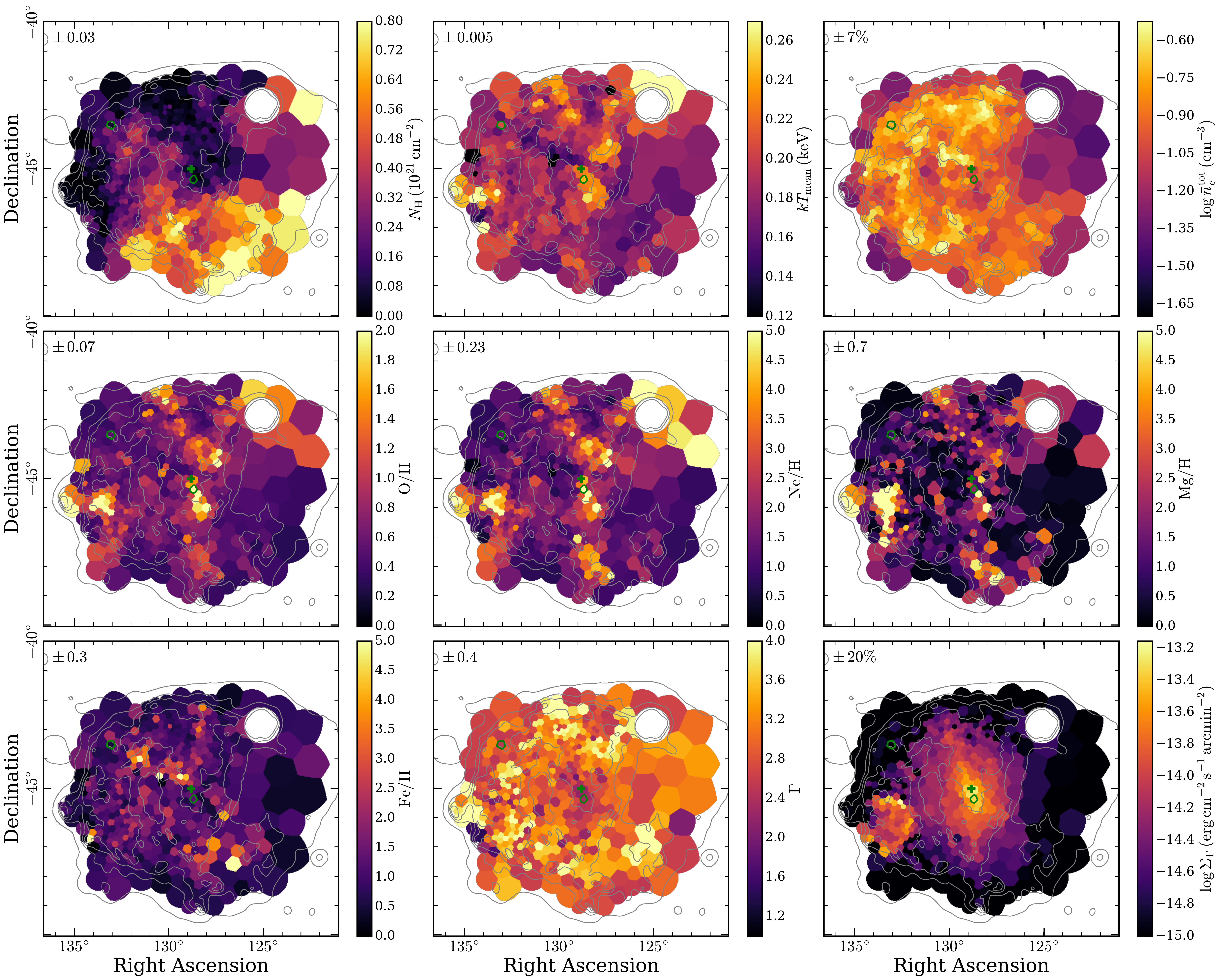} 
\caption{Parameter maps from spatially resolved spectroscopy of Vela, using the 2TNT model. 
We display the distribution of the following physical quantities: absorption column density $N_{\rm H}$, mean temperature $kT_{\rm mean}$, total electron density $\bar{n}_{e}^{\rm tot}$, relative abundances of O, Ne, Mg, Fe, normalized to solar values, of the thermal components, as well as spectral index $\Gamma$ and surface brightness $\Sigma_{\Gamma}$ in the $1.0-5.0\,\si{keV}$ range of the non-thermal component. 
To provide a rough estimate of the typical noise level, the upper left corner of each panel indicates the median uncertainty for the respective parameter across all bins.
The gray contours reflect the broad-band count rate of Vela, and are identical to those displayed in Fig.~\ref{VelaImage}. 
The green polygons mark the two regions whose spectra are displayed in Fig.~\ref{SampleSpectra}, and the green $+$ sign marks the position of the Vela pulsar.
}
\label{Spectroscopy_2TNT}
\end{figure*}

\subsubsection{Binning and modelling}
In this section, we aim to disentangle and characterize the thermal and non-thermal contributions to the observed X-ray emission in different regions of Vela by performing spatially resolved spectroscopy. We followed the general approach of subdividing the emission into spatial bins using Voronoi tessellation \citep{Vorbin} with the modification proposed by \citet{Diehl06}, as described in \citet{Mayer22}. Subsequently, we fitted the spectrum of each region using Xspec \citep[version 12.11.0,][]{Arnaud96}, in order to extract physically meaningful parameters. 

In order to prevent contamination by the emission from bright fore- and background sources, 
we masked out all highly significant point-like sources (i.e., those sources with {\tt DET\_LIKE\_0 $> 50$} and {\tt EXT\_LIKE $= 0$}) in the eRASS1 catalog (A. Merloni et al., in prep.) in the input broadband ($0.2-2.3\,\si{keV}$) image, prior to performing the binning. In addition, the region of Puppis A and a $3\arcmin$ radius around the Vela pulsar were excluded. 
In order to avoid obtaining unnecessarily many bins dominated by background counts alone, we subtracted an estimated background count map from the input signal image. This map was created by multiplying a flat background count rate, measured in an ``empty'' region far from the shell of Vela, with the broadband exposure map.  
We decided to use a target signal-to-noise threshold of $S/N = 100$ for the tessellation, which provides decent spatial resolution and around 500 resulting bins across the SNR, while retaining sufficient statistics to perform meaningful spectral fits in each bin.

Similarly to \citet{Mayer22}, we extracted all spectra from TMs $1-4$ and $6$ using {\tt srctool}, and subsequently fitted them with a physical source model, combined with several background templates. The latter consisted of a model of the instrumental background, determined from filter-wheel-closed data in the {\tt c020} processing \citep{Yeung23}, a fixed absorbed extragalactic X-ray background \citep{XRB}, and a thermal background component. The thermal component was constrained by fitting the 
spectrum of an empty region northeast
of the SNR shell with a model consisting of the above components and a model expressed as {\tt acx+apec+TBabs*(apec+apec)} \citep{ACX,APEC,Wilms00}. The individual components reflect charge exchange emission originating within the heliosphere,\footnote{While solar wind charge exchange is not actually a thermal process, we include it in this ``thermal'' background, as the phenomenology of a line-dominated soft spectrum is quite similar to the truly thermal components.} the unabsorbed contribution of thermal plasma in the local hot bubble, as well as absorbed thermal emission from the Galactic halo and a possible hot component from unresolved stars in the Galactic plane \citep[see][]{Wulf19} or a Galactic ``corona'' \citep{Ponti22}, respectively. 
The best-fit shape of the thermal background was fixed and used as a template in our modelling of the source spectra, where only its global normalization was allowed to vary by up to a factor two. 
Generally, it should be noted that, due to the bright soft thermal emission of Vela, this thermal background component has a relatively minor effect on our spectral modelling, as, in most regions, it is outshone by source emission at all relevant energies. 

As indicated in Sect.~\ref{Broadband}, both thermal and non-thermal components contribute to the emission of Vela. In order to reflect this fact, we modelled the source contribution to each spatial bin using a combination of a thermal plane-parallel shocked plasma \citep{Borkowski01} with non-equilibrium ionization (NEI) and a power law model, with foreground absorption following the T\"ubingen-Boulder model \citep{Wilms00}. In the following, we refer to this thermal-nonthermal model, which is expressed as {\tt TBabs*(vpshock+powerlaw)} in Xspec, as the ``TNT'' model. 
While this model allows for the realistic possibility of underionized plasma close to shock fronts, it entails a severe degeneracy between plasma temperature and ionization age, especially in the presence of a possible nonthermal component contributing to continuum emission. 
We found that a decent alternative is given by a model consisting of two thermal plasmas in collisional ionization equilibrium \citep[CIE;][]{APEC} and one nonthermal component, expressed as {\tt TBabs*(vapec+vapec+powerlaw)}, and labeled ``2TNT'' in the following. A similar two-component model in CIE has been used previously to describe the thermal emission of Vela \citep[e.g.][]{Miceli08}, and is likely a good approximation of the spectrum inside the SNR shell, as for instance \citet{Slane18} did not detect any clear deviations from CIE around Vela X. 
For both models, the abundances of N, O, Ne, Mg, and Fe were allowed to vary, as these are the elements with the highest impact on line emission in the observed spectra, whereas all other abundances were fixed to the reference values of \citet{Wilms00}. For the 2TNT model, the abundances of all elements were tied between the two thermal components, in order to reduce the number of degenerate model parameters. 
Initially, we attempted to also leave the silicon abundance free to vary, in order to quantify the presence of \ion{Si}{xiii} He$\alpha$ emission at around $\sim 1.85\,\si{keV}$. However, we found that, at the relevant temperatures, this parameter would mostly affect L-shell ionization states of Si emitting at energies below $0.3\,\si{keV}$. This resulted in unrealistically high abundances in absorbed regions, which is why we decided to fix Si to the solar value. 

In order to make optimal usage of the available statistics, and in particular to explore parameter degeneracies, we decided to use a Markov chain Monte Carlo (MCMC) approach to constrain the physical parameters for each of our spectra: 
after an initial minimization of the fit statistic within Xspec in the $0.2-8.5\,\si{keV}$ range, we ran the affine-invariant ensemble sampler {\tt emcee} \citep{Foreman13, Goodman10}, using a likelihood given by $\ln \mathcal{L} = -\,\mathcal{C}/2$, where $\mathcal{C}$ is the Cash statistic \citep{Cash} evaluated for a given set of model parameters. The 50 walkers were initialized in a ``ball'' around the best fit, following a multivariate normal distribution, and run for 1000 burn-in and 2000 sampling steps.   

We used a uniform prior on the absorption column density $N_{\rm H}$,  
and logarithmically uniform priors on all other parameters. 
After extensive testing, we found that the power-law spectral index $\Gamma$ requires very careful treatment. This is because, in regions without detectable non-thermal emission, it is essentially an ill-defined quantity, which in our tests would often hit the upper limit under a uniform prior, contributing to the spectrum only at very soft energies. We therefore fixed $\Gamma$ to a value of $2.5$ during our initial Xspec fit, and applied a Gaussian prior centered on this value with a width of $0.5$ during our MCMC run. This approach has negligible impact on regions with bright nonthermal emission, and reduces the contamination of our model at soft energies in purely thermal regions.     

Figure \ref{SampleSpectra} illustrates the motivation for our complex approach by displaying spectra and fitted 2TNT models of two representative regions, one dominated by thermal emission, and one with a strong non-thermal contribution. In both cases, one can observe that model components which are not required to satisfactorily fit the spectrum are unconstrained within a large range, as intended. Furthermore, degeneracies between the thermal and non-thermal model components in certain spectral ranges are reflected in their increased uncertainties.   
After performing our MCMC sampling procedure, for each parameter, we used the median and the $68\%$ central interval of the marginalized posterior as an estimate of its most probable value and approximate error. By repeating this for all spectral extraction regions with the 2TNT and TNT models, we created the physical parameter maps shown in Figs.~\ref{Spectroscopy_2TNT} and \ref{Spectroscopy_TNT}, respectively.

\subsubsection{Distribution of physical parameters across Vela}

\begin{figure}
\centering
\includegraphics[width=1.0\linewidth]{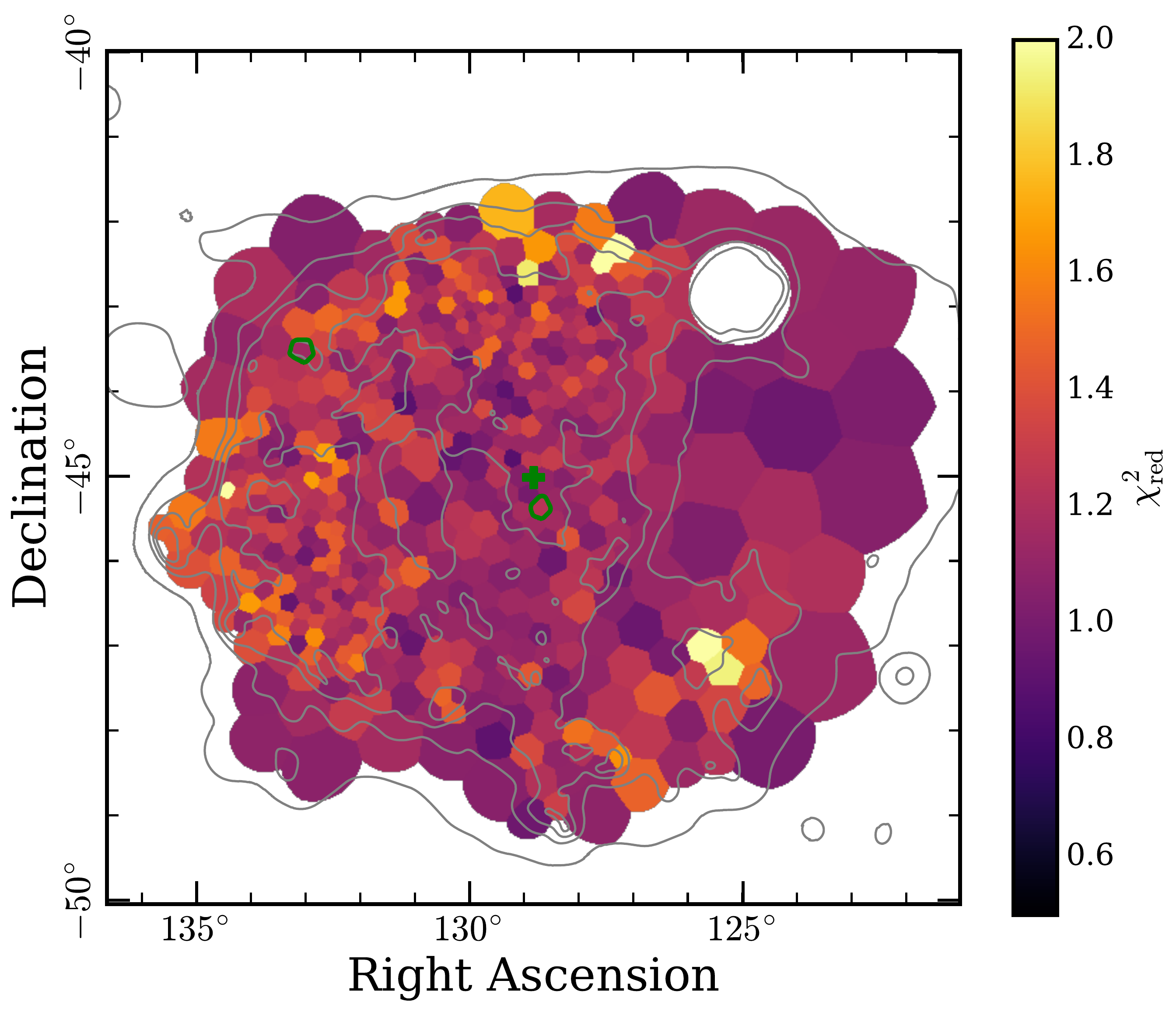} 
\caption{``Reduced $\chi^2$'' statistic computed for each Voronoi bin across the Vela SNR for the fits with the 2TNT model. Contours and markers are as in Fig.~\ref{Spectroscopy_2TNT}.}
\label{RedChi2}
\end{figure}

For the vast majority of Voronoi bins, the 2TNT model was found to yield satisfactory fits from a statistical standpoint. This is illustrated in Fig.~\ref{RedChi2}, which displays rough estimates of the ``reduced $\chi^2$'' statistic of each region, based on observed and median model spectra rebinned to a $5\sigma$ significance in each spectral bin. While this quantity is technically inapplicable to our Bayesian methodology, it may still serve as an estimator of the typical deviation of the observed spectrum from our model. The median value of $\chi^2_{\rm red} \sim 1.2$ and the negligible number of outliers make us confident in the statistical quality of our parameter constraints. 
Generally, we did not find evidence for a significant global difference between the statistical quality of fits of the two used models across the SNR. 
Nonetheless, in the following, we mainly discuss the parameter maps determined via the 2TNT model (Fig.~\ref{Spectroscopy_2TNT}), for the reason that these are somewhat less affected by parameter degeneracies, thus appearing less noisy. Wherever relevant, we point out differences between the global results of the two models.

The map of the equivalent hydrogen column density $N_{\rm H}$ shows that most of the Vela SNR experiences quite little absorption, with the vast majority of regions exhibiting $N_{\rm H} < 8 \times10^{20}\,\si{cm^{-2}}$. This is not surprising given the small distance to Vela of only $290 \,\si{pc}$.
We observe a clear structure in the distribution of $N_{\rm H}$ toward Vela, with the southern part of the SNR experiencing much higher absorption than the north. The north exhibits regions which appear entirely unabsorbed, in the very east and in the center, whereas the far west and the region around a right ascension of $\alpha \sim 133^{\circ}$ display values in the range $2-5\times10^{20}\,\si{cm^{-2}}$. 
On smaller scales, a few filaments and clumps of enhanced absorption seem to be present, for instance around $(\alpha, \delta) \sim (133^{\circ}, -44^{\circ})$.
It is striking that the $N_{\rm H}$ map from the TNT model (Fig.~\ref{Spectroscopy_TNT}) exhibits the same qualitative features as in the 2TNT model, but appears systematically shifted to lower values, by a factor $\sim 1.5$. This is most likely caused by the larger intrinsic soft flux in the 2TNT model from the cooler thermal component, and demonstrates that the quantitative determination of the degree of X-ray absorption is strongly model-dependent. 

We define the mean temperature of our two-component thermal model as the average of the individual component temperatures $kT_{i}$, weighted by their respective emission measures $\mathrm{EM}_{i}$: 
\begin{equation}
    kT_{\rm mean} = \frac{\mathrm{EM}_1 \,kT_1 + \mathrm{EM}_2 \, kT_2}{\mathrm{EM}_1 + \mathrm{EM}_2}.
\end{equation}
The mean plasma temperature spans a quite narrow range across Vela, with virtually all regions in the range $0.15\,\si{keV} < kT_{\rm mean} < 0.25\,\si{keV}$ and a median value around $0.19_{-0.02}^{+0.03}\,\si{keV}$ (errors representing the 68\% central interval of the distribution). 
This quantity is dominated by the temperature of the cold component, as the respective median temperatures are $kT_1 \sim 0.18_{-0.02}^{+0.02}\,\si{keV}$ and $kT_2 \sim 0.60_{-0.15}^{+0.50}\,\si{keV}$ 
for the cold and hot components, respectively.
A few regions stand out in our mean temperature map: we find that a filament of very low-temperature plasma runs from the central pulsar in a northeastern direction, which naturally manifests itself in the extremely soft emission visible from this feature (Fig.~\ref{VelaImage_Linear}). 
In contrast, several coherent structures of comparatively hot plasma are visible. This includes the region of the cocoon, as already indicated by the temperature maps of \citet{Slane18}, as well as blobs of high-temperature material in the direction of shrapnel D, and toward the northern rim. Finally, enhanced temperatures are visible also in the northwestern periphery of the pulsar, apparently pointing in a direction roughly consistent with that of the pulsar jet \citep{Helfand01, Pavlov03}.   

Our proxy for the total electron density is computed as $\bar{n}^{\rm tot}_{e} = \bar{n}_{e,1} + \bar{n}_{e,2}$, where the density estimates of the individual components $\bar{n}_{e,i}$ are calculated from the corresponding emission measures $\mathrm{EM}_{i}$ as in \citet{Mayer22}, assuming a distance of $290\,\si{pc}$ and a shell diameter of $8^{\circ}$.
The distribution of $\bar{n}^{\rm tot}_{e}$ is highly inhomogeneous across Vela, with many thick filaments and clumps standing out, similar to those visible in the soft band in Figs.~\ref{VelaImage_Linear} and \ref{NarrowBand}. The measured density estimates reach up to $0.3\,\si{cm^{-3}}$, with thinner intervening regions down to $0.07\,\si{cm^{-3}}$. The south and especially the west of Vela seem to exhibit a lower average density in their emitting material, consistent with the suspected expansion of these regions into a thinner ISM \citep{Sushch11}.
Generally, while useful for comparing relative densities, it should be emphasized that $\bar{n}^{\rm tot}_{e}$ is computed by assuming a perfectly homogeneous density distribution over a known volume, without accounting for the unknown volume filling factor. This means that the quantitative values given here should be seen as a lower limit to the true characteristic emitting density of the X-ray-luminous material.

The abundance maps of oxygen, neon, and magnesium in Fig.~\ref{Spectroscopy_2TNT} show several prominent peaks, mostly consistent in position between the individual elements, which we interpret as evidence for the presence of ejecta contributing to the X-ray emission. 
The locations of these putative ejecta enhancements -- toward shrapnel D, around the cocoon \citep[see][]{Slane18}, and to the north and northwest of the pulsar -- seem to agree quite well with the regions exhibiting elevated temperatures. 
Generally, the observed range of neon  
abundances is higher than that of oxygen by around a factor of two. This is roughly consistent with abundance patterns typically identified in ejecta in the Vela shrapnels \citep{Miyata01,Katsuda05,Katsuda06,Yamaguchi09}, and therefore may be a general property of the X-ray emitting ejecta in Vela.
In contrast, the map of iron abundance does not appear correlated with the distribution of the lighter elements. We observe an apparent iron peak at the location of a soft filament around two degrees northeast of the pulsar. We investigate the spectrum of this region in detail in Sect.~\ref{DetailedFitSection}, in order to evaluate whether this clump indeed contains a physical iron enhancement. 

The lower right panel in Fig.~\ref{Spectroscopy_2TNT} displays the non-thermal surface brightness $\Sigma_{\Gamma}$, that is, the flux per unit area of the power-law component, evaluated in the comparatively hard energy range $1.0-5.0\,\si{keV}$. 
This energy range was chosen in order to avoid any bias caused by unphysically large fluxes from extremely steep power laws, which may mimic a very soft thermal continuum in regions without a significant nonthermal contribution to the spectrum.
Apart from the clear identification of the non-thermal shell associated to Vela Jr. in the southeast \citep{Camilloni23}, the  distribution of $\Sigma_{\Gamma}$ indicates an extremely large region of significant non-thermal emission in Vela. Its surface brightness is highest at the location of the pulsar and the cocoon, consistent with previous observations \citep{Markwardt95, Slane18}. However, the emission of what appears to be an extended PWN seems to reach up to distances around two to three degrees from the pulsar. 
This extended non-thermal plerion seems to show an asymmetric structure in emission, with the largest extent along an axis at about $15^{\circ}$ east of north, and a narrower profile in the perpendicular direction.
These findings of an extremely large and likely asymmetric X-ray PWN are broadly consistent with what can be observed directly in the images: in the energy range $1.4-2.3\,\si{keV}$, Fig.~\ref{NarrowBand} displays a structure quite similar in size and morphology to the one found in spectral modelling. This makes a dominant role of systematic errors in either of the two methods appear unlikely.

The spectral index $\Gamma$ of our power law component, on the other hand, is difficult to constrain and prone to systematic and statistical errors, as the data and our instrument only offer a relatively short spectral range as a baseline for its measurement. Below $1.0\,\si{keV}$, most spectra are strongly dominated by thermal emission from Vela, while above $2.3\,\si{keV}$, the eROSITA effective area decreases sharply, so that the instrumental background quickly becomes dominant.
Nonetheless, we observe realistic, albeit significantly scattered, spectral indices of $\Gamma \sim 2.2$ in the vicinity of the pulsar, consistent with the range observed there by \citet{Slane18}. 
Furthermore, an apparent increase in $\Gamma$ with distance from the central pulsar is visible. Even though the rate of increase is quantitatively uncertain, an intriguing physical explanation for this phenomenon could be that radiative losses of the underlying electron population lead to a gradual steepening of the non-thermal spectrum toward larger distances from the source.  

Our spectral fits with the TNT model (see Fig.~\ref{Spectroscopy_TNT}) reveal patterns which are mostly qualitatively consistent with those discussed above. Even though the values of certain parameters, such as absorption or plasma temperature, appear to quantitatively disagree with our measurements above, we believe the observed similarity to be encouraging, since two significantly differing physical models were applied to describe the thermal component. In particular, the physical interpretation of the ``temperature'' of a plasma with NEI is fundamentally different from that of a plasma in CIE, where electrons, ions, and protons have fully equilibrated with each other \citep{Borkowski01}.
The ionization age fitted in the TNT model, which is the product of post-shock density and shock age, $\tau=n_{e}\,t_{s}$ is poorly constrained across the extent of Vela. Generally, the observed median value around $2\times10^{12}\,\si{s.cm^{-3}}$ suggests weak overall departure from CIE, meaning there is no ubiquitous evidence for NEI across the SNR. In a few of the hotter regions, however, departures from CIE with $\tau < 10^{11}\,\si{s.cm^{-3}}$ seem to be present. 
However, given the strong model degeneracies involved, in particular that between $kT$ and $\tau$, we believe that these findings of possible localized departures from CIE should be taken with caution.

Our spatially resolved spectral analysis permits the computation of the total intrinsic flux of Vela, by integrating the unabsorbed background-subtracted flux of the 2TNT model over all bins within or overlapping the shell.
We obtain $F_{X} = 2.84\times10^{-8}\,\si{erg.s^{-1}.cm^{-2}}$ in the $0.2-5.0\,\si{keV}$ range, corresponding to an X-ray luminosity of $L_{X} = 2.9\times10^{35}\,\si{erg.s^{-1}}$ at a distance of $290\,\si{pc}$.  
The formal statistical uncertainty of this result is at a sub-percent level, but a relative systematic error of at least $15\%$ due to effective area uncertainties and the assumed extent of Vela is to be expected \citep[see also][]{Mayer22}.
A further systematic uncertainty is illustrated by the fact that the TNT model yields a total intrinsic flux of $F_{X} = 2.48\times10^{-8}\,\si{erg.s^{-1}.cm^{-2}}$,  
which is around $13\%$ lower than the value given by the 2TNT model. The origin of this discrepancy lies mainly in the systematically lower absorption column required by the TNT model, since this quantity sets the fraction of intrinsic flux which is ultimately detected by the telescope.

\begin{figure*}
\centering
\vspace{-0.2cm}
\includegraphics[width=18.4cm]{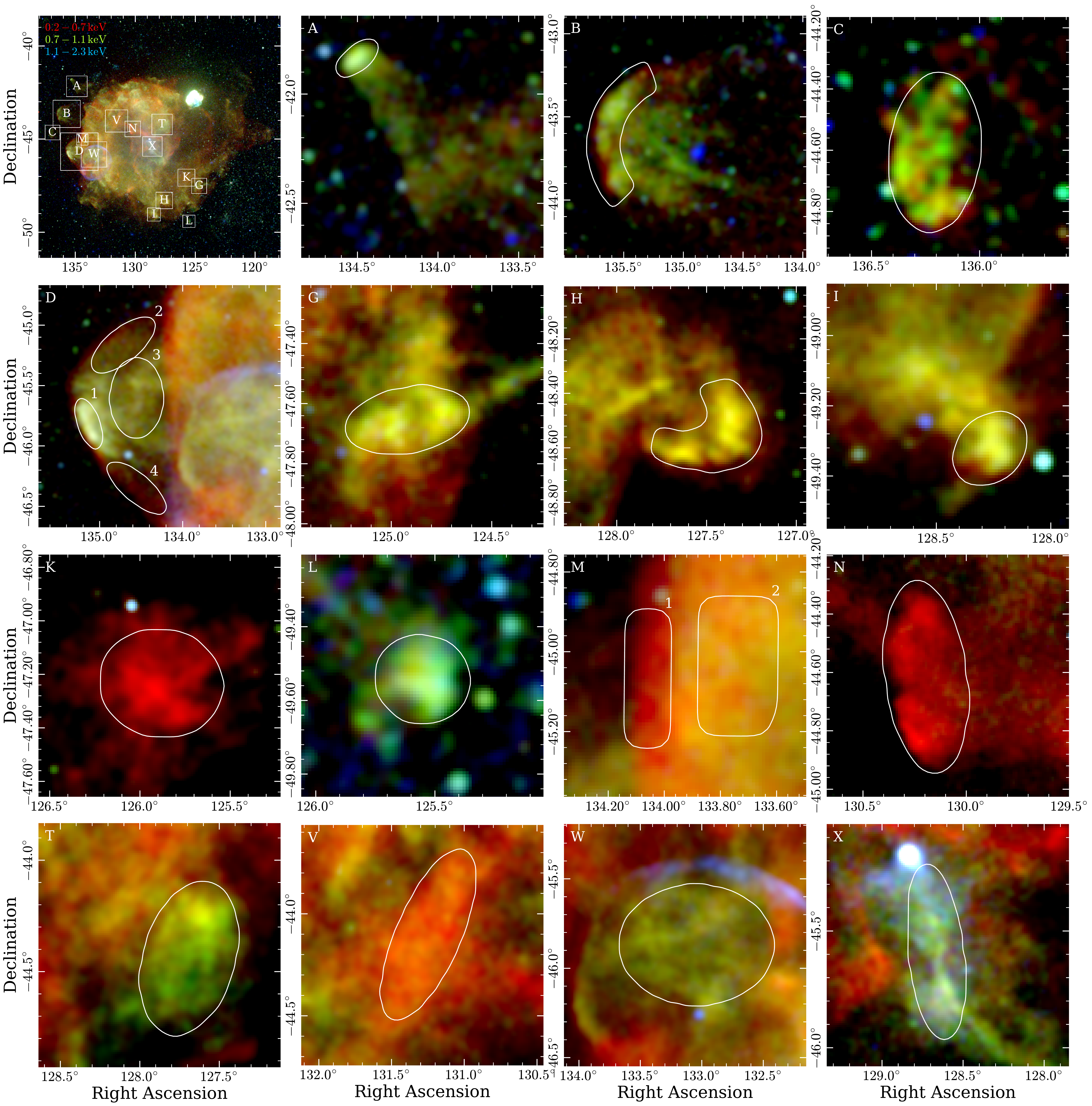} 
\caption{Morphological features of Vela. The upper left panel displays the false-color image of Vela shown in Fig.~\ref{VelaImage}, with white boxes  outlining the regions depicted by each of the other panels, which are identified with the corresponding letter.
The brightness scale used in each sub-panel preserves the relative colors of the original image, via a truncation of the original brightness scale at equal quantiles for the soft, medium, and hard bands, with the goal of optimally displaying the morphology of the features.
In each panel, we indicate the regions used for extraction of the spectra shown in Fig.~\ref{DetailedFits} in white, with numbers distinguishing the individual spectra in case of multiple regions. 
}
\label{Shrapnels}
\vspace{-0.2cm}
\end{figure*}

\subsection{A closer look at prominent features of Vela \label{DetailedFitSection}}
The analysis presented in the previous section has allowed us to obtain an impression of the global properties of shocked thermal plasma, as well as the distribution of relativistic particles from the pulsar wind throughout Vela, in a way that is completely agnostic toward the underlying shapes of features. 
In this section, we complement this with the imaging and spectroscopic investigation of selected features defined by morphological, rather than purely statistical, criteria.

\begin{figure*}
\centering
\includegraphics[width=9.0cm]{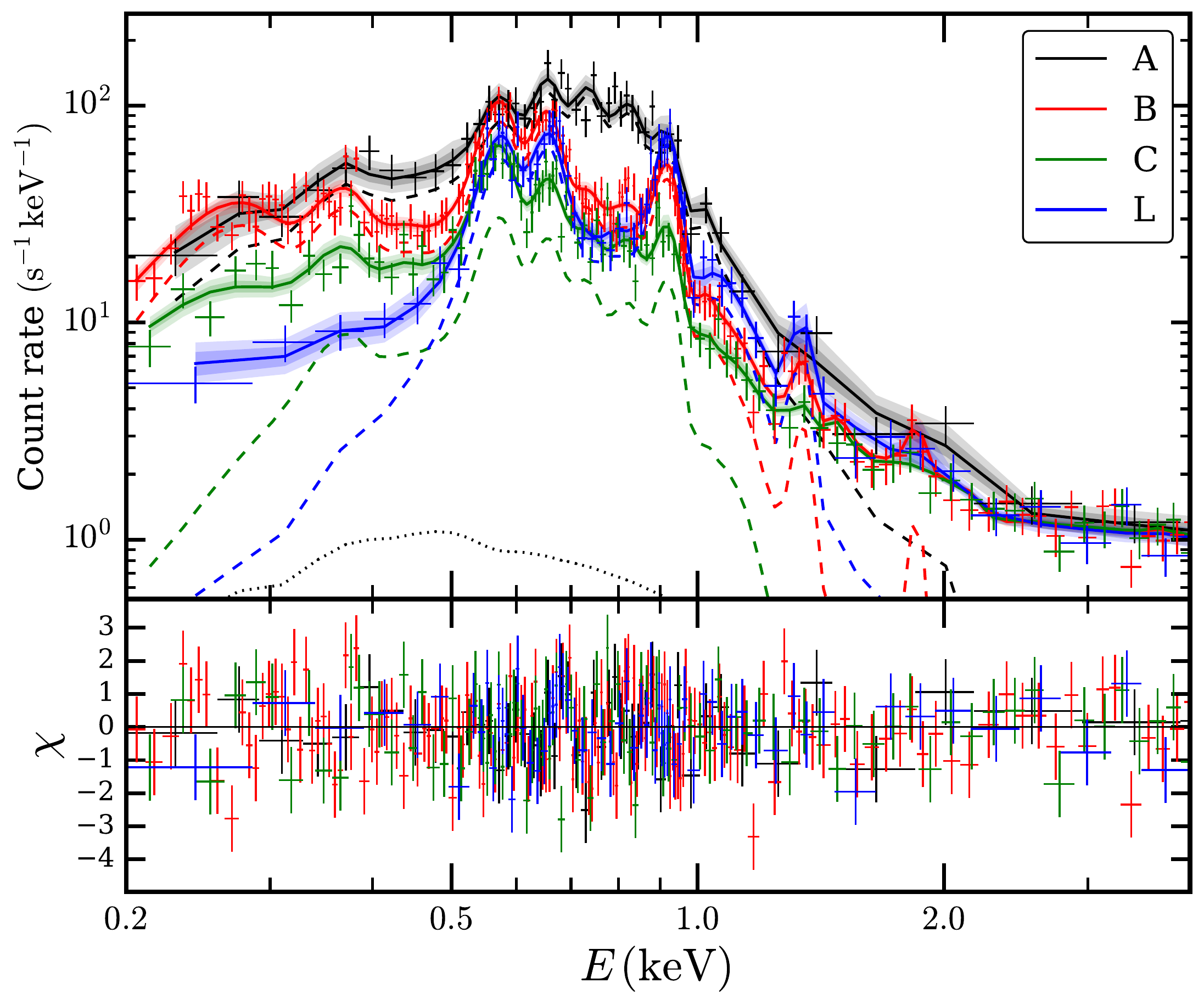} 
\includegraphics[width=9.0cm]{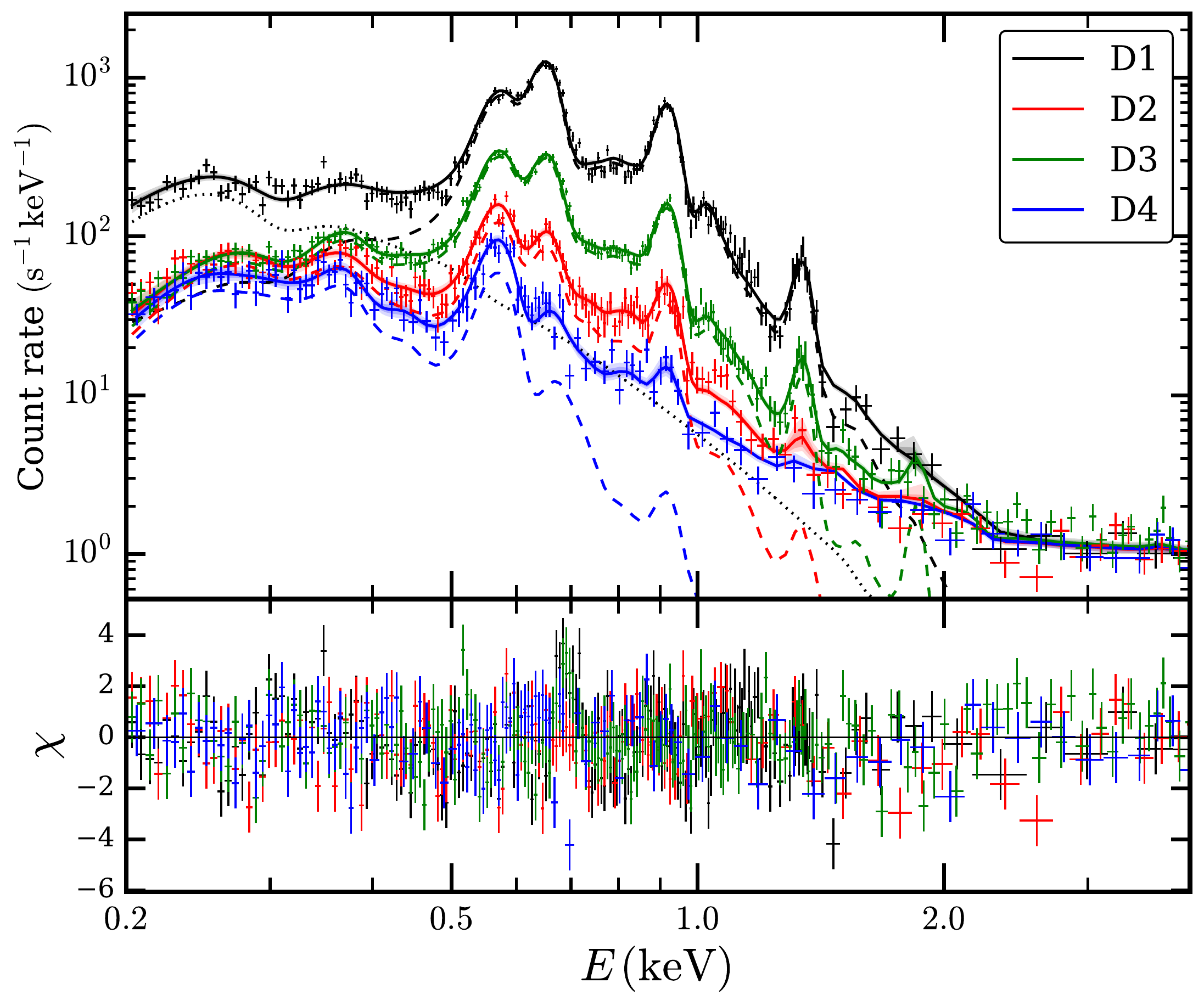} \\  
\includegraphics[width=9.0cm]{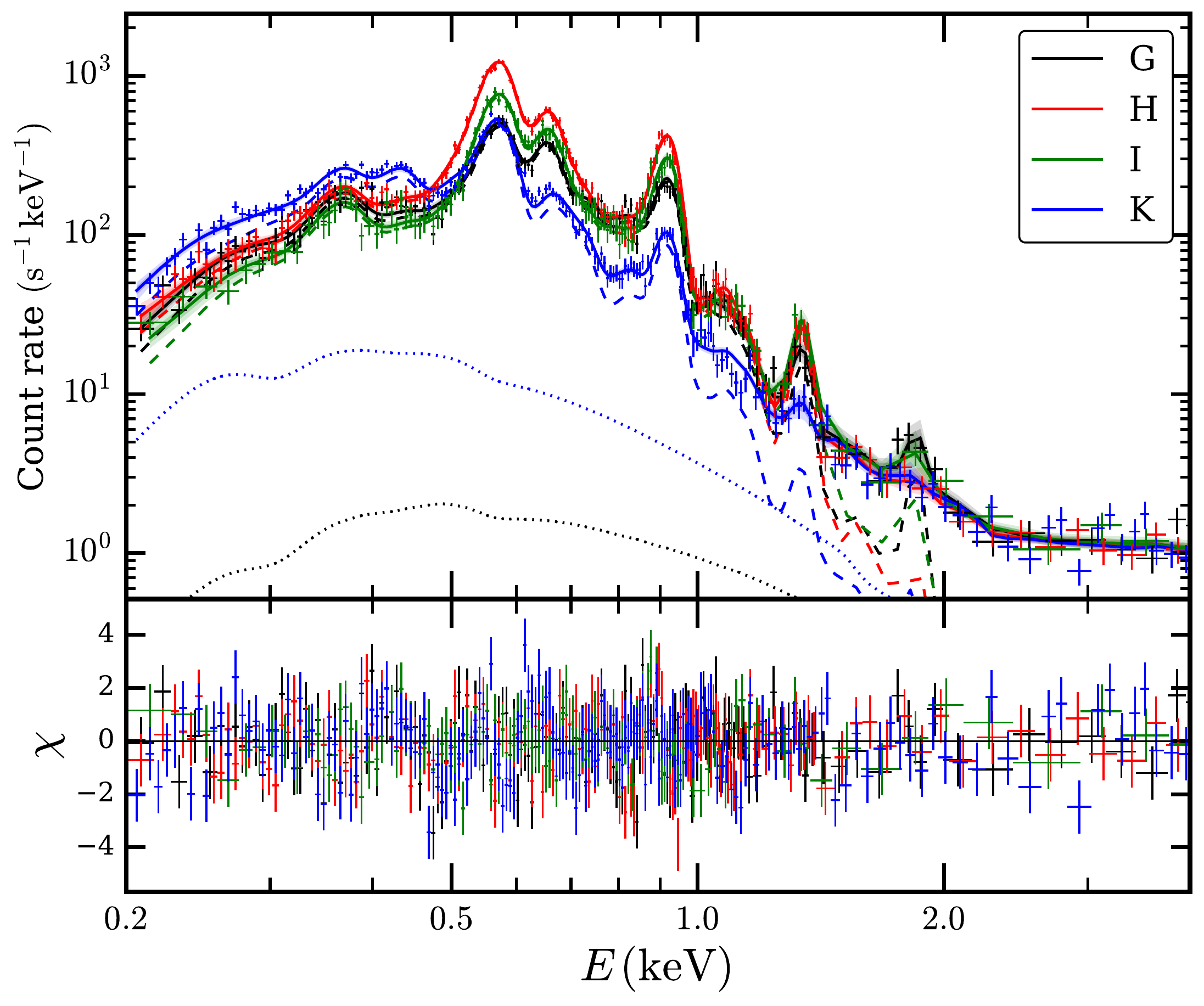} 
\includegraphics[width=9.0cm]{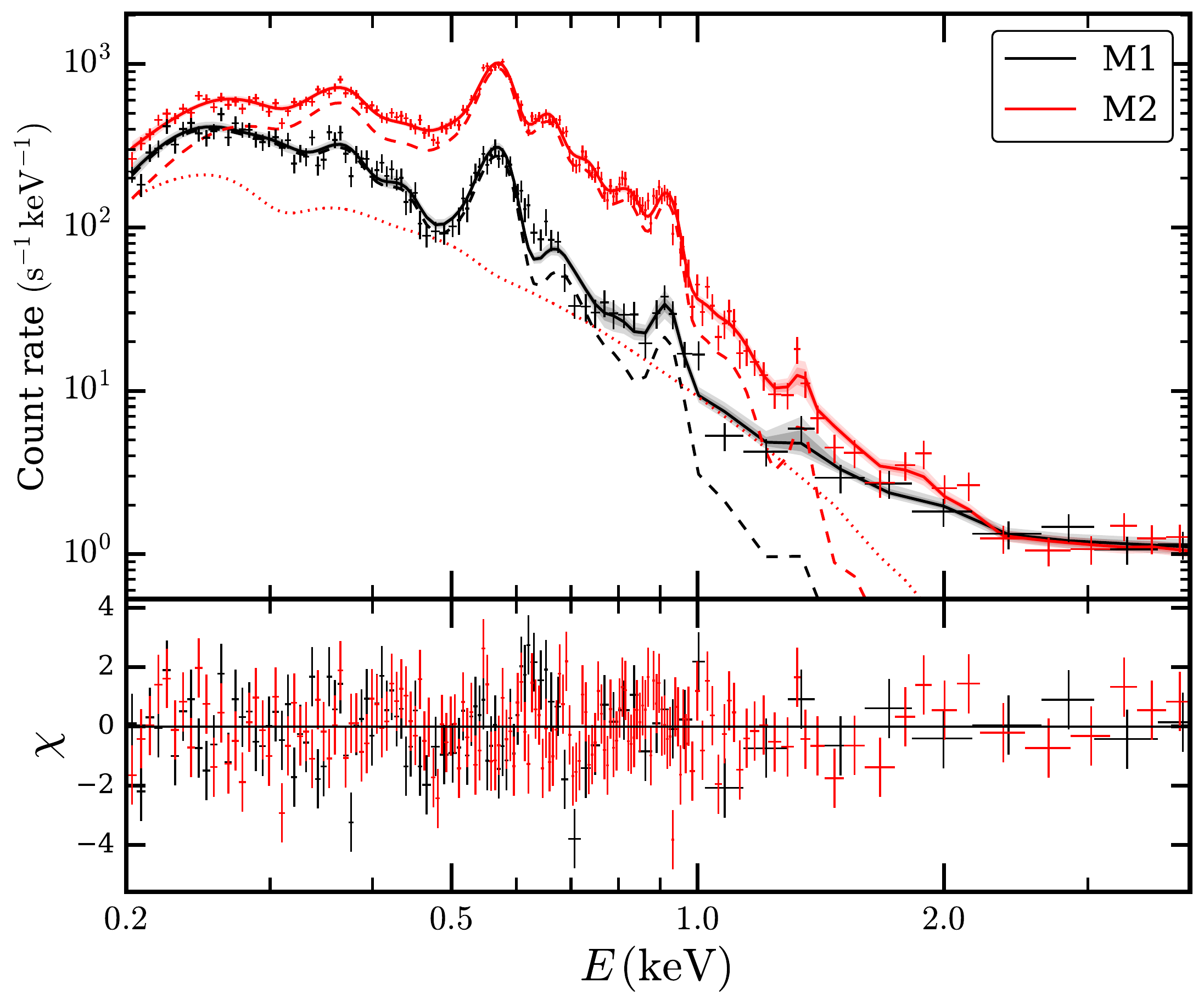} \\ 
\includegraphics[width=9.0cm]{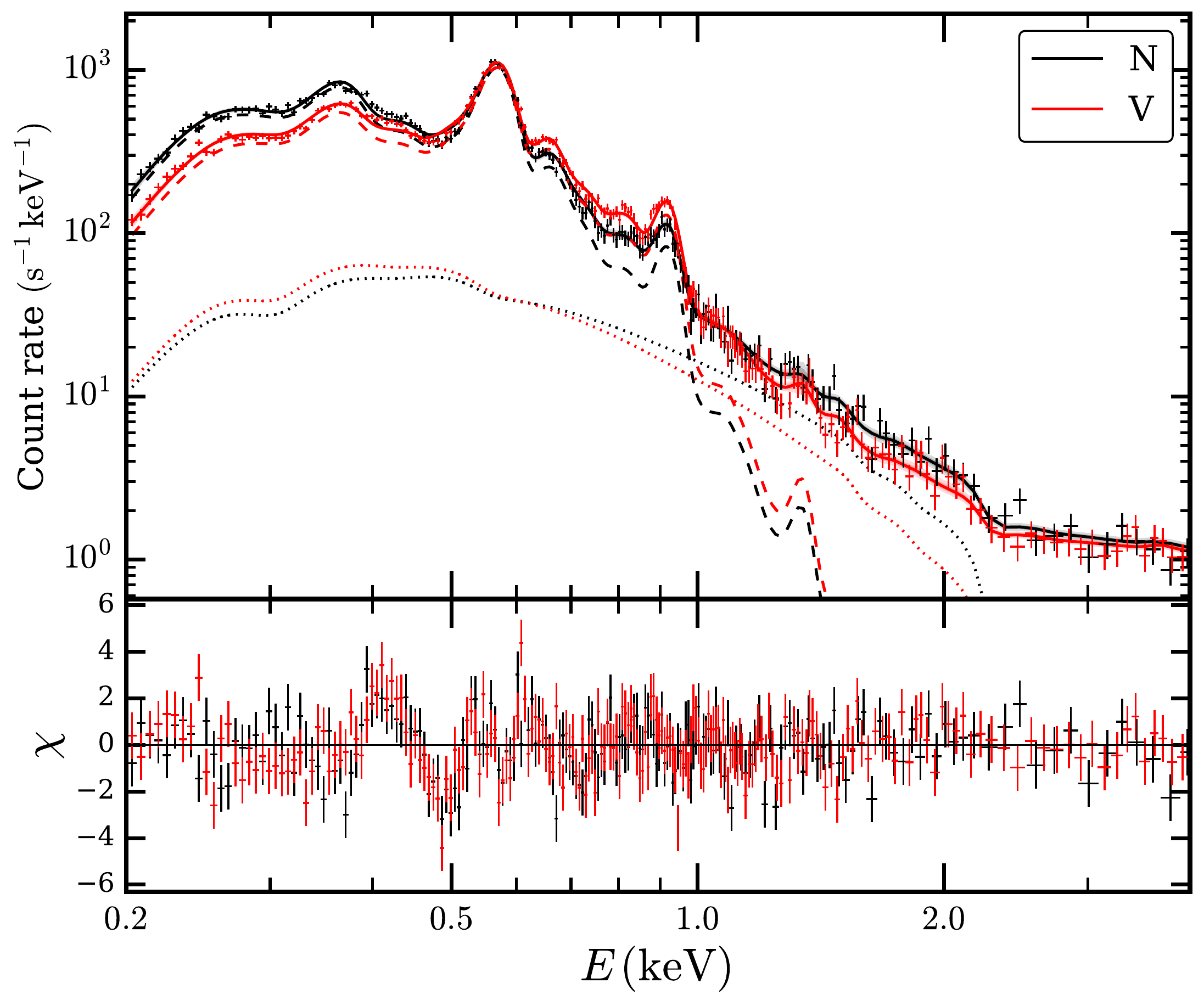} 
\includegraphics[width=9.0cm]{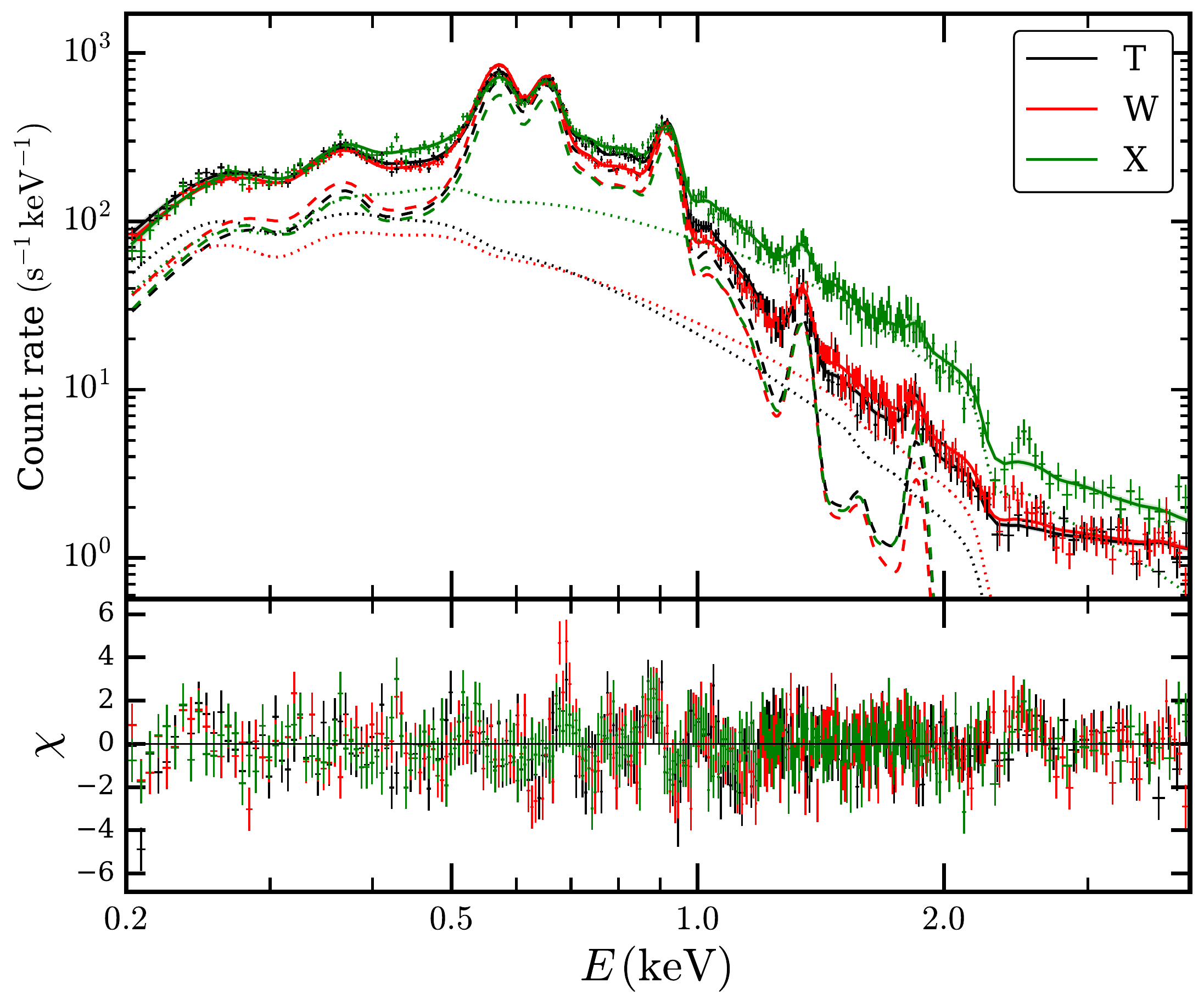} \\
\caption{ 
Spectra of prominent morphological features of Vela. The extraction regions of all spectra are outlined in Fig.~\ref{Shrapnels} and identified according to their assigned labels in the figure legends.
For all panels, we display the results of the TNT spectral model, which in a few cases gave considerably better fits than the 2TNT model. 
In each subplot, the displayed spectral model corresponds to the median of the prediction obtained from the posterior sample of our MCMC procedure. The dashed (dotted) line represents the median contribution from the thermal (nonthermal) source model component. For the sake of comparability, all spectra and models have been normalized dividing by their {\tt BACKSCAL} keyword, resulting in equal levels of particle background.  
}
\label{DetailedFits}
\end{figure*}

\begin{table*}[t]
\renewcommand{\arraystretch}{1.5}
\caption{Parameter constraints from modelling of the spectra shown in Fig.~\ref{DetailedFits} with the TNT model. \label{DetailedFitsTable}}
\centering
\resizebox{18.4cm}{!}{
\begin{tabular}{ccccccccccccccc}
\hline
\hline
Feature&$N_{\rm H}$&$kT$&$\log\,\tau$&$\log\,\mathrm{EM}$&$\rm N/H$&$\rm O/H$&$\rm Ne/H$&$\rm Mg/H$&$\rm Si/H$&$\rm Fe/H$&$\Gamma$&$\log\,F_{\Gamma}$&$\ln \,\mathcal{B}$\tablefootmark{(a)}\\%
&$10^{21}\,\rm{cm}^{-2}$&$\rm{keV}$&$\rm{cm^{-3}\,s}$&$\mathrm{cm}^{-5}$&&&&&&&&$\mathrm{erg\,cm^{-2}\,s^{-1}}$&\\%
\hline
\bf A&$\mathbf{0.19_{-0.12}^{+0.15}}$&$\mathbf{0.43_{-0.09}^{+0.11}}$&$\mathbf{11.59_{-0.38}^{+0.20}}$&$\mathbf{11.06_{-0.27}^{+0.16}}$&$\mathbf{0.4_{-0.3}^{+0.6}}$&$\mathbf{0.56_{-0.14}^{+0.25}}$&$\mathbf{1.0_{-0.3}^{+0.5}}$&$\mathbf{0.5_{-0.4}^{+0.6}}$&$\mathbf{3.5_{-2.1}^{+3.7}}$&$\mathbf{0.80_{-0.24}^{+0.47}}$&$\mathbf{2.6_{-0.5}^{+0.6}}$&$\mathbf{-13.65_{-1.02}^{+0.28}}$&$\mathbf{0.4}$\\%
\bf B&$\mathbf{0.040_{-0.029}^{+0.052}}$&$\mathbf{0.37_{-0.06}^{+0.06}}$&$\mathbf{11.39_{-0.22}^{+0.14}}$&$\mathbf{11.55_{-0.10}^{+0.08}}$&$\mathbf{0.089_{-0.029}^{+0.069}}$&$\mathbf{0.82_{-0.08}^{+0.12}}$&$\mathbf{1.96_{-0.24}^{+0.38}}$&$\mathbf{1.4_{-0.4}^{+0.6}}$&$\mathbf{6.4_{-1.3}^{+1.9}}$&$\mathbf{0.48_{-0.07}^{+0.10}}$&$\mathbf{2.6_{-0.5}^{+0.5}}$&$\mathbf{-14.17_{-0.75}^{+0.26}}$&$\mathbf{0.9}$\\%
\bf C&$\mathbf{0.6_{-0.3}^{+0.4}}$&$\mathbf{0.30_{-0.09}^{+0.25}}$&$\mathbf{11.45_{-0.74}^{+0.26}}$&$\mathbf{11.17_{-0.49}^{+0.22}}$&$\mathbf{0.18_{-0.11}^{+0.36}}$&$\mathbf{0.69_{-0.20}^{+0.30}}$&$\mathbf{1.8_{-0.6}^{+0.9}}$&$\mathbf{0.19_{-0.12}^{+0.46}}$&$\mathbf{1.4_{-1.2}^{+6.3}}$&$\mathbf{1.0_{-0.4}^{+0.6}}$&$\mathbf{2.5_{-0.5}^{+0.5}}$&$\mathbf{-14.07_{-0.77}^{+0.26}}$&$\mathbf{1.5}$\\%
\bf L&$\mathbf{2.2_{-0.7}^{+0.5}}$&$\mathbf{0.80_{-0.17}^{+0.14}}$&$\mathbf{10.53_{-0.13}^{+0.10}}$&$\mathbf{10.59_{-0.39}^{+0.20}}$&$\mathbf{0.31_{-0.22}^{+0.92}}$&$\mathbf{2.1_{-1.0}^{+2.5}}$&$\mathbf{4.9_{-2.4}^{+6.5}}$&$\mathbf{3.0_{-1.6}^{+4.0}}$&$\mathbf{1.1_{-1.0}^{+3.5}}$&$\mathbf{0.32_{-0.20}^{+0.56}}$&$\mathbf{2.5_{-0.5}^{+0.5}}$&$\mathbf{-13.82_{-0.94}^{+0.28}}$&$\mathbf{9.5}$\\%
\hline%
D1&$0.08_{-0.04}^{+0.04}$&$0.285_{-0.003}^{+0.003}$&$12.92_{-0.10}^{+0.08}$&$11.72_{-0.09}^{+0.07}$&$0.11_{-0.04}^{+0.13}$&$8.5_{-2.2}^{+2.0}$&$15_{-4}^{+3}$&$15_{-4}^{+3}$&$1.2_{-1.0}^{+2.6}$&$1.11_{-0.27}^{+0.26}$&$4.72_{-0.28}^{+0.28}$&$-12.59_{-0.15}^{+0.11}$&$-2.0$\\%
D2&$0.009_{-0.007}^{+0.014}$&$0.210_{-0.007}^{+0.024}$&$12.64_{-0.61}^{+0.24}$&$11.98_{-0.11}^{+0.09}$&$0.09_{-0.03}^{+0.08}$&$0.69_{-0.06}^{+0.07}$&$1.76_{-0.26}^{+0.25}$&$1.2_{-0.9}^{+1.2}$&$4.3_{-1.1}^{+1.8}$&$0.53_{-0.18}^{+0.17}$&$2.8_{-0.6}^{+0.9}$&$-13.61_{-1.12}^{+0.28}$&$-2.5$\\%
\bf D3&$\mathbf{0.007_{-0.006}^{+0.012}}$&$\mathbf{0.330_{-0.023}^{+0.018}}$&$\mathbf{11.62_{-0.11}^{+0.09}}$&$\mathbf{12.20_{-0.05}^{+0.05}}$&$\mathbf{0.13_{-0.06}^{+0.09}}$&$\mathbf{1.14_{-0.07}^{+0.09}}$&$\mathbf{2.41_{-0.17}^{+0.20}}$&$\mathbf{2.30_{-0.29}^{+0.34}}$&$\mathbf{4.1_{-0.5}^{+0.5}}$&$\mathbf{0.271_{-0.029}^{+0.035}}$&$\mathbf{2.8_{-0.6}^{+0.8}}$&$\mathbf{-13.15_{-1.32}^{+0.29}}$&$\mathbf{4.1}$\\%
D4&$0.13_{-0.10}^{+0.15}$&$0.23_{-0.07}^{+0.07}$&$10.37_{-0.28}^{+0.17}$&$11.33_{-0.19}^{+0.13}$&$0.35_{-0.19}^{+0.17}$&$0.46_{-0.07}^{+0.08}$&$0.36_{-0.23}^{+0.37}$&$0.4_{-0.3}^{+1.9}$&$5.1_{-2.7}^{+4.7}$&$0.8_{-0.7}^{+4.0}$&$2.6_{-0.5}^{+0.6}$&$-14.28_{-0.64}^{+0.25}$&$-1.1$\\%
\hline%
G&$0.35_{-0.09}^{+0.10}$&$0.37_{-0.05}^{+0.06}$&$11.12_{-0.17}^{+0.12}$&$11.91_{-0.12}^{+0.10}$&$0.45_{-0.15}^{+0.15}$&$0.68_{-0.06}^{+0.07}$&$1.64_{-0.15}^{+0.16}$&$1.20_{-0.20}^{+0.24}$&$3.6_{-1.0}^{+1.1}$&$0.43_{-0.05}^{+0.06}$&$2.7_{-0.5}^{+0.6}$&$-12.85_{-1.45}^{+0.29}$&$-1.3$\\%
\bf H&$\mathbf{0.31_{-0.09}^{+0.11}}$&$\mathbf{0.336_{-0.026}^{+0.020}}$&$\mathbf{10.97_{-0.08}^{+0.07}}$&$\mathbf{11.85_{-0.08}^{+0.07}}$&$\mathbf{0.71_{-0.14}^{+0.15}}$&$\mathbf{1.91_{-0.16}^{+0.17}}$&$\mathbf{5.3_{-0.5}^{+0.6}}$&$\mathbf{4.1_{-0.6}^{+0.6}}$&$\mathbf{2.2_{-1.2}^{+1.4}}$&$\mathbf{0.48_{-0.07}^{+0.08}}$&$\mathbf{2.6_{-0.5}^{+0.5}}$&$\mathbf{-13.47_{-1.18}^{+0.29}}$&$\mathbf{28.0}$\\%
\bf I&$\mathbf{0.54_{-0.14}^{+0.13}}$&$\mathbf{0.37_{-0.04}^{+0.06}}$&$\mathbf{10.96_{-0.16}^{+0.12}}$&$\mathbf{11.56_{-0.15}^{+0.11}}$&$\mathbf{0.15_{-0.08}^{+0.16}}$&$\mathbf{1.14_{-0.12}^{+0.17}}$&$\mathbf{2.8_{-0.3}^{+0.4}}$&$\mathbf{2.9_{-0.5}^{+0.6}}$&$\mathbf{4.2_{-1.7}^{+2.1}}$&$\mathbf{0.29_{-0.06}^{+0.07}}$&$\mathbf{2.5_{-0.5}^{+0.5}}$&$\mathbf{-13.93_{-0.88}^{+0.27}}$&$\mathbf{16.0}$\\%
K&$0.28_{-0.09}^{+0.11}$&$0.47_{-0.05}^{+0.04}$&$10.07_{-0.05}^{+0.05}$&$11.83_{-0.07}^{+0.06}$&$1.57_{-0.14}^{+0.14}$&$0.72_{-0.06}^{+0.07}$&$1.37_{-0.16}^{+0.19}$&$0.8_{-0.4}^{+0.4}$&$2.8_{-1.6}^{+2.3}$&$1.5_{-0.4}^{+0.5}$&$3.6_{-0.7}^{+0.5}$&$-12.19_{-0.15}^{+0.11}$&$-8.2$\\%
\hline%
M1&$0.30_{-0.06}^{+0.05}$&$0.221_{-0.029}^{+0.036}$&$10.16_{-0.11}^{+0.09}$&$11.79_{-0.13}^{+0.10}$&$0.57_{-0.07}^{+0.08}$&$0.330_{-0.028}^{+0.037}$&$0.70_{-0.21}^{+0.28}$&$2.2_{-2.0}^{+6.7}$&$15.5_{-3.7}^{+3.0}$&$8_{-5}^{+6}$&$2.6_{-0.5}^{+0.5}$&$-13.93_{-0.84}^{+0.27}$&$-12.5$\\%
M2&$0.033_{-0.026}^{+0.051}$&$0.220_{-0.015}^{+0.016}$&$11.76_{-0.18}^{+0.13}$&$12.48_{-0.07}^{+0.06}$&$0.45_{-0.11}^{+0.11}$&$0.498_{-0.027}^{+0.031}$&$0.85_{-0.08}^{+0.08}$&$0.8_{-0.3}^{+0.4}$&$2.4_{-0.6}^{+0.6}$&$0.69_{-0.12}^{+0.20}$&$4.32_{-0.30}^{+0.29}$&$-12.28_{-0.10}^{+0.08}$&$-7.9$\\%
\hline%
N&$0.27_{-0.05}^{+0.06}$&$0.164_{-0.008}^{+0.007}$&$11.88_{-0.13}^{+0.10}$&$13.15_{-0.08}^{+0.07}$&$0.41_{-0.05}^{+0.05}$&$0.412_{-0.030}^{+0.029}$&$0.72_{-0.06}^{+0.06}$&$0.6_{-0.4}^{+0.6}$&$2.7_{-0.6}^{+0.7}$&$1.5_{-0.4}^{+0.8}$&$3.06_{-0.24}^{+0.25}$&$-11.44_{-0.04}^{+0.03}$&$-18.0$\\%
\bf V&$\mathbf{0.37_{-0.04}^{+0.06}}$&$\mathbf{0.164_{-0.005}^{+0.006}}$&$\mathbf{12.23_{-0.15}^{+0.11}}$&$\mathbf{13.35_{-0.08}^{+0.06}}$&$\mathbf{0.55_{-0.07}^{+0.07}}$&$\mathbf{0.49_{-0.03}^{+0.04}}$&$\mathbf{1.15_{-0.09}^{+0.09}}$&$\mathbf{0.8_{-0.4}^{+0.4}}$&$\mathbf{2.5_{-0.5}^{+0.7}}$&$\mathbf{2.7_{-0.7}^{+0.9}}$&$\mathbf{3.7_{-0.4}^{+0.4}}$&$\mathbf{-11.48_{-0.04}^{+0.03}}$&$\mathbf{5.2}$\\%
\hline%
\bf T&$\mathbf{0.14_{-0.04}^{+0.03}}$&$\mathbf{0.402_{-0.031}^{+0.026}}$&$\mathbf{11.24_{-0.08}^{+0.07}}$&$\mathbf{12.29_{-0.08}^{+0.06}}$&$\mathbf{0.22_{-0.12}^{+0.14}}$&$\mathbf{1.40_{-0.17}^{+0.21}}$&$\mathbf{3.1_{-0.4}^{+0.5}}$&$\mathbf{2.25_{-0.29}^{+0.40}}$&$\mathbf{4.1_{-0.6}^{+0.8}}$&$\mathbf{0.75_{-0.08}^{+0.11}}$&$\mathbf{3.41_{-0.16}^{+0.13}}$&$\mathbf{-11.214_{-0.031}^{+0.029}}$&$\mathbf{6.0}$\\%
\bf W&$\mathbf{0.095_{-0.016}^{+0.017}}$&$\mathbf{0.335_{-0.015}^{+0.009}}$&$\mathbf{11.43_{-0.05}^{+0.04}}$&$\mathbf{12.70_{-0.04}^{+0.03}}$&$\mathbf{0.082_{-0.023}^{+0.050}}$&$\mathbf{1.42_{-0.07}^{+0.08}}$&$\mathbf{3.18_{-0.17}^{+0.19}}$&$\mathbf{2.77_{-0.22}^{+0.23}}$&$\mathbf{3.6_{-0.4}^{+0.5}}$&$\mathbf{0.468_{-0.028}^{+0.031}}$&$\mathbf{2.94_{-0.08}^{+0.07}}$&$\mathbf{-10.733_{-0.016}^{+0.016}}$&$\mathbf{43.4}$\\%
\bf X&$\mathbf{0.130_{-0.028}^{+0.027}}$&$\mathbf{0.40_{-0.04}^{+0.05}}$&$\mathbf{11.29_{-0.14}^{+0.11}}$&$\mathbf{12.05_{-0.09}^{+0.07}}$&$\mathbf{0.34_{-0.21}^{+0.22}}$&$\mathbf{1.31_{-0.15}^{+0.20}}$&$\mathbf{2.8_{-0.3}^{+0.5}}$&$\mathbf{2.3_{-0.4}^{+0.5}}$&$\mathbf{5.8_{-1.2}^{+1.7}}$&$\mathbf{0.56_{-0.06}^{+0.08}}$&$\mathbf{2.30_{-0.08}^{+0.08}}$&$\mathbf{-10.569_{-0.014}^{+0.013}}$&$\mathbf{6.9}$\\%
\hline
\end{tabular}%
}
\tablefoot{All parameters are defined as in Sect.~\ref{Spectroscopy}. The emission measure $\rm EM$ quantifies the normalization of the thermal component, whereas $F_{\Gamma}$ indicates the unabsorbed nonthermal flux integrated over the range $1.0-5.0\,\si{keV}$.
The given values and uncertainties correspond to the median and $68\%$ central interval of the posterior parameter distribution. 
The parameters corresponding to spectra shown in the same panel in Fig.~\ref{DetailedFits} are delimited by horizontal lines. \\
\tablefoottext{a}{Approximate Bayes factor from the ratio of model evidence estimates 
$\mathcal{B} = \mathcal{Z}_{\rm TNT}/\mathcal{Z}_{\rm 2TNT}$.
Regions with a positive value of $\ln \,\mathcal{B}$, indicating statistical preference for the TNT model, are highlighted in bold.}
}
\end{table*}

\begin{table*}[t]
\renewcommand{\arraystretch}{1.5}
\caption{Same as Table \ref{DetailedFitsTable}, but for the 2TNT model. \label{DetailedFitsTable_2TNT}}
\centering
\resizebox{18.4cm}{!}{
\begin{tabular}{cccccccccccccc}
\hline%
\hline%
Feature&$N_{\rm H}$&$kT_1$&$\log\,\mathrm{EM}_1$&$kT_2$&$\log\,\mathrm{EM}_2$&$\rm N/H$&$\rm O/H$&$\rm Ne/H$&$\rm Mg/H$&$\rm Si/H$&$\rm Fe/H$&$\Gamma$&$\log\,F_{\Gamma}$\\%
&$10^{21}\,\rm{cm}^{-2}$&$\rm{keV}$&$\mathrm{cm}^{-5}$&$\rm{keV}$&$\mathrm{cm}^{-5}$&&&&&&&&$\mathrm{erg\,cm^{-2}\,s^{-1}}$\\%
\hline%
A&$0.30_{-0.14}^{+0.18}$&$0.269_{-0.024}^{+0.025}$&$11.24_{-0.16}^{+0.12}$&$0.54_{-0.23}^{+0.26}$&$10.34_{-0.48}^{+0.22}$&$0.35_{-0.25}^{+0.99}$&$0.78_{-0.17}^{+0.27}$&$1.3_{-0.4}^{+0.6}$&$0.7_{-0.5}^{+0.9}$&$5_{-3}^{+6}$&$0.90_{-0.26}^{+0.42}$&$2.6_{-0.5}^{+0.5}$&$-13.75_{-0.94}^{+0.28}$\\%
B&$0.07_{-0.05}^{+0.08}$&$0.227_{-0.006}^{+0.007}$&$11.79_{-0.06}^{+0.06}$&$0.62_{-0.23}^{+0.26}$&$10.38_{-0.42}^{+0.21}$&$0.11_{-0.05}^{+0.13}$&$1.22_{-0.11}^{+0.14}$&$3.3_{-0.4}^{+0.4}$&$2.9_{-1.2}^{+1.4}$&$5.8_{-1.9}^{+2.9}$&$0.87_{-0.14}^{+0.18}$&$2.6_{-0.5}^{+0.5}$&$-14.10_{-0.74}^{+0.26}$\\%
C&$0.9_{-0.3}^{+0.4}$&$0.194_{-0.011}^{+0.012}$&$11.69_{-0.20}^{+0.14}$&$0.52_{-0.26}^{+0.45}$&$10.03_{-0.35}^{+0.19}$&$0.14_{-0.07}^{+0.26}$&$0.61_{-0.18}^{+0.25}$&$1.9_{-0.7}^{+0.9}$&$0.25_{-0.17}^{+0.60}$&$2.2_{-1.9}^{+8.4}$&$1.5_{-0.5}^{+1.0}$&$2.5_{-0.5}^{+0.6}$&$-14.05_{-0.80}^{+0.27}$\\%
L&$2.60_{-0.62}^{+0.29}$&$0.203_{-0.015}^{+0.017}$&$11.44_{-0.36}^{+0.19}$&$0.36_{-0.07}^{+0.19}$&$10.57_{-0.45}^{+0.22}$&$0.22_{-0.14}^{+0.77}$&$2.0_{-0.9}^{+1.9}$&$5.7_{-2.7}^{+6.0}$&$5.1_{-2.6}^{+5.6}$&$1.5_{-1.4}^{+7.3}$&$0.39_{-0.23}^{+0.52}$&$2.5_{-0.5}^{+0.5}$&$-13.37_{-1.19}^{+0.29}$\\%
\hline%
\bf D1&$\mathbf{0.04_{-0.03}^{+0.04}}$&$\mathbf{0.2736_{-0.0024}^{+0.0026}}$&$\mathbf{11.79_{-0.11}^{+0.09}}$&$\mathbf{0.31_{-0.06}^{+0.42}}$&$\mathbf{9.97_{-0.33}^{+0.19}}$&$\mathbf{0.10_{-0.04}^{+0.12}}$&$\mathbf{7.2_{-1.5}^{+1.9}}$&$\mathbf{12.8_{-2.7}^{+3.2}}$&$\mathbf{14.0_{-2.9}^{+3.4}}$&$\mathbf{2.3_{-1.9}^{+2.5}}$&$\mathbf{1.01_{-0.20}^{+0.25}}$&$\mathbf{4.54_{-0.22}^{+0.26}}$&$\mathbf{-12.56_{-0.15}^{+0.11}}$\\%
\bf D2&$\mathbf{0.009_{-0.007}^{+0.015}}$&$\mathbf{0.195_{-0.005}^{+0.004}}$&$\mathbf{12.04_{-0.04}^{+0.04}}$&$\mathbf{0.49_{-0.19}^{+0.66}}$&$\mathbf{10.49_{-0.43}^{+0.21}}$&$\mathbf{0.11_{-0.04}^{+0.10}}$&$\mathbf{0.70_{-0.04}^{+0.05}}$&$\mathbf{1.77_{-0.21}^{+0.21}}$&$\mathbf{1.0_{-0.8}^{+1.1}}$&$\mathbf{2.8_{-0.7}^{+0.7}}$&$\mathbf{0.62_{-0.17}^{+0.24}}$&$\mathbf{2.6_{-0.5}^{+0.6}}$&$\mathbf{-14.08_{-0.76}^{+0.26}}$\\%
D3&$0.013_{-0.009}^{+0.020}$&$0.32_{-0.11}^{+0.05}$&$11.57_{-0.29}^{+0.17}$&$0.223_{-0.005}^{+0.057}$&$12.31_{-0.31}^{+0.18}$&$0.44_{-0.18}^{+0.18}$&$1.68_{-0.07}^{+0.08}$&$3.65_{-0.25}^{+0.25}$&$4.3_{-0.6}^{+0.6}$&$2.7_{-0.5}^{+0.5}$&$0.54_{-0.05}^{+0.05}$&$2.5_{-0.5}^{+0.5}$&$-13.33_{-1.25}^{+0.29}$\\%
\bf D4&$\mathbf{0.14_{-0.07}^{+0.19}}$&$\mathbf{0.075_{-0.005}^{+0.005}}$&$\mathbf{12.53_{-0.15}^{+0.11}}$&$\mathbf{0.231_{-0.022}^{+0.043}}$&$\mathbf{10.83_{-0.16}^{+0.12}}$&$\mathbf{0.76_{-0.24}^{+0.25}}$&$\mathbf{1.7_{-0.3}^{+0.4}}$&$\mathbf{1.4_{-1.0}^{+1.2}}$&$\mathbf{0.4_{-0.3}^{+1.6}}$&$\mathbf{0.5_{-0.4}^{+3.8}}$&$\mathbf{0.6_{-0.4}^{+0.8}}$&$\mathbf{2.6_{-0.5}^{+0.5}}$&$\mathbf{-14.41_{-0.57}^{+0.24}}$\\%
\hline%
\bf G&$\mathbf{0.84_{-0.09}^{+0.08}}$&$\mathbf{0.163_{-0.019}^{+0.016}}$&$\mathbf{12.51_{-0.07}^{+0.06}}$&$\mathbf{0.257_{-0.013}^{+0.020}}$&$\mathbf{12.25_{-0.23}^{+0.15}}$&$\mathbf{0.15_{-0.07}^{+0.11}}$&$\mathbf{0.63_{-0.04}^{+0.05}}$&$\mathbf{1.54_{-0.12}^{+0.15}}$&$\mathbf{1.46_{-0.28}^{+0.35}}$&$\mathbf{6.9_{-1.9}^{+2.4}}$&$\mathbf{0.47_{-0.07}^{+0.08}}$&$\mathbf{2.5_{-0.5}^{+0.5}}$&$\mathbf{-13.60_{-1.09}^{+0.28}}$\\%
H&$1.14_{-0.11}^{+0.09}$&$0.1671_{-0.0019}^{+0.0017}$&$12.88_{-0.04}^{+0.04}$&$0.47_{-0.20}^{+1.24}$&$10.38_{-0.44}^{+0.21}$&$0.12_{-0.05}^{+0.08}$&$1.24_{-0.09}^{+0.11}$&$4.7_{-0.4}^{+0.4}$&$6.7_{-1.0}^{+1.1}$&$11_{-4}^{+5}$&$1.5_{-0.3}^{+0.4}$&$2.6_{-0.5}^{+0.5}$&$-13.74_{-0.95}^{+0.28}$\\%
I&$1.09_{-0.15}^{+0.11}$&$0.180_{-0.005}^{+0.004}$&$12.33_{-0.08}^{+0.07}$&$0.26_{-0.05}^{+0.45}$&$10.49_{-0.69}^{+0.25}$&$0.077_{-0.021}^{+0.054}$&$1.06_{-0.11}^{+0.11}$&$3.6_{-0.4}^{+0.5}$&$7.2_{-1.3}^{+1.5}$&$12_{-6}^{+5}$&$0.84_{-0.19}^{+0.23}$&$2.4_{-0.5}^{+0.5}$&$-13.00_{-1.22}^{+0.29}$\\%
\bf K&$\mathbf{0.49_{-0.06}^{+0.06}}$&$\mathbf{0.083_{-0.005}^{+0.007}}$&$\mathbf{13.15_{-0.11}^{+0.09}}$&$\mathbf{0.205_{-0.004}^{+0.008}}$&$\mathbf{11.80_{-0.11}^{+0.09}}$&$\mathbf{2.25_{-0.25}^{+0.27}}$&$\mathbf{2.4_{-0.6}^{+0.6}}$&$\mathbf{7.3_{-1.8}^{+2.0}}$&$\mathbf{3.5_{-2.5}^{+2.6}}$&$\mathbf{0.21_{-0.13}^{+0.63}}$&$\mathbf{2.9_{-0.7}^{+0.8}}$&$\mathbf{4.32_{-0.32}^{+0.25}}$&$\mathbf{-12.00_{-0.07}^{+0.06}}$\\%
\hline%
\bf M1&$\mathbf{0.460_{-0.035}^{+0.025}}$&$\mathbf{0.0677_{-0.0024}^{+0.0028}}$&$\mathbf{13.29_{-0.09}^{+0.07}}$&$\mathbf{0.240_{-0.025}^{+0.025}}$&$\mathbf{11.17_{-0.09}^{+0.07}}$&$\mathbf{0.61_{-0.12}^{+0.13}}$&$\mathbf{1.24_{-0.19}^{+0.20}}$&$\mathbf{1.6_{-0.4}^{+0.5}}$&$\mathbf{1.4_{-1.2}^{+2.7}}$&$\mathbf{17.3_{-3.3}^{+1.9}}$&$\mathbf{0.50_{-0.23}^{+0.34}}$&$\mathbf{2.5_{-0.5}^{+0.5}}$&$\mathbf{-14.22_{-0.67}^{+0.25}}$\\%
\bf M2&$\mathbf{0.21_{-0.10}^{+0.09}}$&$\mathbf{0.115_{-0.008}^{+0.008}}$&$\mathbf{12.87_{-0.09}^{+0.07}}$&$\mathbf{0.212_{-0.008}^{+0.013}}$&$\mathbf{12.49_{-0.06}^{+0.05}}$&$\mathbf{0.68_{-0.09}^{+0.09}}$&$\mathbf{0.55_{-0.04}^{+0.05}}$&$\mathbf{0.91_{-0.11}^{+0.12}}$&$\mathbf{1.0_{-0.3}^{+0.4}}$&$\mathbf{2.4_{-1.2}^{+1.5}}$&$\mathbf{0.84_{-0.19}^{+0.19}}$&$\mathbf{2.4_{-0.5}^{+0.5}}$&$\mathbf{-12.45_{-0.54}^{+0.23}}$\\%
\hline%
\bf N&$\mathbf{0.242_{-0.019}^{+0.026}}$&$\mathbf{0.0833_{-0.0018}^{+0.0019}}$&$\mathbf{13.64_{-0.04}^{+0.04}}$&$\mathbf{0.204_{-0.003}^{+0.007}}$&$\mathbf{12.08_{-0.05}^{+0.05}}$&$\mathbf{0.83_{-0.07}^{+0.07}}$&$\mathbf{1.55_{-0.12}^{+0.12}}$&$\mathbf{2.87_{-0.29}^{+0.31}}$&$\mathbf{0.29_{-0.21}^{+0.90}}$&$\mathbf{0.10_{-0.04}^{+0.13}}$&$\mathbf{1.50_{-0.24}^{+0.26}}$&$\mathbf{3.81_{-0.15}^{+0.14}}$&$\mathbf{-11.509_{-0.029}^{+0.027}}$\\%
V&$0.48_{-0.04}^{+0.05}$&$0.1502_{-0.0013}^{+0.0012}$&$13.57_{-0.05}^{+0.04}$&$0.40_{-0.17}^{+0.36}$&$10.05_{-0.37}^{+0.20}$&$0.42_{-0.06}^{+0.05}$&$0.410_{-0.023}^{+0.037}$&$1.07_{-0.09}^{+0.12}$&$1.0_{-0.5}^{+0.5}$&$2.1_{-0.5}^{+0.7}$&$5.2_{-0.8}^{+0.8}$&$3.6_{-0.5}^{+0.5}$&$-11.49_{-0.04}^{+0.04}$\\%
\hline%
T&$0.18_{-0.03}^{+0.03}$&$0.2163_{-0.0015}^{+0.0014}$&$12.66_{-0.06}^{+0.05}$&$0.53_{-0.03}^{+0.03}$&$11.59_{-0.08}^{+0.07}$&$0.56_{-0.27}^{+0.24}$&$1.94_{-0.23}^{+0.31}$&$5.1_{-0.6}^{+0.8}$&$3.9_{-0.5}^{+0.7}$&$4.4_{-0.8}^{+1.1}$&$1.23_{-0.15}^{+0.20}$&$3.46_{-0.19}^{+0.16}$&$-11.28_{-0.04}^{+0.04}$\\%
W&$0.140_{-0.014}^{+0.014}$&$0.2128_{-0.0010}^{+0.0010}$&$12.969_{-0.025}^{+0.023}$&$0.48_{-0.09}^{+0.07}$&$11.41_{-0.12}^{+0.09}$&$0.10_{-0.04}^{+0.08}$&$2.05_{-0.11}^{+0.12}$&$5.3_{-0.3}^{+0.3}$&$5.9_{-0.6}^{+0.6}$&$2.4_{-0.5}^{+0.5}$&$1.08_{-0.07}^{+0.08}$&$2.99_{-0.07}^{+0.06}$&$-10.726_{-0.022}^{+0.021}$\\%
X&$0.157_{-0.025}^{+0.026}$&$0.2205_{-0.0022}^{+0.0022}$&$12.35_{-0.07}^{+0.06}$&$0.64_{-0.07}^{+0.07}$&$11.17_{-0.17}^{+0.12}$&$0.6_{-0.4}^{+0.4}$&$2.03_{-0.25}^{+0.34}$&$4.9_{-0.7}^{+0.9}$&$4.7_{-1.0}^{+1.3}$&$5.2_{-1.3}^{+1.6}$&$1.08_{-0.17}^{+0.21}$&$2.34_{-0.08}^{+0.07}$&$-10.576_{-0.015}^{+0.014}$\\%
\hline%
\end{tabular}%
}
\tablefoot{The parameters $kT_{1}$ ($kT_{2}$) and $\rm EM_{1}$ ($\rm EM_{2}$) refer to the temperatures and emission measures of the cooler (hotter) thermal component, respectively.
}
\end{table*}

Figure \ref{Shrapnels} displays the location and energy-dependent morphology of prominent features selected for further study. 
These include the Vela shrapnels A to D \citep{Aschenbach95}, as well as potentially similar features discussed in later studies \citep{Garcia17, Sapienza21}, labelled G, H, I, and K. Furthermore, we follow up on several regions that stood out in the previous sections, due to high elemental abundances (T, V, W, X), peculiar temperatures (N, V), or morphological criteria (L, M).

In order to investigate the physical conditions within our features, we extracted spectra from the regions indicated in Fig.~\ref{Shrapnels}. These spectra were modelled in analogous fashion to the previous section. This concerns the choice of background and source models, as well as the usage of an MCMC sampler for obtaining parameter constraints. 
However, here, we allowed the silicon abundance parameter to vary, as the presence or absence of silicon is an interesting aspect to explore in the many ejecta-rich regions we investigate. 
In order to evaluate which of our models describes the individual spectra better, we computed approximate Bayesian model evidences from the probabilities of our posterior sample. We accomplished this by using the modified harmonic mean approximation outlined by \citet{Robert09}. 
We display the investigated spectra, along with their fitted TNT models, in Fig.~\ref{DetailedFits}. The corresponding constraints on physical parameters are detailed in Tables \ref{DetailedFitsTable} and \ref{DetailedFitsTable_2TNT} for the TNT and 2TNT models, respectively.

Our data set allows us to resolve for the first time the energy dependence of the morphology of the Vela shrapnels labelled A to D by \citet{Aschenbach95}. Features B and D clearly show spatially varying hardness in their X-ray emission, with the hardest emission originating close to the apex of the broad bow-shock, and much softer emission (visible in red in Fig.~\ref{Shrapnels}) along the sides. While the spectra at the head of the shrapnels A, B, D have been studied in detail elsewhere \citep[e.g.,][]{Katsuda05,Katsuda06,Yamaguchi09}, Fig.~\ref{DetailedFits} illustrates very clearly that feature A exhibits a fundamentally different spectrum from the other shrapnels. This applies in particular at $0.7-0.9\,\si{keV}$, where line emission of \ion{Fe}{xvii} likely contributes. While the statistics of our data set for shrapnel A are limited, 
we find enhanced abundance ratios of iron relative to lighter elements (e.g.~$\rm Fe/O=1.4_{-0.4}^{+0.6}$),
similarly to \citet{Katsuda06}. This may be a signature either of a strong contribution of shocked ISM of at least solar iron content to the emission or of enhanced mixing of explosively synthesized iron into this silicon-rich ejecta clump. 
In shrapnel B, we have detected significant silicon line emission, which may indicate a larger abundance thereof than previously established  for this feature \citep{Yamaguchi09}.

During the inspection of broad-band images of Vela, we discovered a previously unnoticed compact clump southwest of the SNR shell, which we have labelled feature L.  
Due to its apparent faint tail in the northeast direction and the clump's highest brightness in the medium energy band, it seems reasonable to propose that this feature may be a potential analog to shrapnel A. Its observed spectrum is quite curious, however: it exhibits a signature of a comparatively high temperature, as well as very high absorption, with a fitted value around $2\times10^{21}\,\si{cm^{-2}}$. The latter fact makes a reconciliation with the distance of Vela problematic, as one would probably expect much less intervening material. 
Nonetheless, to our knowledge, there is no known potentially X-ray emitting object that would appear extended (e.g., galaxy cluster or star cluster) at the location of our feature. Therefore, an association with the Vela SNR cannot be ruled out, in particular since the feature is located close to the most absorbed part of the SNR. Follow-up observations of this clump may be desirable in order to infer both its composition and its distance, similarly to \citet{Garcia17}.     

Spectra extracted from multiple regions across shrapnel D exhibit a vast range of electron temperatures and ionization ages, particularly visible in the varying ratio of \ion{O}{viii} to \ion{O}{vii} line emission. The highest ratio is visible at the apex of the bow shock (D1), while it is lowest along the sides (D2 and D4), and exhibits intermediate values further downstream (D3). 
Our detection of extreme abundances of oxygen, neon, and magnesium at the apex, with $\rm Ne/O$ and $\rm Mg/O$ ratios around two, and no strong iron enhancement, is completely consistent with the results of \citet{Katsuda05}, cementing the interpretation of this hot feature as a light-element ejecta-rich clump. Interestingly, the plasma in this region shows no convincing signature for departure from CIE, which is somewhat unexpected for material which should have recently been shocked by interaction with the ISM.    
The spectra of the lateral regions D2 and D4 generally exhibit much lower light-element abundances, consistent with the dominant contribution coming from shock-heated ISM. Furthermore, the low observed temperature  
appears consistent with an interaction with a shock that is likely moving more slowly than at the apex of the structure.  

The features G, H, I, and K were identified in \citet{Garcia17} and \citet{Sapienza21} with G, H, and I proposed as potential analogues to the established shrapnels. Our imaging demonstrates that their energy-dependent morphologies show similarities with feature D: they exhibit clumpy emission mainly in the $0.7-1.1\,\si{keV}$ band which may be protruding outward, surrounded by diffuse, softer emission, likely from shocked ISM. 
Spectral modelling indicates that all three clumps exhibit enhanced elemental abundances, as expected if the features correspond to ejecta clumps. It is interesting to observe that the spectra of two features, G and I, exhibit a clear signature of \ion{Si}{xiii} emission at around $1.85\,\si{keV}$, indicating a significant amount of silicon mixed into the ejecta  \citep[see also][]{Garcia17}, which implies an origin at a deeper layer in the progenitor star than for instance feature H. 
Feature K exhibits a quite unique spectrum. It can be fitted either with a strong departure from CIE, at an ionization age of $\tau \sim 10^{10}\,\si{s.cm^{-3}}$, or with a combination of low-temperature plasmas. In addition, a pronounced emission line at $0.43\,\si{keV}$ is present. Our modelling attributes this line to a super-solar nitrogen abundance, which we do not observe in any single other investigated region. 
Feature K appears to intersect with an extended soft filament, interpreted as residual stellar wind material by \citet{Sapienza21}. Therefore, one may suspect that the observed clump consists of nitrogen-rich material from the progenitor's stellar wind, which was only recently overrun by the blast wave.

Region M presents a close look at thermal emission close to the forward shock, focussing on a relatively ``clean'' location, where the shock appears close to plane-parallel. Our imaging reveals a clear hardness gradient, with the soft band becoming increasingly dominant close to the shock front. 
This is confirmed by the spectra extracted from regions M1 and M2, with the former exhibiting its peak count rate at the extremely low energy of $0.25\,\si{keV}$. 
While the 2TNT model is preferred for both regions from a statistical point of view, physically, one would expect NEI plasma this close to the shock front. 
Intriguingly, our spectral fits with the TNT model attribute the observed difference in hardness mostly to different ionization ages in the two regions. While the almost perfect agreement in plasma temperature is probably partly coincidental, an inward increase of the ionization age would be perfectly consistent with expectations, as material located further downstream should have interacted with the forward shock at an earlier time. 
An important point to make is that the extremely high silicon abundance measured in region M1 is most likely not physical, as we expect this region to be dominated by shocked ISM. Instead, it probably originates from the highly statistically significant downward curvature of the spectrum below $0.25\,\si{keV}$, which our model can only reproduce by increasing both the emission from the silicon L-shell and the absorption column density. In reality, this curvature could be caused for instance by an overestimated response of eROSITA at these very low energies, or by the true energy dependence of the absorption cross-section differing from the {\tt TBabs} model. Generally, we argue that any potentially enhanced silicon abundance is only credible in combination with the detection of \ion{Si}{xiii} emission at $1.85\,\si{keV}$.

Regions N and V were selected for further study due to their appearance as broad and cold filamentary structures, and, in the case of region V, for the apparent iron enrichment encountered in Sect.~\ref{Spectroscopy}. Feature N is found to exhibit an intriguing morphology in the soft band below $0.7 \,\si{keV}$, as it appears as a clumpy structure, with a well-defined brightened edge at its eastern boundary. 
The spectra of the two regions appear quite similar, revealing soft thermal components with little departure from CIE. This is superimposed by significant non-thermal emission from the PWN.  
In both regions, our spectral fits argue in favor of super-solar abundances of iron.  
One may thus speculate that relatively cool heavy-element ejecta may be present in these thick filamentary structures. However, we found that, in our modelling, the iron abundance 
is heavily correlated with other model parameters, such as plasma temperature or ionization age (for the TNT model), and thus we cannot exclude that the detected overabundances might also be a signature of systematic effects not considered by our models.

Finally, features T, W, and X were defined around regions found to be rich in typical ejecta elements in Sect.~\ref{Spectroscopy}, those being the northwest periphery of the pulsar (T), the northern interior of the shell of Vela Jr. (W), and the Vela cocoon (X). All three regions are characterized by a relative high flux in the intermediate energy band in imaging, which is the energy band that includes neon or iron emission lines. 
First, we note that the power law spectral index $\Gamma$ found in region X agrees very well with the typical value $\Gamma \approx 2.2-2.3$ observed elsewhere for the inner portion of the Vela cocoon \citep{LaMassa08,Slane18, HESS19}. This shows that our data set allows us to characterize bright non-thermal emission reasonably well. 
All three investigated regions exhibit similar thermal plasma properties, with moderate NEI at $\tau \sim 2\times 10^{11}\,\si{s.cm^{-3}}$. Also, they appear to show comparable abundance patterns, with enhancements in the light ejecta elements oxygen, neon, and magnesium, at $\mathrm{O/H} \sim 1.4$, $\mathrm{Ne/H} \sim 3$, $\mathrm{Mg/H} \sim 2.5$ (for the TNT model). Intriguingly, all three regions exhibit significant \ion{Si}{xiii} line emission, consistent with a strong contribution of silicon to the emission of these ejecta-rich clumps. In contrast, no indication of a supersolar iron enrichment is observed.

\section{Discussion \label{Discussion}}

\subsection{X-ray absorption toward Vela in a multiwavelength context}

\begin{figure}
\centering
\includegraphics[width=1.0\linewidth]{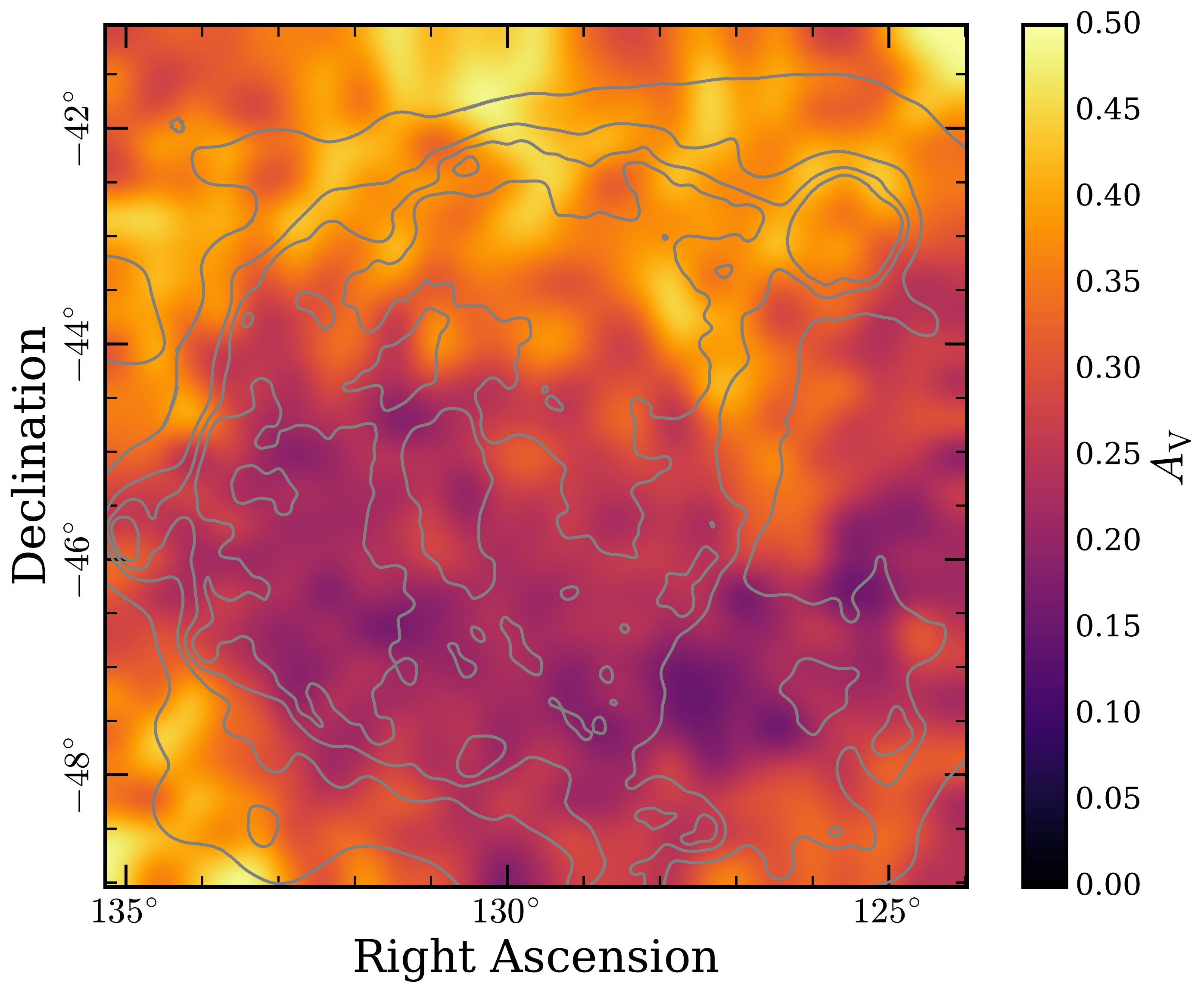} 
\caption{Spatially resolved integrated optical extinction in the direction of Vela, assumed to lie at a distance of $290\,\si{pc}$, using the catalog based on the {\tt StarHorse} code \citep{Anders22}. The color bar indicates the value of $A_{\rm V}$ in units of magnitude. Contours as in Fig.~\ref{Spectroscopy_2TNT}. 
}
\label{AV_StarHorse}
\end{figure}

\begin{figure*}
\centering
\includegraphics[width=18cm]{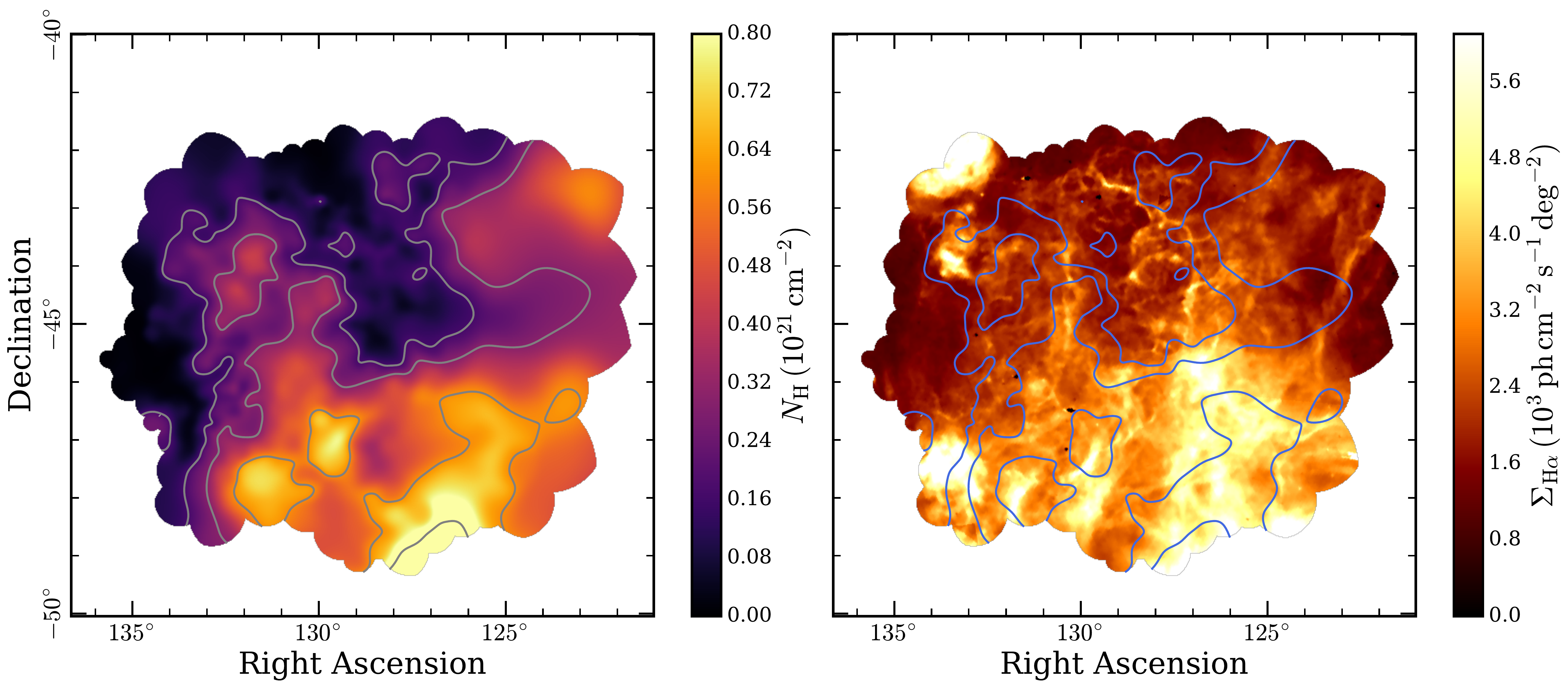} 
\caption{Comparison between X-ray absorption and H$\alpha$ emission. The left panel displays a smoothed map of our measured $N_{\rm H}$ values toward Vela, with contours at the level of $N_{\rm H} = 1.7, 3.0, 6.0, 10.0 \times 10^{20}\,\si{cm^{-2}}$. The same contours are overlaid on a map of the surface brightness of H$\alpha$ emission in the region in the right panel. 
Both panels employ a linear color scale, and have been masked as in Fig.~\ref{Spectroscopy_2TNT}.
}
\label{Halpha}
\end{figure*}

The absorption column density toward Vela (Fig.~\ref{Spectroscopy_2TNT}) exhibits a rich structure with a strong north-south asymmetry and several clumps or filaments of enhanced absorption, tracing the inhomogeneous distribution of foreground material. 
Our absorption map strongly resembles that based on ROSAT data \citep{LuAschenbach00}, even though we seem to recover a systematically larger absorption column. This may be due to differences in the model choice, such as different thermal plasma or absorption models.  
As noted already by \citet{LuAschenbach00}, the large-scale distribution of $N_{\rm H}$ appears to be anti-correlated with the observed distribution of neutral hydrogen and CO in the direction of Vela: while the latter two appear most concentrated toward the north and east \citep{Dubner98,Moriguchi01}, we observe the highest X-ray absorption in the south.   
However, even though tentative signs of interaction between the SNR and neutral material are found, the major drawback of the maps of \citet{Dubner98} and \citet{Moriguchi01} is that they may include material that is in reality located behind Vela. 
In contrast, the three-dimensional dust-extinction information extracted from  the photometric and astrometric data analyzed with the {\tt StarHorse} SED fitting code \citep{Queiroz18, Anders22} allows us to predict the amount of intervening material in a distance-resolved manner. 
We achieved this based on the integrated optical extinction $A_{\rm V}$ in the catalog, which we evaluated in a slice around the precisely known distance of $290\,\si{pc}$ \citep{Dodson03} to Vela. Using the fact that optical extinction and X-ray absorption are expected to correlate with an approximate relation of $N_{\rm H} = 2.08 \times10^{21}\,A_{\rm V}\,\si{cm^{-2}}$ \citep{Zhu17}, 
a comparison between the two is expected to reveal similar spatial distributions of intervening material. 

As displayed in Fig.~\ref{AV_StarHorse}, also the prediction based on {\tt StarHorse} 
is opposite to what is seen in X-ray absorption. Similarly to the neutral gas tracers, it exhibits the largest column density of absorbing material in the north of Vela, contrasting the observed peak of X-ray absorption in its south.
Even though optical and X-ray absorption mostly trace components of the ISM that are not necessarily identical, dust grains and metals, respectively, such a strong disagreement between the two is quite unexpected, as their tight correlation is well established on larger scales \citep{Predehl95, Guever09, Zhu17}.

A possible avenue toward the resolution of this conundrum may lie in dust destruction: 
given the fact that the massive progenitor of the Vela SNR likely exploded in a star-forming region which may have been dust-rich, and that the immediate neighborhood of the sun appears to be almost dust-free \citep{Lallement19}, it seems reasonable to assume that the majority of intervening dust was located close to the SNR. Assuming a shock velocity $\gtrsim 200 \,\si{km.s^{-1}}$ \citep{Slavin15}, it is thus imaginable that the destruction of dust grains within the reach of the SNR shock wave has significantly reduced the integrated dust budget along the entire line of sight to Vela \citep{Zhu19}.    
In this scenario, the X-ray absorption would be unaffected, as heavy elements would still be present, but optical light would be less attenuated. Figure \ref{AV_StarHorse} may support this idea, as the depression in $A_{\rm V}$ appears to coincide mostly with the southern portion of the shell. An analogous scenario has been invoked to explain the apparent lack of optical extinction toward Cas A \citep{Predehl95, Hartmann97}.  
Alternatively, one could also imagine that dust destruction has occurred within the local hot bubble, which itself is believed to be the signature of heating by supernovae which occurred millions of years ago \citep{Zucker22}.
However, in any case, it is unclear why the destruction of dust would show an asymmetry along our line of sight, and preferentially occur in the southern portion of Vela, unless a larger fraction of dust were to be located out of reach of the shock wave in the north. A brief inspection of the local mid-infrared emission at $22\,\si{\mu m}$ surveyed by AllWISE \citep{AllWISE}, which could trace shock-heated dust in the process of being destroyed \citep[e.g.,][]{Arendt10}, reveals no clear correlation with the observed X-ray emission.    

A further serious discrepancy exists between the dust extinction within the surroundings of the sun predicted by {\tt StarHorse} and the results of three-dimensional inversion techniques \citep{Lallement19, Lallement22, Vergely22}, with the latter predicting a smaller integrated extinction $A_{\rm V} < 0.1$ up to $290\,\si{pc}$ in the direction of Vela. This disagreement may be caused by the large statistical scatter for individual stars in {\tt StarHorse}, which may drive the estimated local average extinction to unrealistically large values, in comparison with the hierarchical spatial inversion technique which reconstructs the differential extinction in each volume element. As an example, in the catalog of \citet{Anders22}, the average extinction toward stars within $100\,\si{pc}$ of the sun in the direction of Vela is around $0.15$, which seems significantly too large for objects within the local hot bubble. 

As noted by \citet{LuAschenbach00}, the X-ray absorption column density appears to correlate well with optical H$\alpha$ line emission. 
In Fig.~\ref{Halpha}, we visualize this by comparing a smoothed version of our $N_{\rm H}$ map (Fig.~\ref{Spectroscopy_2TNT}) to the smoothed continuum-subtracted H$\alpha$ map of the region, adapted from the public Southern H-Alpha Sky Survey Atlas \citep[SHASSA;][]{SHASSA}. An almost step-like increase from north to south is clearly visible in both X-ray absorption and optical line intensity. 
The H$\alpha$ emission in an evolved SNR is expected to be caused by  
low-velocity shocks interacting with a high-density, or partially neutral, medium, resulting in the observed filamentary structure. The presence of the higher ISM density in the south appears to be contradicted by both our density map, computed from the emission measure of the spectra (Fig.~\ref{Spectroscopy_2TNT}), and the larger shell radius there \citep{Slane18}, both of which imply the densest ISM in the northeast. One may however speculate that the ISM could be, on average, more clumpy in the south, despite exhibiting a lower mean density. This would explain the larger shock radius there, as the expansion behavior would be determined by the least dense component \citep{Sushch11}. Furthermore, this scenario could be reconciled with the lower X-ray emission measure in the south, if the volume filling factor is much smaller than in the north, as would be expected for a clumpy medium. 
If the X-ray emission in the south really originates preferentially from dense filaments and clumps, it is imaginable that these locally contribute a large amount of additional X-ray absorption, on top of what would be expected from the ``spatial average'' given by optical extinction or neutral gas tracers. 

Finally, despite the observation of enhanced X-ray absorption in the south both here and in the ROSAT data \citep{LuAschenbach00}, one should keep in mind that our measurements of $N_{\rm H}$ are model-dependent, and that the true unabsorbed spectrum emitted by the SNR is not perfectly known. This is somewhat concerning, as we do observe a slight anticorrelation between $N_{\rm H}$ and the mean plasma temperature $kT_{\rm mean}$, both when looking at the covariances of our constraints within individual regions (typical correlation coefficient $\rho \sim -0.5$), and at the global distribution of values across different regions (Fig.~\ref{Spectroscopy_2TNT}).      
While we believe it is unlikely that the observed distributions of $N_{\rm H}$ or $kT_{\rm mean}$ are entirely spurious, this demonstrates our dependence on the uncertain intrinsic shape of the source spectrum when modelling X-ray absorption in CCD-resolution spectra.

\subsection{Thermal emission from the Vela SNR}
\subsubsection{Temperature distribution and shocks}

Our eRASS:4 data set has allowed us to identify the multi-component nature of the thermally emitting plasma in the Vela SNR in imaging. Cooler material seems to be concentrated in several thick shells and filaments, whereas the hotter component appears to dominate in thin radially oriented structures. 
This complex morphology can clearly not be described in its entirety with spherically symmetric models of the density structure of SNRs. Nonetheless, it is tempting to identify the colder thermal component as representative of a ``typical'' SNR shell, as it exhibits several smooth structures which appear to delineate the local boundary between shocked, X-ray-emitting and unshocked, X-ray-dark ISM, in particular in the northeast. The thinner radial filaments making up the hotter component could be associated to overdense clumps originating from deeper layers, which are now penetrating outward, due to the smaller deceleration they experience. This would be akin to the more pronounced shrapnels, which have already protruded out further into the unshocked ISM. 

Our spatially resolved spectroscopy in Sect.~\ref{Spectroscopy} demonstrates that the median temperature of X-ray emitting plasma in Vela is very low, at $kT \sim 0.19\,\si{keV}$. Assuming that this value is approximately representative of the equilibrium temperature reached by the plasma in the shell a sufficiently long time after its forward shock interaction, this implies a late-time shock velocity of \citep{Vink12}
\begin{equation}
    v_{s} = \left (\frac{16}{3}\frac{kT}{\bar{m}}   \right )^{1/2} \sim 400\,\si{km.s^{-1}}. 
\end{equation}
Here, we have assumed an average particle mass of $\bar{m} = 0.61\,m_{p}$, typical for a fully ionized plasma of ISM composition.
If we further assume that the forward shock expansion of Vela has been approximately following the expectation for a Sedov-Taylor blast wave throughout the majority of its lifetime, we obtain a crude estimate for the SNR age of $t = (2r_{s})/(5v_s) \sim 20\,\si{kyr}$, given a current shell radius $r_{s} \approx 20\,\si{pc}$. 
While this value is around a factor of two larger than the characteristic  age of the Vela pulsar \citep{Manchester05}, a pulsar age up to $30\,\si{kyr}$ is considered plausible, given the uncertain spin-down history of the frequently glitching object \citep{Espinoza17}. Therefore, an age around $20\,\si{kyr}$ for the Vela SNR and its pulsar is certainly reasonable, for instance if the pulsar was born with a non-negligible fraction of its present-day spin period, or if it has exhibited a higher braking index in the past. 

A close look at an almost plane-parallel portion of the forward shock (region M in Sect.~\ref{DetailedFitSection}) has revealed a drastic softening of the thermal X-ray emission toward the shock front. 
This could point toward the strong underionization of the plasma right behind the shock, suppressing the line emission from high ionization states, whereas plasma further downstream likely has almost reached CIE. 
While our spectroscopic ionization age measurements are extremely uncertain, in principle, they can be related to the expected amount of time passed between shock interaction in the two regions. Assuming a shock velocity of $400\,\si{km.s^{-1}}$, and a compression factor of $4$, one expects shocked material to move downstream at around $100\,\si{km.s^{-1}}$ in the rest frame of the shock. Given the angular separation of around $15\arcmin$ between the two fitted regions, this implies a difference in shock ages on the order of $\Delta t_{s}\sim 10\,\si{kyr}$.
This can be compared with the (quite uncertain) ionization age difference $\Delta \tau \sim 6\times10^{11}\,\si{cm^{-3}.s}$, to estimate the post-shock electron density $n_{e} \sim  \Delta \tau / \Delta t_{s} \sim 2\,\si{cm^{-3}}$,
corresponding to a proton density around $n_{\rm H} \sim 0.4\,\si{cm^{-3}}$ in the unshocked ISM. This value is a factor of a few higher than ISM densities typically inferred from X-ray emission in the east of Vela \citep{Katsuda05,Yamaguchi09}, which one may justifiably attribute to the large uncertainties involved, for instance when estimating the past velocity of the blast wave. 
However, alternatively, a resolution might also be given by a clumpy ISM, in which a low-density component controls the expansion behavior, while a denser component acts as the main source of X-ray emission \citep{Sushch11, Slane18}, biasing the spectroscopically estimated parameters.

Shrapnel D is an archetypical target for studying the temperature and morphology of an overdense clump, which has overtaken the main blast wave and is penetrating the unshocked ISM. In Sect.~\ref{DetailedFitSection}, we have found a strong temperature gradient across the feature: the hottest material is encountered at the presumable apex of the structure, where the highest concentration of ejecta material is also observed, whereas much colder plasma is found in the outer portions of the bow shock.
\citet{Miceli13} successfully reproduced the observed X-ray brightness distribution of shrapnel D, by modelling the evolution of a moderately overdense ejecta clump, initially located at $1/3$ of the ejecta radius, throughout the lifetime of an SNR. While they did not publish a prediction on the observed spectrum in different regions of the shrapnel, Fig.~5 in their paper seems to indicate that the densest region, which contributes the brightest X-ray emission, should exhibit a lower temperature than the rest of the bow shock.
This appears to be in tension with our observation of comparatively hard emission originating from a hot plasma at the X-ray-brightest portion of the shrapnel.
This could indicate that an additional source of heating, for instance by the reverse shock, may have affected the ejecta overdensity during its outward propagation. Alternatively, the initial conditions of \citet{Miceli13}, where, to maintain pressure equilibrium, the overdensity has a much lower initial temperature than ambient ejecta material, might be violated in practice. 

\subsubsection{Ejecta inside and outside the SNR shell}
\paragraph{Spatial distribution}

Throughout this work, we have encountered signatures of ejecta of the Vela SNR both in spatially resolved spectroscopy, agnostic to feature morphologies (Sect.~\ref{Spectroscopy}), and in the dedicated spectral analysis of interesting structures (Sect.~\ref{DetailedFitSection}). 
As previous works have found, X-ray emitting ejecta are present both inside  and outside the limits of the SNR shell. Clumpy features located outside the SNR shell in the plane of the sky are relatively easily identifiable as such, as there is no background emission from swept-up ISM. This is why the original six shrapnels were the first identified ejecta clumps in Vela \citep{Aschenbach95,Miyata01,Katsuda05,Katsuda06,Yamaguchi09}.
It has been shown, here and in \citet{Garcia17}, that several features emanating from the shell in the southward direction are rich in neon, magnesium, and silicon, and may thus correspond to dense ejecta clumps, too. Their less prominent appearance may be linked to a thinner ISM in the southwest direction \citep{Sushch11,Slane18}, leading to a later breakout from the forward shock. 
\citet{Miceli08} found signatures of several clumps of neon- and magnesium-rich material close to the northern rim of Vela, which they interpreted as ejecta shrapnels located inside the SNR shell in projection. Our analysis has revealed several peaks in metal abundances inside the shell, which may constitute similar cases, for instance close to the eastern rim  (region W in Sect.~\ref{DetailedFitSection}).  
This may indicate that in the evolved Vela SNR, a significant fraction of ejecta heated to X-ray-emitting temperatures is located in outward-protruding overdense clumps, many of which have overtaken the main SNR blast wave already. 

On the other hand, \citet{Slane18} observed a concentration of X-ray emitting ejecta also at the very center of Vela. They provided an interesting model to explain the morphology of the ejecta-infused cocoon, in which the northeast portion of the reverse shock arrives at the pulsar's location first, asymmetrically crushing the PWN. This leads to a shift in the peak of the distribution of both relativistic electrons and ejecta toward the southwest, consistent with the current location of the cocoon. Our analysis clearly confirms this ejecta enhancement inside the cocoon. Furthermore, our maps in Fig.~\ref{Spectroscopy_2TNT} show that the associated peak in oxygen and neon abundance is indeed confined to a narrow stripe extending southward from the pulsar.

In addition, we have discovered a further abundance enhancement about one degree northwest of the pulsar (region T in Sect.~\ref{DetailedFitSection}). While one may naturally ascribe this to a further isolated ejecta clump seen merely in projection, we believe that our X-ray images may indicate a direct association between the pulsar and this feature. In particular, at intermediate energies in Fig.~\ref{VelaImage_Linear}, one can see a thin filamentary structure emanating from the vicinity of the pulsar and connecting to our ejecta-rich feature. While significantly fainter, its thin, curved morphology appears similar to that of the outer portion of the cocoon. 
Even though this apparent connection with the pulsar may be merely coincidental, it is intriguing to note that the apparent base of our supposed filament is located in the northwest of the pulsar, approximately in the direction in which the jet of the Vela pulsar is being launched \citep{Helfand01,Pavlov03}. 
It may therefore be possible that energy input from mildly relativistic particles in the polar outflow of the pulsar plays a role in powering thermal emission in this region. 
Alternatively, observing the morphological similarity with the cocoon, one may suspect a similar scenario as in \citet{Slane18}: the thin filamentary structure could have been created by anisotropic crushing of ejecta-rich material by a secondary shock, causing the observed elongated shape. 

A final interesting observation regarding the distribution of ejecta in Vela is that regions rich in ejecta seem to exhibit systematically higher temperatures, irrespective of the exact model setup (see Figs.~\ref{Spectroscopy_2TNT} and \ref{Spectroscopy_TNT}).  
The observation of this trend with two separate models makes us confident in its physical origin, rather than an origin in a systematic modelling issue. 
While the majority of X-ray emitting ISM in Vela has likely been heated by a decelerated forward shock, one may thus conclude that the interaction between ejecta clumps and the reverse shock has heated the ejecta to systematically higher temperatures. If we, conversely, assume that hot thermal emission in Vela tends to be linked to ejecta material, we can speculate that most structures that appear dominant in imaging in the $0.7-1.1\,\si{keV}$ energy band (Fig.~\ref{VelaImage_Linear}) are rich in ejecta. This scenario is especially intriguing because of the many radially oriented features visible in this energy band, which can be interpreted as ejecta clumps protruding outward, either through the decelerated shell associated to the forward shock or through the unshocked ISM.

\paragraph{Composition}

\begin{figure}
\centering
\includegraphics[width=1.0\linewidth]{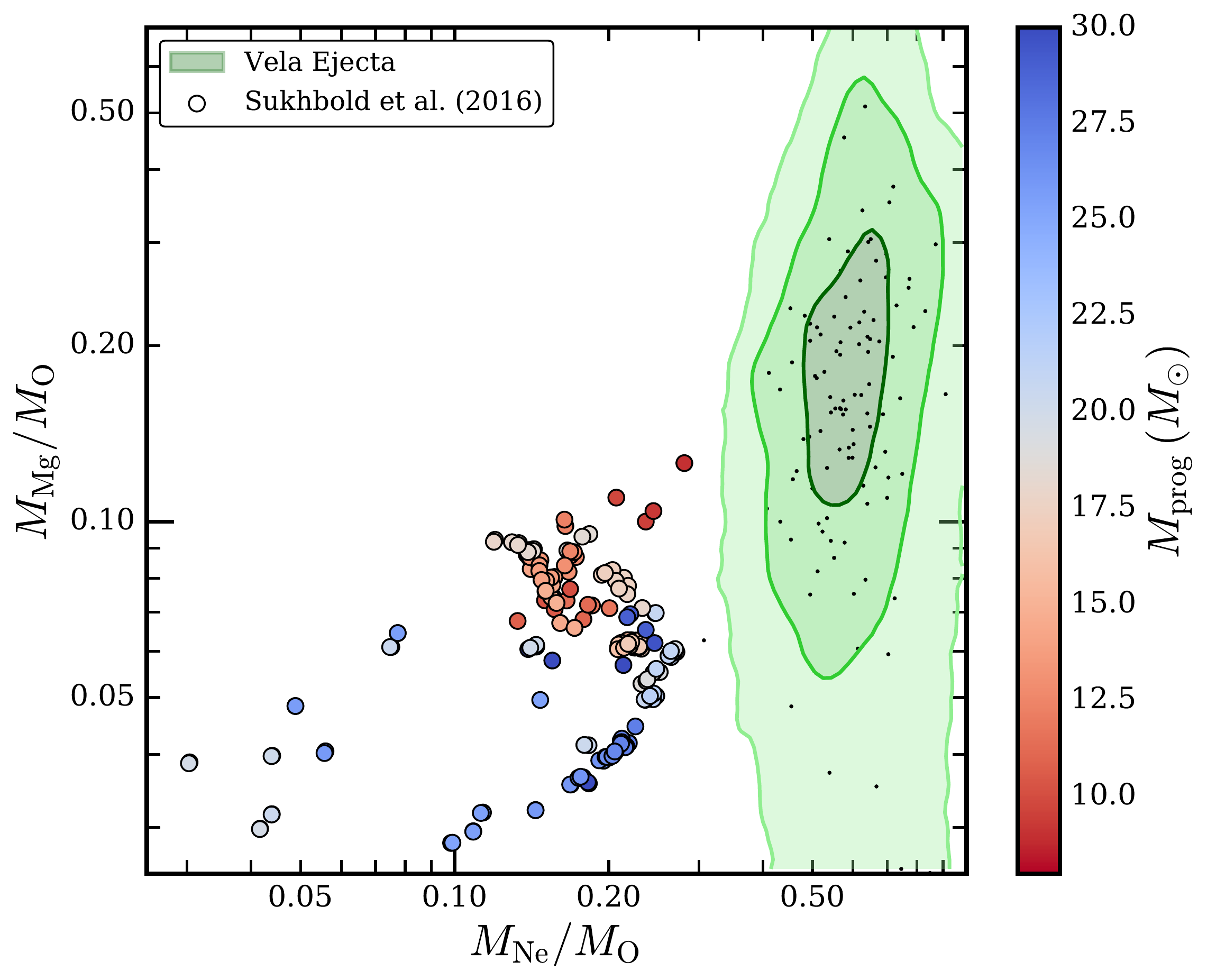} 
\caption{Comparison of observed light-element ejecta to theoretical predictions. 
In green, we show the one, two, and three sigma contours of the distribution of neon-to-oxygen and magnesium-to-oxygen mass ratios in the ejecta of Vela, based on the 2TNT model. The locations of the individual bins in this parameter space are indicated with black dots.
This is compared to the predicted total element ratios for individual explosions in the simulations by \citet{Sukhbold16}, where the color coding of the circular markers indicates the simulated progenitor mass. 
}
\label{Abundance}
\end{figure}

Several patterns emerge, concerning elemental abundances in the ejecta of Vela: throughout the SNR, the distributions of oxygen, neon, and magnesium show peaks which are generally well correlated, as expected for a common origin of these elements in the progenitor star. However, heavier, explosively synthesized elements, such as iron and silicon, seem to not follow this correlation. 
This can be observed in particular for the Vela shrapnels: while the head of shrapnel D exhibits a large concentration of oxygen, neon, and magnesium, little evidence for silicon or iron emission is present. This is in contrast to shrapnel A, which exhibits a large amount of silicon ejecta \citep{Miyata01,Katsuda06}, with little evidence for an enrichment in lighter elements. This feature is particularly intriguing, since the narrow opening angle of the associated Mach cone indicates a higher current velocity than for the other shrapnels. This implies that shrapnel A has overtaken clumps formed in outer ejecta layers due to either an extremely high density contrast or a large initial velocity, possibly as part of a silicon-rich jet from deeper ejecta layers \citep{Garcia17}.
The observed dichotomy in composition is also found for the ejecta clumps studied in the south of Vela (see Sect.~\ref{DetailedFitSection}): the relative concentration of silicon with respect to neon and magnesium is much higher for features G and I than for feature H, implying an origin at different depths in the progenitor despite their similar angular distances from the center.  

While silicon ejecta are clearly present in several clumps inside and outside the shell, evidence for X-ray-emitting iron ejecta is sparse. Shrapnel A 
does exhibit significant iron L-shell emission. However, given the small absolute abundances of all typical metals except silicon \citep{Katsuda06}, one might ascribe this to a strong contribution of ISM with about solar iron abundance to the emission of the ejecta clump. 
The soft filament labelled feature V is a further location where tentative enhanced emission from \ion{Fe}{xvii} is observed. However, detailed follow-up analysis would be necessary to clarify whether this is actually caused by enhanced iron abundances, or may be explained by a particular superposition of plasma temperatures and/or ionization states. Furthermore, one should keep in mind that numerous atomic transitions contribute to the observed iron L-shell emission \citep{AtomDB}, with uncertain emissivities at the low fitted temperatures \citep[e.g.,][]{Heuer21}. Hence, the model itself may constitute a systematic error source when searching for cool iron ejecta.  
In principle, consulting the comprehensive simulations of core-collapse supernova nucleosynthesis by \citet{Sukhbold16}, a comparable concentration of iron and oxygen ejecta (relative to the solar composition) should be present in the integrated yield of the supernova, at least for relatively light progenitors ($\lesssim 12\,M_{\odot}$). Given the fact that the reverse shock has likely traversed and reheated the entire ejecta material in the northeastern portion of Vela \citep{Slane18}, one would probably expect to observe some X-ray emitting iron-rich ejecta there. 

An interesting observation regarding the composition of ejecta in Vela is that neon and magnesium seem to be consistently enriched with respect to oxygen, when compared to solar abundances. Typical abundance ratios in ejecta-rich regions appear to be around $\langle \mathrm{Ne/O} \rangle \sim 2.5$ and $\langle \mathrm{Mg/O} \rangle \sim 2.0$. The corresponding typical mass ratios $M_{\rm Ne}/M_{\rm O} \sim 0.6$ and $M_{\rm Mg}/M_{\rm O} \sim 0.15$ may thus be representative of the composition of the outer ejecta layers during the explosion. 
Similar compositions were observed by other authors for the ejecta located inside the shell in projection, and in several shrapnels outside the shell \citep{Miceli08,Katsuda05,Katsuda06,Yamaguchi09}.
Considering this ubiquitous trend, it is important to note that most studies of core-collapse supernova nucleosynthesis do not predict strongly super-solar concentrations of neon or magnesium with respect to oxygen for any progenitor mass 
\citep[e.g.,][]{Woosley95,Rauscher02,Sukhbold16}, at least when integrating over the total ejecta yield. 
In Fig.~\ref{Abundance}, we indicate the observed distribution of O-Ne-Mg abundance ratios in the ejecta-rich regions of Vela, defined to include all those Voronoi bins in Fig.~\ref{Spectroscopy_2TNT} with super-solar oxygen content, $\mathrm{O/H} > 1$. This is compared to explosion models of different progenitor masses by \citet{Sukhbold16}, none of which seem to reach the relative neon- and magnesium-concentration which we observe throughout Vela. A very similar trend is observed when using the results of the TNT model instead. 
We note that, for several $15\,M_{\odot}$-progenitor models, the simulations of \citet{Fryer18}, which assume a broad parametrization of the ``supernova engine'', seem to be able to produce similar abundance patterns as observed here. 
However, given the wide range of input values assumed for the explosion energy and timescale, and the resulting extremely wide spread of element abundances in their models, we believe that some skepticism is warranted concerning the applicability of their results to our case.

Qualitatively, the low relative oxygen concentration might be understood as a signature of a relatively light progenitor, as the implied high central density during helium burning would tend to disfavor the production of oxygen compared to carbon \citep{Woosley02}. This could ultimately lead to a lack of oxygen ejecta with respect to the products of carbon burning, including neon and magnesium.    
Nonetheless, a major caveat of performing such comparisons is that, to our knowledge, all currently available nucleosynthesis predictions for a large sample of progenitors are based on one-dimensional supernova models, due to computational limitations. Naturally, these do not capture three-dimensional effects during the explosion, and their consequences on nucleosynthesis in individual ejecta clumps. We therefore hope to learn in the future, if the predictions of potentially more realistic arrays of two- or three-dimensional models of supernova nucleosynthesis differ substantially from the studies discussed above. 
owever, it is also imaginable that the observed unusual patterns in X-ray abundances are caused by a systematic observational issue, such as the preferential cooling of oxygen ejecta out of the X-ray regime, or a spectral modelling issue, like an inadequate model or an insufficient number of model components. 
A further possibility to consider is that the depletion of ejecta material onto dust grains may also modify the observed composition in X-rays \citep[e.g.,][]{Hwang08}, and may enhance the inferred relative abundance of neon, which, due to its noble gas nature, is not depleted. However, this effect cannot be invoked to explain the enhanced $\rm Mg/O$ ratio, as magnesium would be strongly depleted.

\subsection{The vast extent of non-thermal X-ray emission from Vela X} \label{DiscVelaX} 

\subsubsection{X-ray emission from arcsecond to degree scales}
\paragraph{Morphology}

\begin{figure}
\centering
\includegraphics[width=\linewidth]{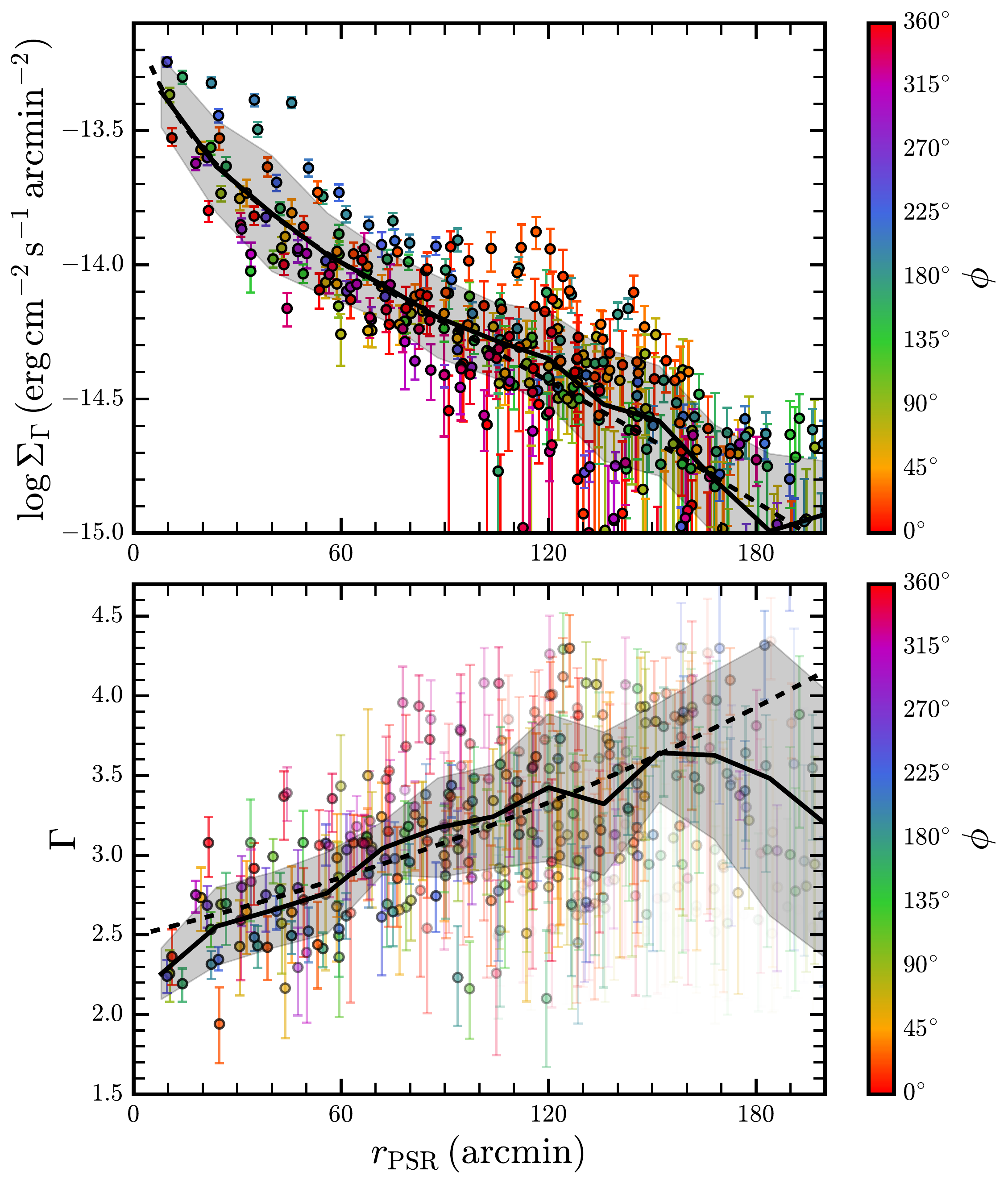} 
\caption{Radial evolution of non-thermal emission from Vela X. The top panel depicts the non-thermal surface brightness $\Sigma_{\Gamma}$ of each Voronoi bin in the $1.0-5.0 \,\si{keV}$ range from the 2TNT model, plotted against its angular distance $r_{\rm PSR}$ from the pulsar. The bottom panel shows the measured spectral index $\Gamma$ versus $r_{\rm PSR}$, where the transparency of the individual markers indicates the relative contribution of the non-thermal component to the hard spectrum. 
In each panel, the marker color indicates the direction of the bin, that is, its relative angle $\phi$ (east of north) with respect to the pulsar position.
In both plots, the solid black line and shaded area indicate the weighted average and associated standard deviation of the measured quantities, determined within narrow radial bins. 
The dashed lines indicate the best fit of the brightness profile (top) and radial spectral softening (bottom), using the models described in the text. 
All bins located within $1.2^{\circ}$ from the center of Vela Jr.~have been excluded from this figure to remove any contamination by its non-thermal shell.
}
\label{Nonthermal_Radii}
\end{figure}

The X-ray synchrotron emission associated to the plerion of the Vela pulsar has been investigated extensively in previous studies. On small scales, a complex, spectrally hard structure similar to the one in the Crab nebula is visible \citep{Helfand01,Pavlov03}. A mildly relativistic jet-counterjet structure emanates from the pulsar toward northwest and southeast, approximately parallel to its spin axis and proper motion direction. In the pulsar's equatorial plane, a torus consisting of two arc-like structures, around $50\arcsec$ in diameter, is visible. This likely corresponds to the location of the termination shock of the pulsar wind \citep{Helfand01}, in which the local magnetic field appears to follow a highly ordered toroidal structure \citep{Xie22}. 
This torus is embedded in larger-scale diffuse non-thermal emission, which is brightest toward the southwest. This diffuse component likely constitutes the base of the cocoon which is thought to contain relativistic electrons from the pulsar wind, crushed by a one-sided interaction with the reverse shock \citep{Slane18}. In X-rays, the cocoon extends up until around $1.5^{\circ}$ southwest of the pulsar. 

In this work, we have for the first time surveyed the vast extent of non-thermal X-ray emission beyond the cocoon. While several previous studies have found evidence for a non-thermal X-ray component in other directions from the pulsar \citep{Willmore92,Mattana11,Katsuda11,Slane18}, this study has exposed the entirety of a diffuse synchrotron nebula extending two to three degrees from the Vela pulsar (Fig.~\ref{Spectroscopy_2TNT}). 
The size of our extended PWN is quantified in Fig.~\ref{Nonthermal_Radii}, which displays the evolution of the non-thermal surface brightness $\Sigma_{\Gamma}$ with distance from the pulsar $r_{\rm PSR}$. It reveals a smooth radial decrease of the diffuse nebula's brightness by $1.5$ orders of magnitude across a large portion of the SNR shell. 
Interestingly, the non-thermal emission seems to show a quadrupolar asymmetry, with the largest extent seemingly running along a position angle $\sim 15^{\circ}$ east of north, which, if real, may indicate a preferred direction of particle transport in the nebula.

\paragraph{Size and energetics}
Using our constraints on the brightness of synchrotron emission in the entire Vela SNR (Fig.~\ref{Spectroscopy_2TNT}), we can compute the total non-thermal flux of the plerion. Apart from the central $3\arcmin$ radius around the pulsar which was masked already during spectral fitting, we excluded a $1.2^{\circ}$ region around Vela Jr., and integrated the non-thermal flux within a $4^{\circ}$ radius around the pulsar. 
Since our spectral modelling approach enforces a non-negative value of the power law normalization in each bin, even if the true flux is negligible, it is possible that a simple integration over all bins overestimates the total flux. Therefore, a conservative flux estimate is provided by only including those bins whose flux posterior indicates a nonzero value by at least $5\sigma$ significance. This yields an estimate of  $F_{\Gamma}=4.6\times10^{-10}\,\si{erg.cm^{-2}.s^{-1}}$ 
in the $1.0-5.0\,\si{keV}$ band, and a corresponding synchrotron luminosity of $L_{\Gamma}=4.7\times10^{33}\,\si{erg.s^{-1}}$. 
In order to be able to compare the calculation with published energetics of other X-ray PWNe, we also compute the luminosity in the $0.5-8.0\,\si{keV}$ band, finding $L_{\Gamma}=1.02\times10^{34}\,\si{erg.s^{-1}}$. 
This does not include the emission from the core of the PWN, which would contribute on the level of a few percent only \citep{Helfand01}. While this broad energy band may be considered ``standard'' for the quantification of X-ray fluxes, we note that the extrapolation of the non-thermal component to below $1.0\, \si{keV}$ is somewhat problematic, since at lower energies, its emission is strongly overpowered by the thermal component. Thus, our spectral fits are not sensitive to the true contribution of the non-thermal component there.

Comparing our luminosity with the well-known spin-down power of the pulsar, $\dot{E} = 6.9\times10^{36}\,\si{erg.s^{-1}}$ \citep{Dodson07}, we obtain an approximate PWN efficiency, that is, the ratio of X-ray luminosity to the present-day pulsar spin-down power, 
of $\eta_{\rm PWN} \sim 1.5\times10^{-3}$. 
While this value is still far below the efficiency of young and X-ray-bright PWNe, such as the Crab or PSR B0540$-$69, it is almost two orders of magnitude higher than other estimates for Vela X \citep[$\eta_{\rm PWN} \sim 2\times10^{-5}$,][]{Kargaltsev08}). 
We note however, that this is not really an ``apples to apples'' comparison, as other studies only took the arcminute-size PWN core into account. Our estimate considers a much more extended region and therefore integrates over a much longer history of particle injection from the pulsar wind, with older electrons contributing to the diffuse outer X-ray emission. In any case, our measurement challenges the picture of the Vela plerion being particularly X-ray-underluminous. 

While there does not appear to be any abrupt outer border to the synchrotron emission from our extended PWN, 
we attempted to roughly quantify its size, using a model for the emission profile of a diffusive pulsar halo, assuming a constant magnetic field. 
We modelled the observed radial surface brightness evolution using the brightness profile given by \citet{HAWC17}, for simplicity, evaluated at a single effective energy only (see Appendix \ref{FittingDetails}). 
The best fit is indicated in the upper panel of Fig.~\ref{Nonthermal_Radii}, and corresponds to a characteristic angular size $\theta_{D} = 164\pm6 \,\si{arcmin}$, 
which, at the distance of Vela, corresponds to a physical diffusion radius of $r_{D} = 13.8\pm0.5\,\si{pc}$. 

While the relative size of the nebula in the plane of the sky and compared to its host SNR is astonishing, its absolute physical size is not inexplicably large by itself, considering that TeV PWNe regularly reach tens of parsec in size \citep{Kargaltsev13}. Furthermore, \citet{Bamba10} listed two X-ray PWNe comparable in size to our nebula for pulsars slightly older than the Vela pulsar.
A similar case to Vela X may be given by PSR J1826$-$1334, corresponding to the TeV source HESS J1825$-$137, whose X-ray PWN has been measured to extend up to $17\,\si{pc}$ from the pulsar \citep{Uchiyama09}. It is intriguing that both the characteristic age of $21\,\si{kyr}$ and the spin-down power of $2.8\times10^{36}\,\si{erg.s^{-1}}$ are comparable to the properties of the Vela pulsar. Thus, the energetic PSR J1826$-$1334 and the very extended PWN in HESS J1825$-$137 might be seen as a slightly more evolved analog of the extended PWN in Vela X. By analogy, one would thus expect a rather low magnetic field in Vela X, similar to the measurement $B \sim 4-5 \,\si{\mu G}$ in HESS J1825$-$137, inferred from the relative intensities and extents of X-ray and $\gamma$-ray emission \citep{Principe20}.
Therefore, the measured extent of our PWN is by no means unphysical. In particular, the observation of Vela X presented here seems ideal for the detection of faint, diffuse non-thermal X-ray emission, given its dominant character down to $\sim 1.0 \,\si{keV}$, the physical proximity and negligible foreground absorption of Vela, and the homogeneous and sensitive coverage of the region in the eROSITA all-sky survey. 

Its diffuse and extended nature make it tempting to identify our non-thermal nebula as a ``pulsar halo''. However, it does not technically qualify as such in the evolutionary picture of PWNe presented by \citet{Giacinti20}. Instead, Vela X is classified to be in an intermediate stage \citep[stage 2 in][]{Giacinti20} of its evolution, where a highly irregular relic PWN (i.e., the cocoon) has been created by the interaction between pulsar and SNR, but diffusive escape from the PWN core has become possible. Synchrotron emission from these escaping electrons may provide an explanation for the observed extended X-ray emission of Vela X. This stage differs from the ``true'' halo stage in that the pulsar has not yet left its parent SNR shell, meaning the escaped electrons are not diffusing through the unperturbed ISM, but through a more turbulent medium inside the SNR. Yet, given the large physical size of the X-ray PWN revealed here, it seems quite likely that Vela X is closely related to the population of gamma-ray halos seen around middle-aged pulsars.         

In order to robustly quantify the physical quantities regulating the observed properties of the X-ray PWN, such as the magnetic field and diffusion constant, one would need to model the full X-ray to $\gamma$-ray spectral energy distribution of the region in a spatially resolved manner, which is beyond the scope of this paper. 
However, with the given data set, we can attempt to infer the order of magnitude of the involved quantities in our extended nebula based on the following considerations: 
The ``characteristic'' energy $E_{e}$ of an electron emitting synchrotron emission at an energy $E_{X}$ in a transverse magnetic field $B_{\bot}=B_{\si{\mu G}}\,\si{\mu G}$ can be written as \citep{deJager09}
\begin{equation}
    E_{e} = 220 \times B_{\si{\mu G}}^{-1/2}\,\left (\frac{E_{X}}{1\,\si{keV}} \right )^{1/2}\, \si{TeV}. \label{SyncEnergy}
\end{equation}
The corresponding approximate lifetime of the electron under losses from synchrotron emission and inverse Compton scattering is \citep{Aharonian06}
\begin{equation}
    \tau = 18 \times  \frac{1}{1+0.144\,B_{\si{\mu G}}^2} \,\left (\frac{E_{e}}{100\,\si{TeV}} \right )^{-1}\, \si{kyr}. \label{SyncLifetime}
\end{equation}
Combining this with the expectation for the diffusion constant $D_{B}$ in the Bohm limit 
\begin{equation}
    D_{B} = 3.3\times10^{27} B_{\si{\mu G}}^{-1}\,\left (\frac{E_{e}}{100\,\si{TeV}} \right )\, \si{cm^2.s^{-1}}, \label{BohmDiff}
\end{equation}
one can derive a characteristic energy-independent radius $r_{B} \sim \left( 4 D_{B} \tau \right) ^{1/2}$ covered by an electron transported through Bohm diffusion during its lifetime 
\begin{equation}
    r_{B} \sim 28 \times  \frac{B_{\si{\mu G}}^{-1/2}}{\left ( 1+0.144\,B_{\si{\mu G}}^2 \right )^{1/2}} \, \si{pc}. \label{SyncBohmRadius}
\end{equation}
By equating this with the observed PWN size on the order of $14\,\si{pc}$, 
we can derive an approximate magnetic field strength in the Bohm limit of 
$B_{\bot} \sim 2.3\,\si{\mu G}$. 
This estimate for the extended nebula is less than half of that measured in SED modelling of the cocoon \citep{HESS19}, and slightly below the typical ISM magnetic field strength of $3\,\si{\mu G}$ \citep{Minter96}. 
The lifetime of an electron emitting at  X-ray energies where the non-thermal component is typically best-constrained in our data, $E_{X} \approx 1.6\,\si{keV}$,\footnote{For typical spectra with non-thermal contributions (e.g., regions W and T in Fig.~\ref{DetailedFits}), one can observe that the relative contribution of the non-thermal component is maximal 
at around this energy. } 
is around $\tau \sim 5.2\,\si{kyr}$. 
The corresponding Bohm diffusion coefficient would be 
$D_{B} \sim 2.6\times10^{27}\,\si{cm^2.s^{-1}}$. 

If the above equations are taken at face value, this low magnetic field implies a very high characteristic energy of the electrons responsible for synchrotron emission at $1.6\,\si{keV}$ of 
$E_{e} \sim 180\,\si{TeV}$. 
This energy is more than high enough for escape from the spatial scales of the cocoon, with $B\sim6\,\si{\mu G}$ \citep{HESS19} and a perpendicular extent $\sim1\,\si{pc}$, to be possible if particle transport occurs via diffusion. The escape timescale would be $\sim130\,\si{yr}$ for Bohm diffusion \citep{deJager09, Tang12}, which is much shorter than the expected synchrotron lifetime for this energy in the cocoon around $\tau \sim1.6\,\si{kyr}$. 
However, a fundamental problem with this high electron energy is that it is in apparent contradiction with the energy cutoff at $\sim 100\,\si{TeV}$ inferred for the outer cocoon regions by \citet{HESS19}. If this energy cutoff is indicative of radiative energy losses, rather than the escape of high-energy particles \citep[see][]{Hinton11}, 
it seems doubtful whether more energetic electrons could have survived out to even larger distances.  
Furthermore, our low inferred magnetic field seems questionable given that, even with no energy cutoff, the ratio of the integrated X-ray to the VHE $\gamma$-ray flux in a given region is expected to scale with the magnetic field as $F_{X}/F_{\gamma} \propto B^2$ \citep{Aharonian97}, and that, at this time, most of our extended nebula has not been detected at TeV energies (see below).  

A possible mitigation of the above issues may lie in assuming a higher diffusion coefficient than implied by the Bohm limit, which would increase the inferred magnetic field, and correspondingly reduce the required energy of synchrotron-emitting electrons and their expected brightness in the TeV band. However, the electron lifetime would be reduced accordingly. 
A further fundamental aspect is that, in reality, the synchrotron emissivity of an electron
has a rather wide spectral distribution, which implies that X-ray synchrotron emission is also expected significantly beyond the characteristic energy corresponding to the cutoff.
Hence, it is likely that either the electrons emitting in our extended X-ray PWN exhibit only a small fraction of the age of Vela itself, or that we are observing the significantly steepened part of the non-thermal spectrum located beyond the cutoff energy.

\begin{figure*}
\centering
\includegraphics[width=18cm]{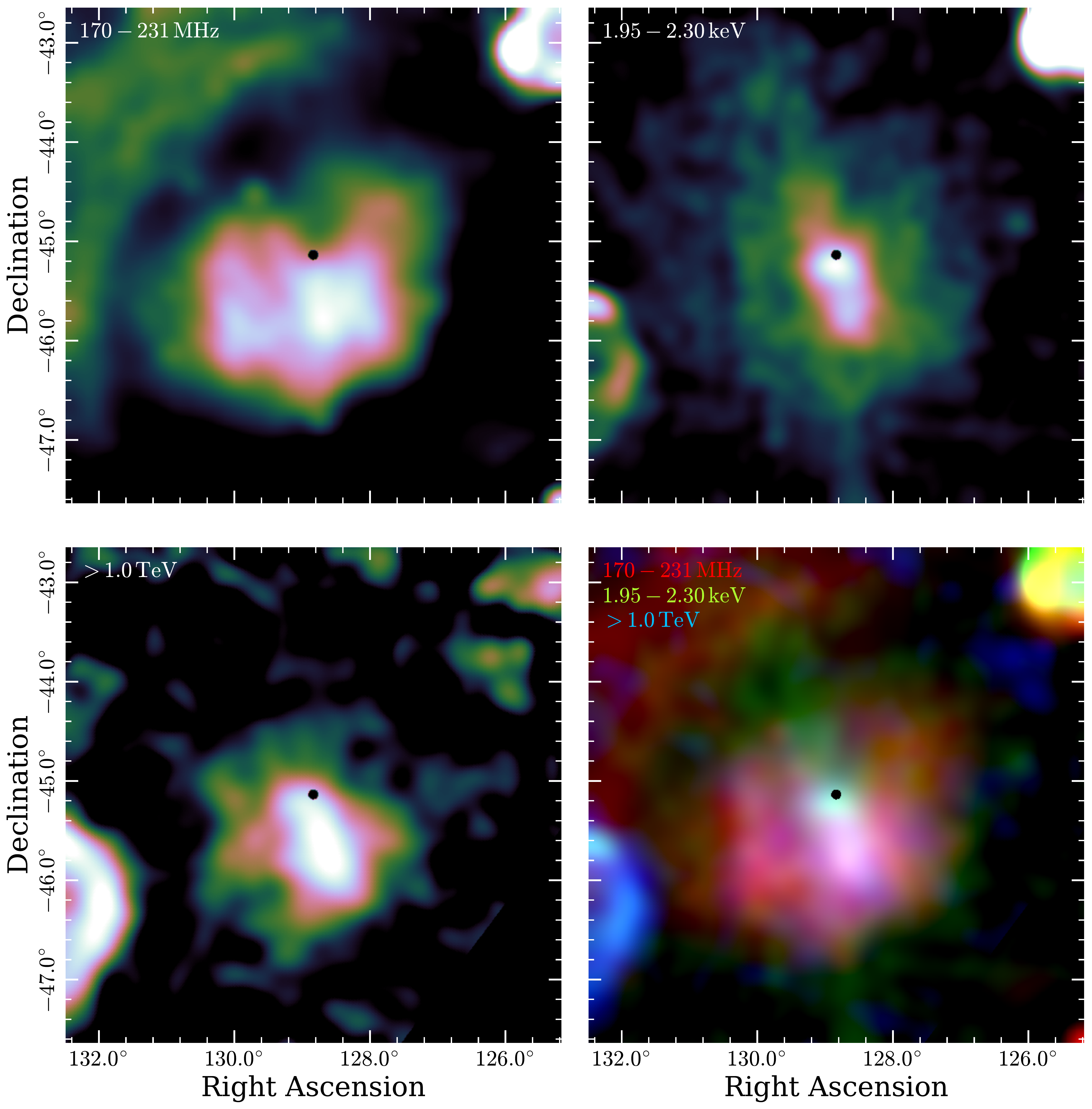} 
\caption{Multiwavelength morphology of Vela X. The individual panels show the morphology of non-thermal emission in a $5^{\circ}\times 5^{\circ}$ region around the Vela pulsar in a low-frequency radio band (top left), in the non-thermal X-ray regime (top right), and at TeV energies (bottom left). The bottom right panel displays a false-color superposition of the three bands. In the radio and X-ray images, a $3\arcmin$ radius around the pulsar was masked, before they were smoothed with a Gaussian kernel of $\sigma=6\arcmin$ to approximately match the resolution of the TeV band. In all three bands, we used a square-root brightness scale, spanning a factor of 30 in dynamic range.}
\label{Multiwav}
\vspace{-0.2cm}
\end{figure*}

\paragraph{Tentative detection of radiative cooling}
Given the vast extent of the detected PWN, it seems natural to expect a significant degree of energy loss via synchrotron and inverse Compton emission affecting the highest-energy electron population, during diffusion away from the pulsar. 
The effect of synchrotron cooling has been directly observed for the electron population in the cocoon, where the energy loss manifests itself in a steepening of the nonthermal X-ray spectrum from a photon index of $\Gamma=2.2$ to $\Gamma=2.6$ within a distance of $100\arcmin$ from the pulsar \citep{Slane18}. 
A similar effect seems to be observed in our data set (see Fig.~\ref{Nonthermal_Radii}), even though our ability to constrain the spectral slope in all but the brightest non-thermal regions is severely hampered by statistical noise. This is due to the relatively short spectral baseline available for constraining $\Gamma$, which is effectively limited by the bright thermal emission from Vela below $1.0\,\si{keV}$ and the sharp drop in instrumental response above $2.3\,\si{keV}$ \citep{Predehl21}, respectively. 
Thus, in order to reduce the potentially spurious impact of soft energies on the measurement of the average non-thermal slope, we introduced a weighting scheme designed to reflect the relative contribution of the non-thermal component to the total observed spectrum above $1.0 \,\si{keV}$ (see Appendix \ref{FittingDetails}). This weighting factor is reflected in the transparency of the data points in the lower panel of Fig.~\ref{Nonthermal_Radii}, and was used to reweight the individual bins to compute the radial averages shown in the figure.    

Despite the large overall noise level in the photon index, a significant outward steepening from $\Gamma \sim 2.2$ in the inner $15\arcmin$ to a maximum of $\Gamma \sim 3.6$ around $160\arcmin$ from the pulsar is apparent in the radial average of the spectral slope. The apparent decrease in $\Gamma$ even further out can likely be explained with our prior becoming dominant, as $\Gamma$ is almost unconstrained there.    
We attempted to test whether the observed increase of $\Gamma$ with $r_{\rm PSR}$ can be reconciled with the expectation from radiative energy loss in a diffusive PWN. To do this, we performed a fit of the pure diffusion model presented by \citet{Tang12} based on \citet{Gratton72} to the observed radial dependence of the photon index. However, given the likely importance of inverse Compton scattering due to the low suspected magnetic fields, we introduced a modification to take into account both synchrotron and inverse Compton losses (see Appendix \ref{FittingDetails}). 
We assumed a distance of $290\,\si{pc}$, a PWN age equal to the approximate age of Vela, $t = 20\,\si{kyr}$, and that our constraints on the X-ray photon index correspond to a measurement at an effective energy of $1.6\,\si{keV}$. The model-predicted power law slope is given by 
\begin{equation}
    \Gamma(r) = 1 - \left. \frac{\mathrm{d} \log P(\nu, r, t)}{\mathrm{d} \log \nu} \right \rvert_{h \nu\,=\,1.6\,\si{keV}} 
\end{equation}
where $P(\nu, r, t)$ corresponds to the synchrotron power radiated at a frequency $\nu$ at a projected radius $r$ from the pulsar, integrated along the line of sight following the modified \citet{Tang12} model. 
In the steady-state regime, where the lifetime of X-ray emitting electrons is shorter than the pulsar age, the predicted slope at a given radius only depends on the energy slope of the injected particles $p$ \mbox{($\mathrm{d} N/\mathrm{d} E_{e} \propto E_{e}^{-p}$)}, and on a combination of diffusion coefficient $D$ and transverse magnetic field $B_{\bot}$, whose values are assumed to be spatially uniform on the relevant scales. The degeneracy between $D$ and $B_{\bot}$ cannot be lifted based on X-ray data alone without further assumptions, but would require the measurement of the level of the corresponding emission at TeV energies. 
The physical parameters constrained by the model are $p$ and a characteristic ``cooling radius'' $r_{C} \coloneqq \left( 4 D \tau \right) ^{1/2}$, describing the degree of radial steepening, where the electron lifetime $\tau$ is computed as in Eqs.~\ref{SyncEnergy} and \ref{SyncLifetime}.  
We extracted the parameter constraints by modelling the range $r_{\rm PSR} < 200\arcmin$ via MCMC sampling with {\tt emcee} \citep{Foreman13}, using a uniform prior on $p$ and a logarithmic prior on $r_{C}$.

The best fit of our radiative cooling model is indicated as a dashed line in the lower panel of Fig.~\ref{Nonthermal_Radii}, and is strongly statistically preferred over the null hypothesis of a radially constant photon index with an estimated Bayes factor \citep[following][]{Robert09} of $\ln \,\mathcal{B} \approx 30$.
We find an electron power law index of $p=4.00\pm0.11$, and a characteristic radius of $r_{C} = 14.7^{+1.3}_{-1.1}\,\si{pc}$.  
This electron index predicts an X-ray photon index of $\Gamma = (p+1)/2 \approx 2.5$ at the center of the PWN, clearly somewhat above the observed value. This may indicate that the electron population powering the extended PWN has already undergone significant losses when leaving the PWN core, in possible conflict with the assumption of an injected electron spectrum without cutoff. 
For a typical ISM magnetic field of $B_{\bot} = 3\,\si{\mu G}$, our measurement implies a diffusion constant of 
$D=3.6^{+0.6}_{-0.5}\times10^{27}\,\si{cm^2.s^{-1}}$. 
However, as shown by \citet{Tang12}, if in reality the diffusion coefficient increases with energy as $D\propto E_{e}^{\alpha}$, the radial spectral index profile would be flattened. In the idealized case of Bohm diffusion ($\alpha=1$), the profile of $\Gamma$ would even become constant, as in the steady-state regime, the cooling radius would be energy-independent.
Therefore, a more realistic constraint on the diffusion constant $D$
at X-ray-emitting electron energies is given by
\begin{equation}
    D = 3.6^{+0.6}_{-0.5}\,\times10^{27} \left( 1-\alpha\right)^2 \,\si{cm^2.s^{-1}}. 
\end{equation}
Given the large statistical and systematic uncertainties of our measurement, and the large number of simplifying assumptions entering the fitted model, this value should be seen only as an order-of-magnitude estimate of the average regime of diffusion in our extended PWN. 
However, our measurement of the radiative cooling radius agrees greatly with the characteristic PWN size of $14\,\si{pc}$, measured via the radial brightness profile, above.
For comparison, the latter value yields a diffusion constant of $D=(3.14\pm0.26)\times10^{27}\,\si{cm^2.s^{-1}}$, 
assuming $B_{\bot} = 3\,\si{\mu G}$. 
Intriguingly, these estimates are on a similar level as the diffusion constant estimated for $100\,\si{TeV}$-electrons in the pulsar halos of Geminga and PSR B0656+14 \citep{HAWC17}.

While the determination of the exact level of radial steepening in the non-thermal spectrum is clearly challenging due to limited statistics, we believe that the observed effect of radiative cooling in our data set is most likely physical. 
We verified this hypothesis by performing a second set of spectral fits only on the hard portion of the observed X-ray spectra, to exclude possible biases introduced in spectral modelling. This allowed us to qualitatively reproduce both the radial brightness profile of the PWN and the radial increase of the non-thermal photon index (see Appendix \ref{OtherMethodSync}).

\subsubsection{The connection to radio, GeV, and TeV emission}

The Vela X PWN exhibits a complex multiwavelength morphology. In order to visualize the contrast between different energy bands, we compare observations of the PWN at radio and TeV energies with the observed non-thermal X-ray emission in Fig.~\ref{Multiwav}. To display the large-scale low-frequency radio emission, we extracted an image in the $170-231\,\si{MHz}$ band from the public data of the Galactic and Extra-Galactic All-sky MWA survey \citep[GLEAM;][]{Hurley17, Hurley19}. TeV observations of the region are available as part of the HESS Galactic plane survey \citep{HGPS}, from which we have used the $> 1\,\si{TeV}$ flux map with a $0.1^{\circ}$ correlation radius. In order to reflect the non-thermal X-ray emission of the plerion, we used an image extracted from our data set in the $1.95-2.30\,\si{keV}$ band, above any significant line emission from Vela, but below the drop in the eROSITA's effective area. 

While at radio energies, an extended filamentary structure extending over $3^{\circ}\times2^{\circ}$ is visible \citep{Bock98}, until now, in X-rays, only the elongated cocoon, extending around $1.5^{\circ}$ south of the pulsar \citep{Slane18} was known to be powered by the pulsar wind. At $\gamma$-ray energies below around $100\,\si{GeV}$, a diffuse structure similar to the radio nebula is visible, together with a (possibly unrelated) point-like source west of the pulsar \citep{Grondin13,Tibaldo18}. In contrast, VHE $\gamma$-ray emission appears to trace the shape of the cocoon, albeit with a larger lateral extent than seen in X-rays \citep{HESS06,HESS12,HESS19}.

A two-zone leptonic model has been invoked in order to explain these multiwavelength observations \citep{deJager08}. In this model, a lower-energy electron population with a cutoff on the order of $100\,\si{GeV}$ \citep{Grondin13} produces the radio synchrotron component, whereas inverse Compton scattering is responsible for the GeV component. Analogously, the TeV and X-ray emission components originate from a more energetic component with a cutoff around $100\,\si{TeV}$ \citep{HESS19}. So far, the somewhat wider extent of TeV emission has been explained by the shorter expected lifetime of the electron population visible through X-ray synchrotron emission, as the TeV emission traces lower energies, where particles are longer-lived. 

This view is now confronted with our detection of a diffuse non-thermal X-ray nebula, which extends far beyond the dimensions of the cocoon and the established TeV emission region. Since the newly detected emission is likely solely due to synchrotron radiation, its explanation requires the presence also of an associated diffuse $\gamma$-ray PWN, which has not been observed at TeV energies, so far \citep{HESS06, HESS12}. Since the diffuse X-ray component exhibits a softer and fainter character than the cocoon, the associated electron population is likely older, and exhibits a lower cutoff energy from radiative losses \citep{HESS19}. 

It is interesting that the extent of our X-ray nebula appears to exceed even that of the radio emission associated to Vela X, which traces a less energetic, hence older, particle population. Furthermore, the diffuse X-ray component exhibits an apparent elongation in the north-south direction. This contrasts the radio emission, which is widest in the east-west direction, and is clearly centered south of the pulsar, likely due to an asymmetric interaction with the reverse shock \citep{Bock98,deJager09,Slane18}. If this discrepancy is truly caused by different spatial distributions of the underlying particle populations, the highest-energy electrons seem to have ``overcome'' the interaction with the reverse shock. This could be attributed to a comparatively recent injection by the pulsar, after the central PWN portion had already been crushed by the reverse shock. Alternatively, this may the signature of an increase in diffusivity with electron energy, preventing the long-term confinement of X-ray emitting electrons in the cocoon.  

An interesting comparison was made between Vela X and the Geminga pulsar by \citet{Fang19}. In their work, they showed that the vast TeV halo of Geminga can be explained by confinement in a turbulent slow-diffusion environment created by the shock wave of its parent SNR. They argued that a similar process may be at play in Vela X, where electrons escaped from the PWN core are diffusing through the turbulent environment inside the SNR shell, producing a relatively smooth extended synchrotron nebula.  
The topic of particle escape was addressed also by \citet{Hinton11}, who argued that the extremely soft GeV spectrum of the radio nebula can be explained by the escape of particles from Vela X. They interpreted the presumed lack of spectral variability of the cocoon in the TeV band as evidence for it being advection-dominated, preventing higher-energy electrons from easily escaping. 
Evidence for spectral variability in the X-ray band along the cocoon has since been found \citep[this work;][]{Slane18, HESS19}, showing that energy losses do in fact play a role.

In summary, there remain two fundamental questions regarding our scenario of an extended diffusive X-ray nebula: first, it is unclear whether sufficiently energetic electrons can escape from the PWN core without losing the majority of their energy. 
Second, the assumption of an ISM-level magnetic field outside the cocoon leads to sufficiently long electron lifetimes and sufficiently large diffusion distances to reach the observed extent. However, it also predicts the presence of similarly extended TeV emission from relatively energetic particles (up to $\sim100\,\si{TeV}$), in order to explain the detection of non-thermal X-ray emission above $1\,\si{keV}$, in contrast with existing data. 

We hope that our discovery of an extended non-thermal X-ray nebula can be reconciled with future observations of Vela X at TeV energies, which might reach a higher sensitivity toward extended emission. Only the combination of TeV and X-ray data \citep[similarly to][]{HESS19} can provide a full picture of the high-energy particle content, particle propagation, and magnetic field structure across the full Vela X PWN, an archetype for a PWN in an evolved SNR.

\section{Summary \label{Summary}}
The observations during the first four eROSITA all-sky surveys constitute by far the deepest X-ray data set of the whole Vela SNR acquired to date, and will likely stay so for the foreseeable future, providing a ``treasure trove'' for scientific exploitation by the community after its public release.  
In this work, we have used this data set to explore the distribution and properties of shock-heated plasma and relativistic electrons throughout the entire remnant at unprecedented spatial and spectral resolution. 
Our analysis included the dissection of emission of Vela into broad and narrow energy bands for imaging, the spatially resolved spectroscopy of over 500 independent regions across the SNR, and the dedicated investigation of several prominent morphological structures, such as the Vela shrapnels.

We found that the energy-dependent morphology of Vela exhibits at least three separate components:
the soft band is dominated by a diffuse shell and thick filaments, which are likely related to heated ISM behind the forward shock. At intermediate energies (i.e., starting at the \ion{O}{viii} lines at $0.65\,\si{keV}$), thin radial structures become visible, which are probably tracing the outward propagation of dense ejecta fragments. In the harder bands, above around $1.4\,\si{keV}$, the emission is dominated by an extended non-thermal nebula, centered on the Vela pulsar. 

Owing to the eight-degree angular extent of Vela, the foreground absorption toward the SNR was found to be inhomogeneous and highly structured. Our study reinforces previous findings of a puzzling anti-correlation between X-ray absorption, which is strongest in the south, and optical extinction as well as neutral hydrogen column density, both of which are highest in the north of Vela. A possible, yet highly speculative, solution to this contradiction may lie in dust destruction in the SNR blast wave. This process may have disintegrated the local dust grains, responsible for optical extinction, while preserving the X-ray absorption. Alternatively, a clumpy ISM in the south, possibly traced also by H$\alpha$ emission, may introduce additional absorption on top of a homogeneous background there. 

The majority of shocked ISM in Vela was found to be relatively cold, at a median temperature of $0.19\,\si{keV}$, and to show no significant deviations from CIE. If this temperature corresponds to full equilibration of the blast wave's kinetic energy, the forward shock is presently expanding into the ISM at a velocity around $400\,\si{km.s^{-1}}$.
However, a close look at the outermost region of the shell and at the bow shock of shrapnel D has revealed large hardness gradients, possibly due to recent shock-heating of the ISM. Dedicated observations of such regions could permit to trace in detail the transition between underionized and equilibrated plasma, and reconstruct the ionization history of material behind a rather slow shock.

We have found ample evidence for the presence of ejecta produced during the explosion. This includes dense ejecta fragments in the shrapnels, as well as several newly detected clumps protruding into the ISM in the south of Vela. Furthermore, significantly enhanced elemental abundances inside the shell indicate the presence of further ejecta-rich features, which are possibly located outside the shell but seen in projection. Interestingly, two regions in the vicinity of the central pulsar show similar signatures, which may point toward a recent crushing of ejecta and relativistic pulsar wind particles by secondary shocks.

The X-ray-detected ejecta signatures appear to be almost universally enriched in oxygen, neon, and magnesium, which are expected to originate from the outer ejecta layers of the progenitor star. Interestingly, in virtually all ejecta clumps, neon and magnesium are found to be strongly enhanced with respect to oxygen at about twice the solar ratio, which cannot be easily reconciled with expectations for supernova nucleosynthesis.
Silicon ejecta are encountered in several clumps (e.g., shrapnel A) inside and outside the shell, but appear to be almost absent in a few cases (e.g., shrapnel D), indicating that the X-ray-bright ejecta trace a varying mix of hydrostatically and explosively synthesized elements released during the supernova. 
In contrast to the lighter elements, no secure signature of iron ejecta was found anywhere in the SNR.

Thanks to the improved sensitivity and spectral resolution of our data set with respect to {\it ROSAT}, we were able to isolate the non-thermal contribution to the X-ray emission, revealing the vast size of the plerion of the Vela pulsar. The extended synchrotron nebula, which extends up to three degrees or $14\,\si{pc}$ from the pulsar, exceeds the PWN core by almost two orders of magnitude, both in total luminosity and size. Thus, the conversion efficiency of spin-down power into PWN X-ray luminosity is much larger than estimated for the core alone, at around $1.5\times10^{-3}$. 

The likely physical origin for this extended non-thermal emission lies in synchrotron emission of relativistic electrons from the pulsar wind. These particles have escaped confinement in the PWN core or the cocoon, and are likely transported via diffusion through the turbulent medium inside the SNR shell. 
In order to be able to reach such a large physical size within the radiative electron lifetime, a rather small magnetic field is required, around $3\,\si{\mu G}$ if diffusion occurs at a similar rate as observed in $\gamma$-ray pulsar halos. Observing non-thermal X-ray emission above $1\,\si{keV}$ thus requires the escape of electrons around $100\,\si{TeV}$ into the diffuse nebula, unless the magnetic field is significantly stronger. 
We have tentatively observed and quantified the effect of radiative energy losses on the electron population, traced by the steepening of the non-thermal X-ray spectrum toward larger distances from the pulsar, which appears to be at a level consistent with the estimated physical parameters. 

It appears puzzling that our X-ray nebula extends further away from the pulsar than the PWN seen at radio and GeV energies, which are generally expected to trace a less energetic, older electron population. Furthermore, up until now, the emission detected in the VHE $\gamma$-ray band is dominated by a comparatively small region around the cocoon, with no detected counterpart to our extended X-ray nebula, in particular in the north of the pulsar. This is potentially problematic since electrons at multi-TeV energies are required for the production of the detected X-ray synchrotron radiation, and should be visible in the TeV band via inverse Compton scattering. 
Therefore, we believe that future observations of the Vela region in the VHE $\gamma$-ray band are likely to detect the suspected population of energetic electrons in the diffuse nebula, in order to independently confirm and complete the picture of one of the largest X-ray PWNe observed to date.

\begin{acknowledgements}
We are grateful to the anonymous referee for their constructive criticism. 
We would like to thank J.~Tr\"umper, H.T.~Janka, M.~Ramos-Ceja, C.~Maitra, J.~Sanders, E.~Gatuzz, N.~Locatelli, and F.~Camilloni for fruitful discussions. 
MGFM acknowledges support by the International Max-Planck Research School (IMPRS) on Astrophysics at the Ludwig-Maximilians University. M.S. acknowledges support from the Deutsche Forschungsgemeinschaft through the grants SA 2131/13-1, SA 2131/14-1, and SA 2131/15-1.
\\
This work is based on data from eROSITA, the soft X-ray instrument aboard SRG, a joint Russian-German science mission supported by the Russian Space Agency (Roskosmos), in the interests of the Russian Academy of Sciences represented by its Space Research Institute (IKI), and the Deutsches Zentrum f\"ur Luft- und Raumfahrt (DLR). The SRG spacecraft was built by Lavochkin Association (NPOL) and its subcontractors, and is operated by NPOL with support from the Max Planck Institute for Extraterrestrial Physics (MPE). The development and construction of the eROSITA X-ray instrument was led by MPE, with contributions from the Dr. Karl Remeis Observatory Bamberg \& ECAP (FAU Erlangen-Nuernberg), the University of Hamburg Observatory, the Leibniz Institute for Astrophysics Potsdam (AIP), and the Institute for Astronomy and Astrophysics of the University of T\"ubingen, with the support of DLR and the Max Planck Society. The Argelander Institute for Astronomy of the University of Bonn and the Ludwig Maximilians Universit\"at Munich also participated in the science preparation for eROSITA.
The eROSITA data shown here were processed using the eSASS software system developed by the German eROSITA consortium.
\\
This research made use of Astropy,\footnote{\url{http://www.astropy.org}} a community-developed core Python package for Astronomy \citep{astropy:2013, astropy:2018}. Further, we acknowledge the use of the Python packages Matplotlib \citep{Hunter:2007}, SciPy \citep{SciPy}, and NumPy \citep{NumPy}. In particular, we acknowledge the use of the {\tt cubehelix} color map by \citet{Green11}.  
\\
\end{acknowledgements}
\vspace{-0.5cm}
\begingroup
\let\clearpage\relax
\bibliographystyle{aa} 
\bibliography{Citations}

\begin{appendix}
\section{Application of physical models to the emission of Vela X \label{FittingDetails}}

Here, we expand on the physically motivated toy models used to characterize the radial profiles of the nonthermal brightness and photon index of Vela X in Sect.~\ref{DiscVelaX}. 
In analogy to \citet{HAWC17}, the intensity of nonthermal emission in a pulsar halo at a single electron energy $E_{e}$, assuming a constant magnetic field and particle transport via isotropic diffusion, is proportional to
\begin{equation}
    F(\theta, E_{e}) \propto \frac{\exp \left(-\theta^2/\theta_{D}^2(E_{e}) \right )}{\theta_{D}(E_{e}) \left( \theta+0.06\,\theta_{D}(E_{e}) \right )}, 
\end{equation}
where $\theta$ describes the angular distance to the pulsar, and $\theta_{D}$ is the ``diffusion angle'', the characteristic angular size of the emission on the sky at the given energy. In the steady-state regime, it is related to the diffusion constant $D$ and electron lifetime $\tau$ following
\begin{equation}
    r_{D} = \theta_{D}\,d = \left ( 4D\tau \right ) ^{1/2}, \label{DiffRadius}
\end{equation}
where $d$ is the distance to the Vela SNR. 
With the main goal of constraining the size of the PWN at an ``effective'' X-ray-emitting energy, we thus fitted the observed nonthermal brightness profile $F$ (Fig.~\ref{Nonthermal_Radii}) with a model of the following form
\begin{equation}
    F(\theta; A, \theta_{D}, C) = \frac{A}{\theta_{D}}\,\frac{\exp \left(-\theta^2/\theta_{D}^2 \right )}{\theta+0.06\,\theta_{D}} + C,
\end{equation}
where $A$ is proportional to the integrated PWN flux and $C$ accounts for a possible spatially uniform background. 
During these fits, we included a systematic relative scatter $s$ around the radial average profile as a free parameter, so that the model likelihood is given by
\begin{equation}
    \log \,\mathcal{L} = -\frac{1}{2}\sum_{i} \left ( \log( 2\pi\rho_{i}^2) + \left ( \frac{F_{i}-F(\theta_{i}; A, \theta_{D}, C)}{\rho_{i}}\right ) ^2 \right ), 
\end{equation}
where the total error $\rho_{i}$ of each bin is obtained from its purely statistical error $\sigma_{i}$ as
\begin{equation}
    \rho_{i}^2 = \sigma_{i}^2 + s^2\,F(\theta_{i}; A, \theta_{D}, C)^2.
\end{equation}

\begin{figure}
\centering
\includegraphics[width=1.0\linewidth]{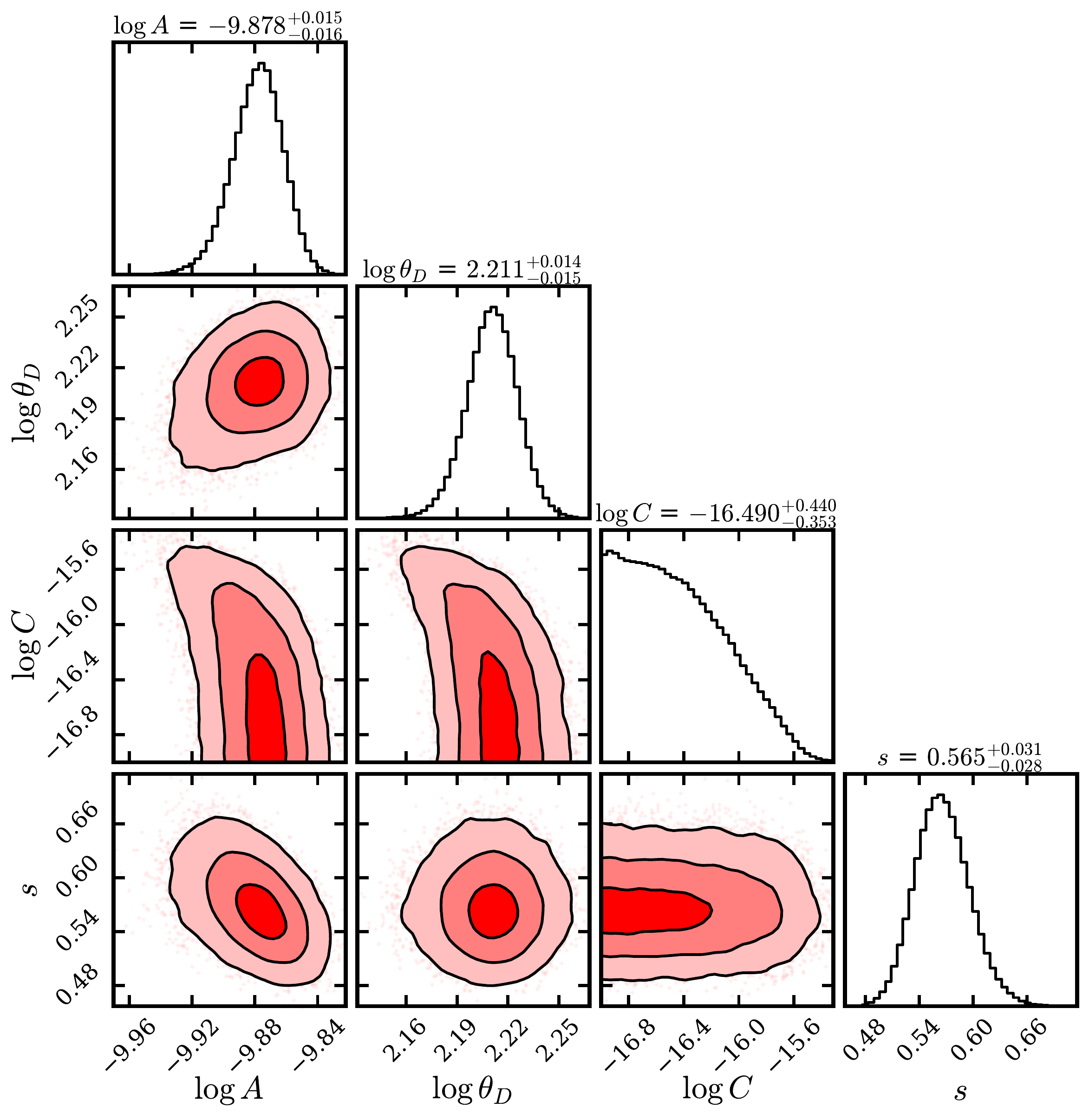}
\caption{Corner plot \citep{Foreman16} displaying the posterior distribution of the parameters of the brightness profile fit shown in Fig.~\ref{Nonthermal_Radii}. The diagonal plots display marginalized posterior distributions of the individual parameters. The contours in the off-diagonal plots correspond to $1\sigma$, $2\sigma$, $3\sigma$ constraints on the joint probability distribution of two parameters.
The units of the parameters $A$, $C$, and $\theta_{D}$ are $\si{erg.s^{-1}.cm^{-2}}$, $\si{erg.s^{-1}.cm^{-2}.arcmin^{-2}}$, and $\si{arcmin}$, respectively.}
\label{SyncFluxCoolCorner}
\end{figure}

The constraints on our model parameters were extracted via MCMC sampling with {\tt emcee} \citep{Foreman13} using logarithmically uniform priors on $A$, $\theta_{D}$, and $C$ and uniform priors on $s$. A total of 100 walkers were run for 5000 burn-in and sampling steps, respectively. This resulted in the posterior distribution displayed in Fig.~\ref{SyncFluxCoolCorner}. 

\begin{figure}
\centering
\includegraphics[width=0.75\linewidth]{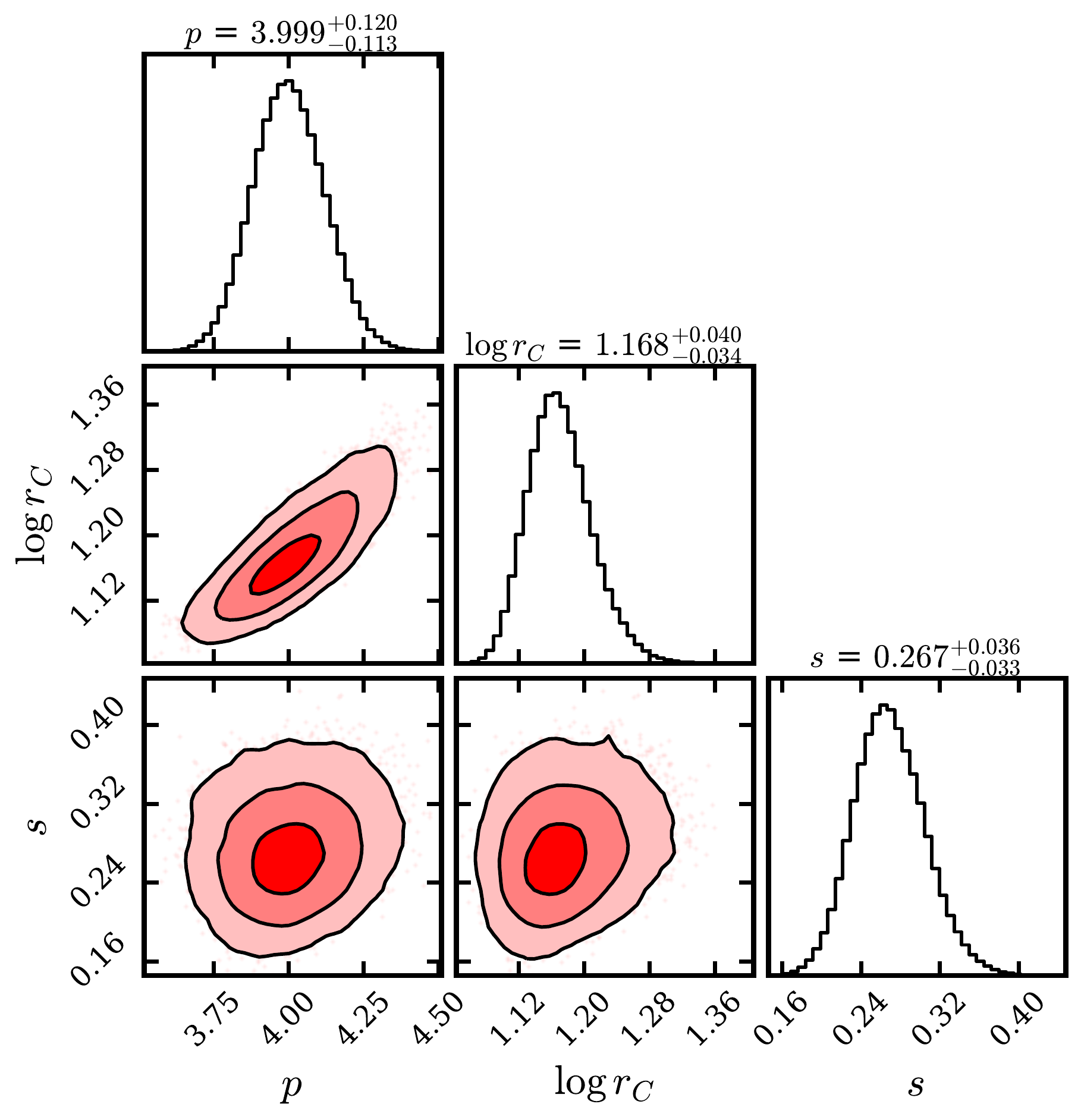} 
\caption{Same as Fig.~\ref{SyncFluxCoolCorner}, but for the fit of the radial photon index evolution shown in Fig.~\ref{Nonthermal_Radii}. 
The cooling radius $r_{C}$ is given in $\si{pc}$, whereas $p$ and $s$ are unitless.
}
\label{SyncCoolCorner}
\end{figure}

When fitting the radial profile of the photon index $\Gamma$, we found it crucial to only include those bins with a significant hard nonthermal contribution, in order to mitigate biases by the prior on $\Gamma$, or by potentially spurious power-law components fitting the soft band only. Thus, we defined the following quantity $g$, with the target of quantifying the relative contribution of the hard nonthermal component to the spectrum:   
\begin{equation}
    g = \left. \int_{1.0\,\si{keV}}^{5.0\,\si{keV}} \mathrm{d} E_{X} \,f(E_{X}) \frac{f_{\rm NT}(E_{X})}{f(E_{X})-f_{\rm NT}(E_{X})} \middle / \int_{1.0\,\si{keV}}^{5.0\,\si{keV}} \mathrm{d} E_{X}\, f(E_{X}) \right. ,
\end{equation}
where $f(E_{X})$ and $f_{\rm NT}(E_{X})$ describe the total (including backgrounds) and nonthermal forward-folded models fitted to the observed X-ray spectrum, respectively. Thus, $g$ is equivalent to a photon-flux-weighted average of the ratio of the nonthermal to all other contributions to the spectrum above $1\,\si{keV}$.
For each spectrum, we defined a corresponding statistical weight $w$ via the transformation 
\begin{equation}
    w = \frac{g}{1+g},
\end{equation}
which ranges from $0$ for no significant hard contribution to $1$ for a very dominant one.

We obtained a rough model of the effect of radiative energy loss on the nonthermal photon index in a diffusive PWN with the following approach: 
We modified the expression for the synchrotron power $P(\nu, r, t)$ at a given frequency $\nu$, radius $r$, and time $t$, described in Eqs.~3 to 5 in \citet{Tang12}, to include the effects of inverse Compton radiation. In the prescription for energy loss $\mathrm{d}E/\mathrm{d}t = -Q\, E^2$, we achieved this by setting
\begin{equation}
    Q = \left ( 2.37\times10^{-15}\,B_{\si{\mu G}}^2 + 1.65 \times10^{-14}\right ) \,\si{erg^{-1}.s^{-1}},
\end{equation}
where the first term accounts for synchrotron losses, and the second term for inverse Compton scattering on the cosmic microwave background in the Klein-Nishina regime \citep{Aharonian06}.
From this, we computed the expected power-law index $\Gamma$ at a given energy as:
\begin{equation}
    \Gamma(\theta, \nu, t) = 1 - \frac{\mathrm{d}}{\mathrm{d} \log \nu} \log \left ( \int_{0}^{\infty} \mathrm{d}z \, P(\nu, \sqrt{\theta^2 d^2+z^2}, t) \right ),
\end{equation}
where we integrated the emission profile along the line of sight $z$, and used the distance to Vela $d$ to convert from physical to angular scales. The degree of radial steepening in this model depends mostly on the quantity $r_{C}^2 \coloneqq 4D\tau \propto D/Q$, meaning a characteristic cooling radius can be defined in analogy to Eq.~\ref{DiffRadius}.
In order to compare the predicted radial profile to our observations, we fixed the PWN age at $t=20\,\si{kyr}$, and assumed $E_{X}=h\nu = 1.6 \,\si{keV}$ as a uniform effective measurement energy, as we found that this is the typical ``pivot'' energy, at which the relative uncertainty on the fitted power law component is minimal. 
The predicted model photon indices $\Gamma(\theta_{i})$ were compared to our measurements $\Gamma_{i}$ with errors $\sigma_{i}$ in each bin, via the likelihood
\begin{eqnarray}
    \log \,\mathcal{L}_{i} &=& -\frac{1}{2}\left ( \log( 2\pi\rho_{i}^2) + \left ( \frac{\Gamma_{i}-\Gamma(\theta_{i}; p, r_{C})}{\rho_{i}}\right ) ^2 \right ), \\
    \rho_{i}^2 &=& \sigma_{i}^2 + s^2.
\end{eqnarray}
This introduces the systematic error scale $s$, added in quadrature to the statistical errors, as an additional free parameter. 
Using the weights $w_{i}$ defined as above, the total model likelihood was computed as 
\begin{equation}
    \log \,\mathcal{L} = \sum_{i} w_{i}\,\log \,\mathcal{L}_{i}.
\end{equation}
In combination with uniform priors on $p$ and $s$ and a logarithmically uniform prior on $r_{C}$, 
we constrained the physical model parameters via the same MCMC approach as above. The resulting posterior distribution of the parameters is illustrated in Fig.~\ref{SyncCoolCorner}.

\section{Characterization of the nonthermal emission of Vela X through spectral fits in the hard band \label{OtherMethodSync}} 

\begin{figure}[t!]
\centering
\includegraphics[width=1.0\linewidth]{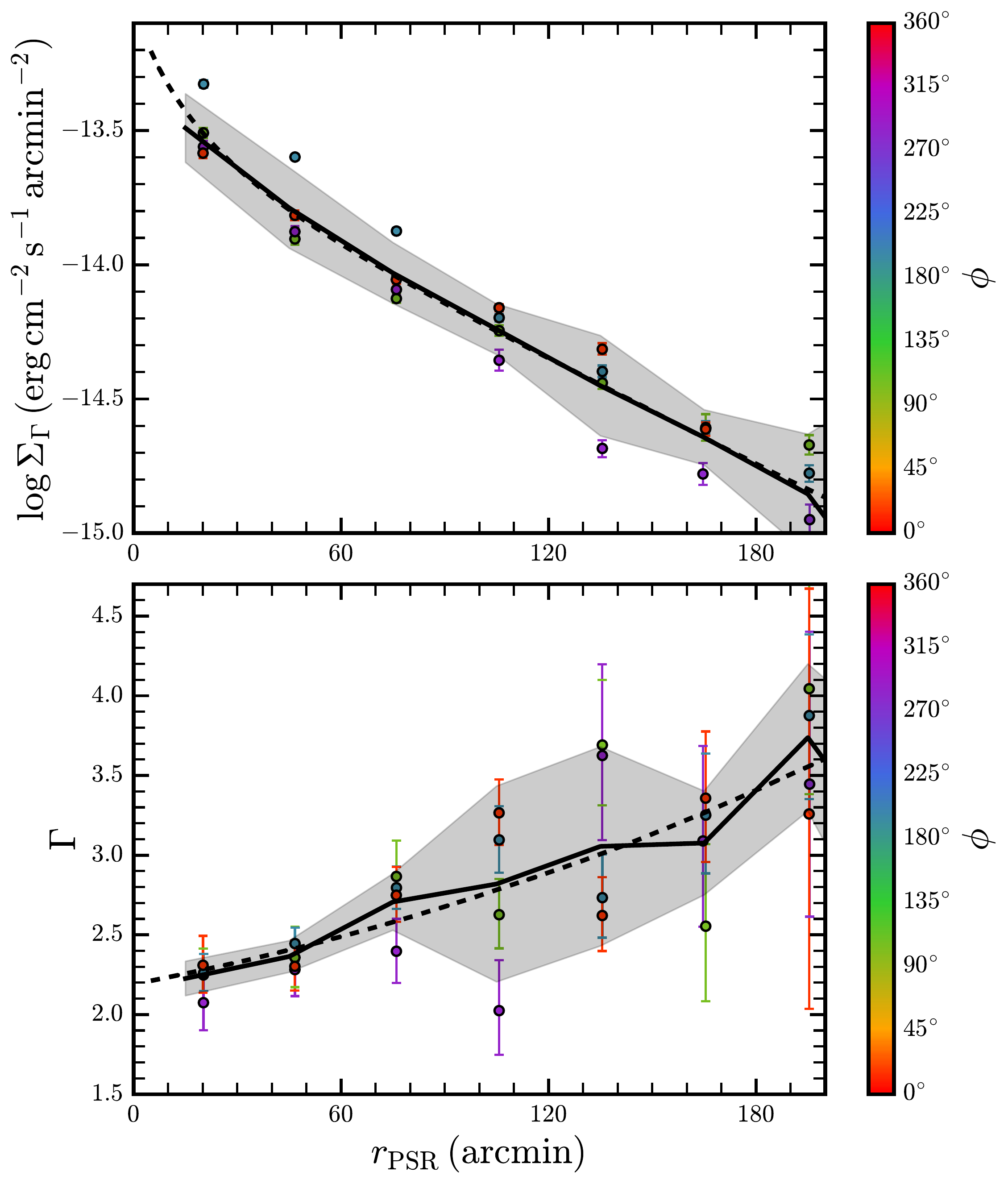} 
\caption{Same as Fig.~\ref{Nonthermal_Radii}, but displaying the results of spectral fits in the $1.43-8.50\,\si{keV}$ band, using the approach described here.}
\label{Nonthermal_Radii_LineTail}
\end{figure}

\renewcommand{\thefigure}{C.\arabic{figure}}
\setcounter{figure}{0}

\begin{figure*}[t!]
\centering
\includegraphics[width=18cm]{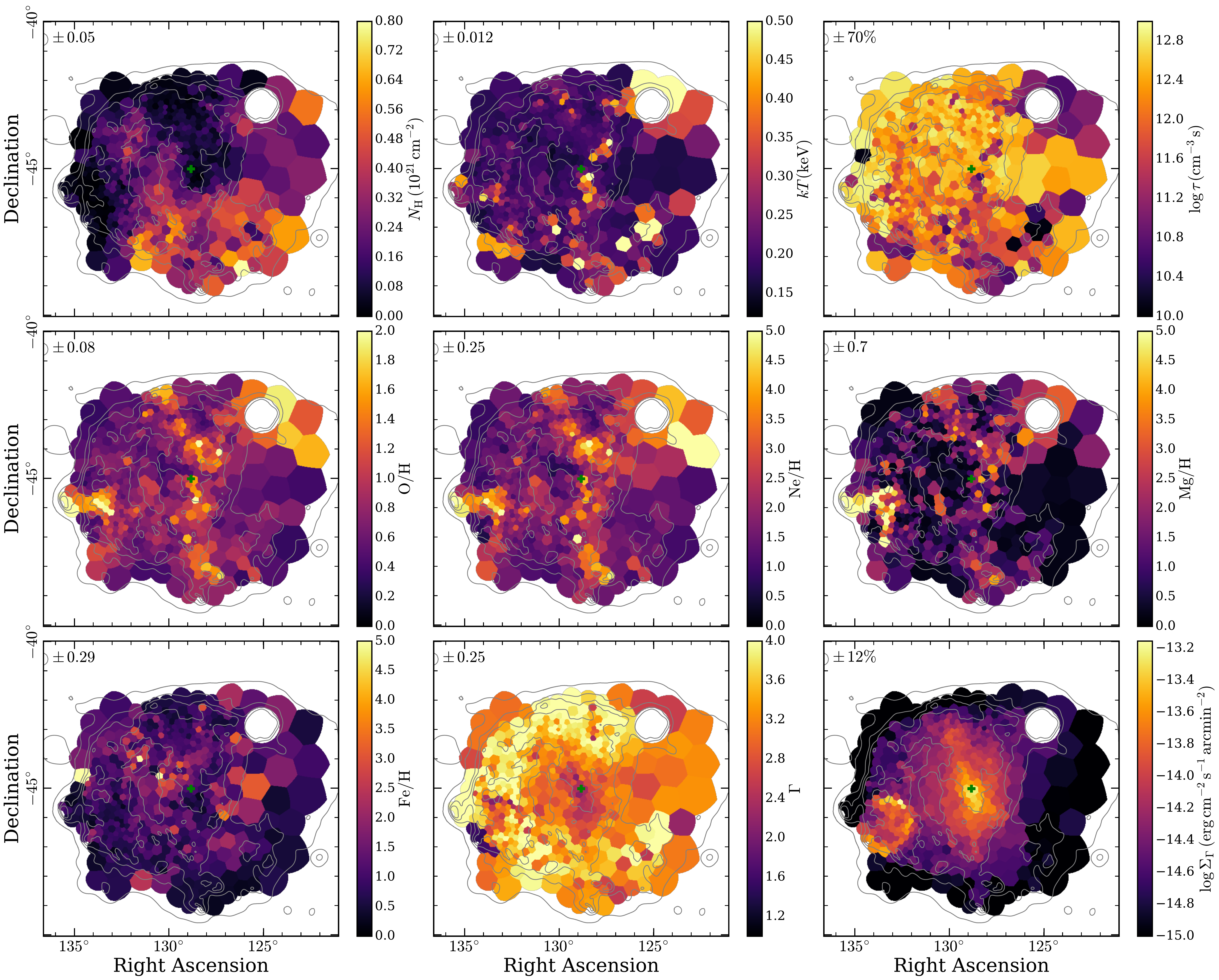} 
\caption{Same as Fig.~\ref{Spectroscopy_2TNT}, but for the TNT model, with most panels displaying analogous quantities. 
However, the upper right panel displays the ionization age $\tau$, corresponding to the degree of departure of the thermal plasma from CIE.
}
\label{Spectroscopy_TNT}
\end{figure*}

Here, we reproduce the results discussed in Sect.~\ref{DiscVelaX} regarding the properties of nonthermal X-ray emission of the extended Vela X PWN, using a more ``traditional'' method: rather than simultaneously modelling the thermal and nonthermal contributions to the broad-band spectra, we restricted our analysis to a relatively simple spectral fit to the hard band, which we defined to begin above the thermal emission lines from the \ion{Mg}{xi} triplet, so that it extends over the range $1.43-8.50\,\si{keV}$. 
In order to obtain similar levels of statistics in each region, we extracted spectra from concentric annuli, centered on the pulsar, radially spaced by $30\arcmin$, and each divided into four sectors of equal size. The same masks as described in Sect.~\ref{Spectroscopy} were applied to the regions, with the additional exclusion of a $1.2^{\circ}$ radius around the center of Vela Jr. 
Despite the weak nature of thermal continuum emission in the employed energy band, we found it necessary to include a Gaussian emission line in our model, to account for \ion{Si}{xiii} line emission around $1.85\,\si{keV}$, such that our complete source model was expressed as {\tt powerlaw+gaussian}. 
Since the soft band was excluded here, a uniform prior on the power-law photon index was found sufficient. The rest of our methodology, including background treatment and parameter estimation, was identical to Sect.~\ref{Spectroscopy}.

Figure \ref{Nonthermal_Radii_LineTail} illustrates the results of this approach, including radial averages and model fits as in Sect.~\ref{DiscVelaX}. 
The radial brightness profile reproduces the results shown in Fig.~\ref{Nonthermal_Radii} quite well, regarding both level and slope of the outward-declining flux profile. The characteristic size of the nebula matches our results from the fits of the full energy band, as is indicated by the best-fit diffusion angle of $\theta_{D}=167_{-25}^{+17}\,\si{arcmin}$.
The photon index $\Gamma$ fitted by our model exhibits a significant radial increase, at a similar average slope as in our fits of the full spectra (Fig.~\ref{Nonthermal_Radii}), 
which is characterized by the resulting characteristic cooling radius of $r_{C} = 15.3^{+2.4}_{-1.6}\,\si{pc}$. 
This confirms our observation of radiative losses in the extended nebula with a completely independent approach applied to the same data set.
It is noteworthy however that, apart from the innermost bin, the fitted power-law slopes are on average lower than in our fits of the full spectral range, as is reflected also by the lower electron spectral index $p=3.38\pm0.14$ in our diffusion model. This is somewhat unexpected, since a possible contaminating high-energy tail of thermal emission from the Vela SNR should lead to an increased inferred spectral slope, as its bremsstrahlung continuum will most likely decrease with energy faster than true nonthermal emission.     
Hence, while the large extent of nonthermal emission and the presence of radiative losses affecting the emitting electron population appear quite certain, a more quantitative characterization of its energy loss likely requires a significantly deeper coverage in the hard band than available here.

\section{Spectral parameter maps for the TNT model}\label{App:TNT}

\end{appendix}
\endgroup

\end{document}